\documentclass[%
 reprint,
superscriptaddress,
 amsmath,amssymb,
 aps,
floatfix
]{revtex4-2}

\usepackage{graphicx}
\usepackage{dcolumn}
\usepackage{bm}
\usepackage{siunitx}
\usepackage{float}
\usepackage{multirow}
\usepackage{color}
\usepackage{enumitem}
\usepackage{mathtools}
\usepackage{hyperref}

\newcommand{\nocontentsline}[3]{}
\newcommand{\tocless}[2]{\vspace{0.1in}\bgroup\let\addcontentsline=\nocontentsline#1{#2}\egroup\vspace{-0.1in}}
\newcommand{\toclesslab}[3]{\vspace{0.1in}\bgroup\let\addcontentsline=\nocontentsline#1{#2\label{#3}}\egroup\vspace{-0.1in}}

\newcommand{\sumsign}[1]{{\textstyle \sum\limits_{#1}}}
\newcolumntype{P}[1]{>{\centering\arraybackslash}p{#1}}
\newcolumntype{C}{>{$}c<{$}} 

\newcommand{\markblue}[1]{{\color{black}#1}}
\newcommand{\am}[1]{{\color{black}#1}}
\newcommand{\jd}[1]{{\color{black}#1}}

\begin{document}

\title{Learning hydrodynamic equations for active matter \\ from particle simulations and experiments}

\author{Rohit Supekar}
\affiliation{
Department of Mechanical Engineering, Massachusetts Institute of Technology, 
77 Massachusetts Avenue, Cambridge, MA 02139
}
\affiliation{
Department of Mathematics, Massachusetts Institute of Technology, 
77 Massachusetts Avenue, Cambridge, MA 02139
}
\author{Boya Song}
\affiliation{
Department of Mathematics, Massachusetts Institute of Technology, 
77 Massachusetts Avenue, Cambridge, MA 02139
}
\author{Alasdair Hastewell}
\affiliation{
Department of Mathematics, Massachusetts Institute of Technology, 
77 Massachusetts Avenue, Cambridge, MA 02139
}
\author{\markblue{Gary P. T. Choi}}
\affiliation{
Department of Mathematics, Massachusetts Institute of Technology, 
77 Massachusetts Avenue, Cambridge, MA 02139
}
\author{Alexander Mietke}
\email{amietke@mit.edu}
\affiliation{
Department of Mathematics, Massachusetts Institute of Technology, 
77 Massachusetts Avenue, Cambridge, MA 02139
}
\author{J\"orn Dunkel}
\email{dunkel@mit.edu}
\affiliation{
Department of Mathematics, Massachusetts Institute of Technology, 
77 Massachusetts Avenue, Cambridge, MA 02139
}

\date{\today}

\begin{abstract}
Recent advances in high-resolution imaging techniques and particle-based simulation methods have enabled the precise microscopic characterization of collective dynamics in various biological and engineered active matter systems. In parallel, data-driven algorithms for learning interpretable continuum models have shown promising potential for the recovery of underlying partial differential equations (PDEs) from continuum simulation data. By contrast, learning macroscopic hydrodynamic equations for active matter directly from experiments or particle simulations remains a major challenge. Here, we present a framework that leverages spectral basis representations and sparse regression algorithms to discover PDE models from microscopic simulation and experimental data, while incorporating the relevant physical symmetries. We illustrate the practical potential through applications to a chiral active particle model mimicking swimming cells and to recent micro-roller experiments. In both cases, our scheme learns hydrodynamic equations that reproduce quantitatively the self-organized collective dynamics observed in the simulations and experiments. This inference framework makes it possible to measure a large number of hydrodynamic parameters in parallel and directly from video data. 
\end{abstract}
\maketitle

\toclesslab\section{Introduction}{Sec:Intro}

Natural and engineered active matter, from cells~\cite{Tambe2011}, tissues~\cite{Heisenberg2013} and organisms~\cite{Tennenbaum2016} to self-propelled particle suspensions~\cite{Geyer2018a,Soni2019} and autonomous robots~\cite{Rubenstein2014, Nash2015, Savoie2019}, exhibits complex dynamics across a wide range of length and time scales.
Predicting the collective self-organization and emergent behaviors of such systems requires extensions of traditional theories that go beyond conventional physical descriptions of non-living matter~\cite{Toner1995,Marchetti2013,Julicher2018}. Due to the inherent complexity of active matter interactions in multi-cellular communities~\cite{Hartmann2019, Li2020} and organisms~\cite{Shah2019}, or even non-equilibrium chemical \cite{Cira2015} or colloidal~\cite{Geyer2018a,Soni2019, Rogers2016} systems, it becomes increasingly difficult and inefficient for humans to formulate and quantitatively validate continuum theories from first principles.  A key question is therefore whether one can utilize computing machines~\cite{Cichos2020} to identify interpretable systems of equations that elucidate the mechanisms underlying collective active matter dynamics. 

\par

Enabled by recent major advances in microscopic imaging~\cite{Stelzer2015,Power2017,Hartmann2019,Shah2019} and agent-based computational modeling~\cite{Shaebani2020}, active matter systems can now be observed and analyzed at unprecedented spatiotemporal~\cite{Jeckel2019, Qin2020, Hartmann2021} resolution. To infer interpretable predictive theories, the high-dimensional data recorded in experiments or simulations have to be compressed and translated into low-dimensional models. Such learned models must faithfully capture the macroscale dynamics of the relevant collective properties. Macroscale properties can be efficiently encoded through hydrodynamic variables, continuous fields that are linked to the symmetries and conservation laws of the underlying microscopic system~\cite{Marchetti2013,Julicher2018}. Although much theoretical progress has been made in the field of dynamical systems learning over the last two decades~\cite{Vallette1997, Bar1999,Bongard2007,Schmidt2009,Brunton2016, Rudy2017a,Maddu2021}, the inference of hydrodynamic models from particle data has remained largely unsuccessful in practice, not least due to severe complications arising from measurement noise, inherent fluctuations and self-organized scale-selection in active systems. Yet, extrapolating the current experimental revolution~\cite{ Stelzer2015,Power2017,Geyer2018a,Soni2019,Hartmann2019, Li2020}, data-driven equation learning will become increasingly more important as simultaneous observations of physical, biological, and chemical properties of individual cells and other active units will become available in the near future~\cite{Linghu2020, Cermak2020}.  

\par 
Learning algorithms for ordinary differential equations (ODEs) and partial differential equations (PDEs) have been proposed and demonstrated based on least-squares fitting~\cite{Vallette1997, Bar1999}, symbolic regression~\cite{Bongard2007, Schmidt2009}, and sparse regression~\cite{Brunton2016, Rudy2017a} combined~\markblue{with weak formulations~\cite{Reinbold2019, Gurevich2019, Reinbold2020, Reinbold2021}, artificial neural networks~\cite{Champion2019b,Both2020,Raissi2019,Rackauckas2020,Shankar2020}}, and stability selection~\cite{Maddu2019, Maddu2021}. These groundbreaking studies, however, focused primarily on synthetic data from \textit{a priori} known continuum models, and recent coarse-graining applications have remained limited to ODEs~\cite{Nardini2021} or one-dimensional PDEs~\cite{Felsberger2019,Bakarji2020}. By contrast, it is still an open challenge to infer higher-dimensional hydrodynamic PDE models directly from microscopic active matter simulations or experiments.

\begin{figure*}
\includegraphics[width=0.95\textwidth]{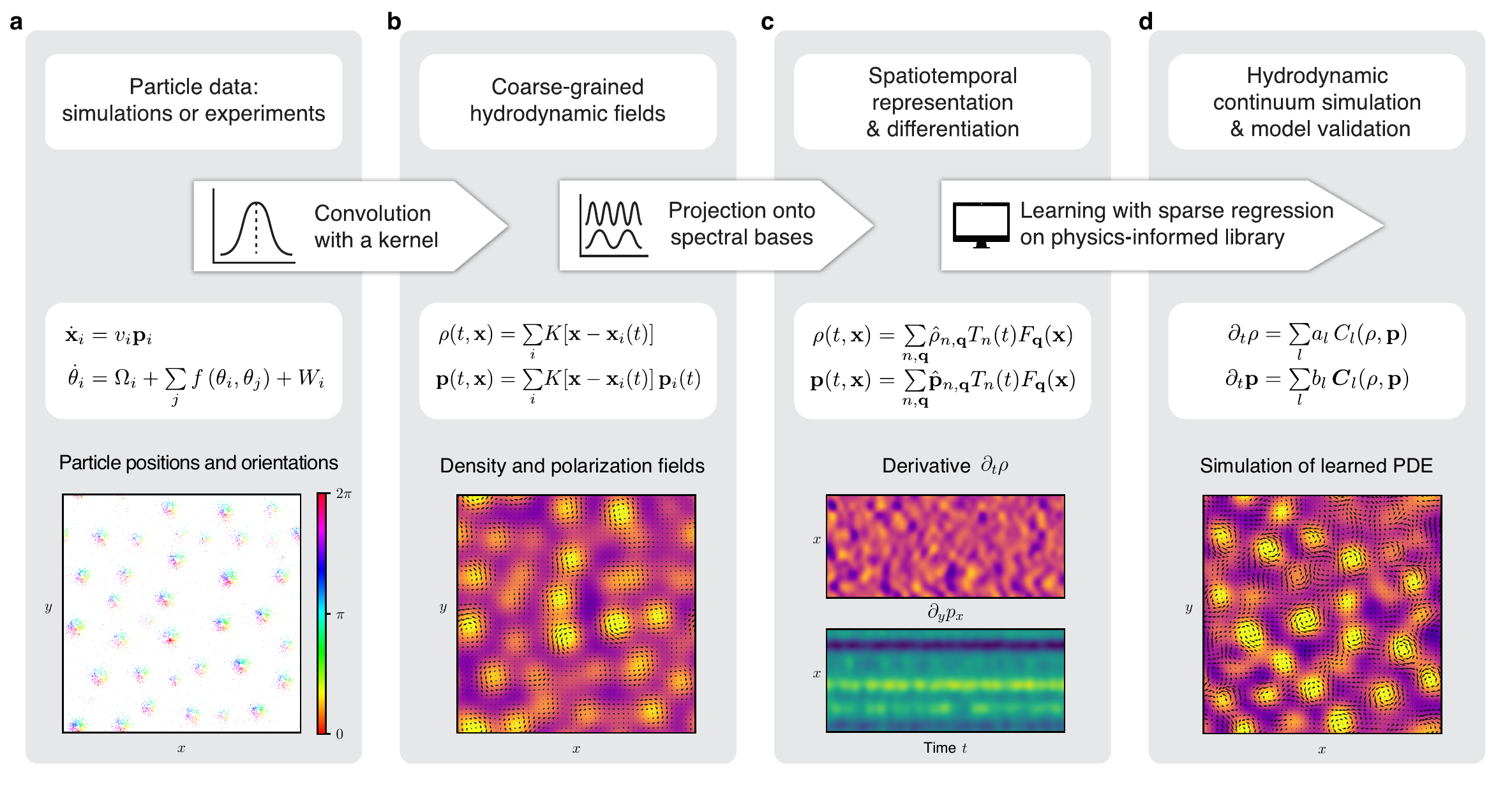}
\caption{Learning hydrodynamic models from particle simulations and experiments. {\textbf{a},}~Inputs are time-series data for particle positions~$\mathbf{x}_i(t)$, particle orientations ~$\mathbf{p}_i(t)=(\cos\theta_i,\sin\theta_i)^\top$, etc., measured in simulations or experiments with microscale resolution~(Sec.~\ref{Sec:MicroModel}). {\textbf{b},}~Spatial kernel coarse-graining of the discrete microscopic variables provides continuous hydrodynamic fields, such as the density $\rho(t,\mathbf{x})$ or the polarization density $\mathbf{p}(t, \mathbf{x})$~(Sec.~\ref{Sec:CoarseGraining}). {\textbf{c},}~Coarse-grained fields are sampled on a spatiotemporal grid and projected onto suitable spectral basis functions. Systematic spectral filtering (compression) ensures smoothly interpolated hydrodynamic fields, enabling efficient and accurate computation of spatiotemporal derivatives~(Sec.~\ref{Sec:Projection}). {\textbf{d},}~Using these derivatives, a library of candidate terms $C_l(\rho,\mathbf{p})$ and $\boldsymbol{C}_l(\rho,\mathbf{p})$ consistent with prior knowledge about conservation laws and broken symmetries is constructed. A sparse regression algorithm determines subsets of relevant phenomenological coefficients~$a_l$~and~$b_l$~(Sec.~\ref{Sec:LearningFramework}). The resulting hydrodynamic models are sparse and interpretable, and their predictions can be directly validated against analytic coarse-graining results~(Sec.~\ref{Sec:LearnedEqs}) or experiments~(Sec.~\ref{sec:AplExpData}). Bottom: Snapshots illustrating the workflow for microscopic data generated from simulations of chiral active Brownian particles~[Eq.~(\ref{eqn:EOM})].}
\label{Fig:Schematic}
\end{figure*}

Here, we present a comprehensive learning framework that takes microscopic particle data as input and generates sparse interpretable predictive hydrodynamic models as output~(Fig.~\ref{Fig:Schematic}). We demonstrate its practical potential in applications to active particle data from simulations and recent experiments~\cite{Geyer2018a}. In both cases, we find that the learned hydrodynamic models predict the emergent collective dynamics not only qualitatively but also quantitatively. Conceptually, this advance is made possible by leveraging spectral basis representations~\cite{Schaeffer2017} for systematic denoising and robust numerical differentiation. Our analysis further shows how insights from analytic coarse-graining calculations and prior knowledge of conservation laws and broken symmetries can enhance the robustness of automated equation discovery from microscopic data. 

\toclesslab\section{Learning framework}{sec:learnmain}

Our model learning approach combines recent advances in sparse PDE recovery~\cite{Rudy2017a, Maddu2019} with spectral filtering and compression~\cite{Smith2013, Driscoll2014, Burns2020}. We first demonstrate the key steps of the general framework~(Fig.~\ref{Fig:Schematic}) for an experimentally motivated chiral active particle model, for which the hydrodynamic continuum equations were not known previously. Later on, we will apply the same methodology to infer a quantitative hydrodynamic model directly from video data recorded in recent microroller experiments~\cite{Geyer2018a}~(Sec.~\ref{sec:AplExpData}). 

\toclesslab\subsection{Active particle simulations}{Sec:MicroModel}

To generate challenging test data for the learning algorithm, we simulated a 2D system of interacting self-propelled chiral particles~\cite{Peruani2008,Farrell2012,Liebchen2016,Liebchen2017,Kruk2020}. Microscopic models of this type are known to capture essential aspects of the experimentally observed self-organization of protein filaments~\cite{Sumino2012,Huber2018}, bacterial swarms~\cite{Jeckel2019,Li2019,Chate2020} and cell monolayers~\cite{Giavazzi2017}. In the simulations, a particle $i$ with orientation \hbox{$\mathbf{p}_i=(\cos \theta_i, \sin \theta_i)^\top$} moved and changed orientation according~to the Brownian dynamics
\begin{subequations}
\begin{eqnarray}
\frac{d\mathbf{x}_i}{dt} & = & v_i \mathbf{p}_i, \label{eqn:drdt} \\
\frac{d\theta_i}{dt} & = & \Omega_i + g\sumsign{j \in \mathcal{N}_i} \sin (\theta_j - \theta_i) + \sqrt{2 D_r} \eta_i, \label{eqn:dthetadt}
\end{eqnarray} \label{eqn:EOM}
\end{subequations}
where $\eta_i(t)$ denotes orientational Gaussian white noise, with zero mean and $\langle\eta_i(t)\eta_j(t')\rangle=\delta_{ij}\delta(t-t')$, modulated by the rotational diffusion constant $D_r$. The parameter $g>0$ determines the alignment interaction strength between particles $i$ and $j$ within a neighbourhood $\mathcal{N}_i$ of radius $R$. The self-propulsion speed $v_i\ge0$ and orientational rotation frequency $\Omega_i\ge0$ were drawn from a joint distribution $p(v_i,\Omega_i)$ (SI~Sec.~\ref{App:PartSims}). This heuristic distribution was chosen such that long-lived vortex states, similar to those observed in swimming sperm cell suspensions~\cite{Riedel2005}, formed spontaneously from arbitrary random initial conditions~(Fig.~\ref{Fig:MassCons}a). Emerging vortices are left-handed for $\Omega_i\ge0$, and their typical size is $\sim\langle v_i\rangle_p/\langle\Omega_i\rangle_p$, where $\langle\cdot\rangle_p$ denotes an average over the parameter distribution $p(v_i,\Omega_i)$. We simulated Eq.~(\ref{eqn:EOM}) in non-dimensionalized form, choosing the interaction radius $R$ as reference length and $R/\langle v_i\rangle_p$ as time scale. Accordingly, we set $R=1$ and $\langle v_i\rangle_p=1$ from now on. Simulations were performed for \hbox{$N=12,000$} particles on a periodic domain of size $100 \times 100$~(Fig.~\ref{Fig:MassCons}a).
\par
From a learning perspective, this model poses many of the typical challenges that one encounters when attempting to infer hydrodynamic equations from active matter experiments: spontaneous symmetry breaking and meso-scale pattern formation, microscopic parameter variability, noisy dynamics, anisotropic interactions, and so on. Indeed, similar to many experimental systems, it is not even clear \textit{a priori} whether or not Eqs.~\eqref{eqn:EOM} permit a quantitative description in terms of a sparse hydrodynamic continuum model.  
 
\toclesslab\subsection{Hydrodynamic fields}{Sec:CoarseGraining}

Given particle-resolved data, hydrodynamic fields are obtained by coarse-graining. A popular coarse-graining approach is based on convolution kernels~\cite{Solon2018,Wallin2020}, weight functions that translate discrete fine-grained particle densities into continuous fields, analogous to the point spread function of a microscope. For example, given the particle positions $\mathbf{x}_i(t)$ and orientations $\mathbf{p}_i(t)$, an associated particle number density field $\rho(t, \mathbf{x})$ and polarization density field $\mathbf{p}(t, \mathbf{x})$ can be defined by 
\begin{subequations}
\begin{eqnarray}
\rho(t, \mathbf{x}) & = & \sumsign{i} K[\mathbf{x} - \mathbf{x}_i(t)],\label{eq:densCG}\\
\mathbf{p}(t, \mathbf{x}) & = &\sumsign{i} K[\mathbf{x} - \mathbf{x}_i(t)]\,  \mathbf{p}_i(t).\label{eq:pCG}
\end{eqnarray}
\label{Eq:CoarseGrain}
\end{subequations}
The symmetric kernel $K(\mathbf{x})$ is centered at $\mathbf{x}=0$ and normalized, $\int d^2 \mathbf{x}\; K(\mathbf{x})  = 1$, so that the total number of particles is recovered from $\int  d^2 \mathbf{x}\, \rho(t, \mathbf{x}) = N$. Equations~\eqref{Eq:CoarseGrain} generalize to higher tensorial density fields in a straightforward manner, and can be readily adapted to accommodate different boundary conditions~(SI~Sec.~\ref{App:CG_nearB}). 
\par
\begin{figure*}
\includegraphics[width=\textwidth]{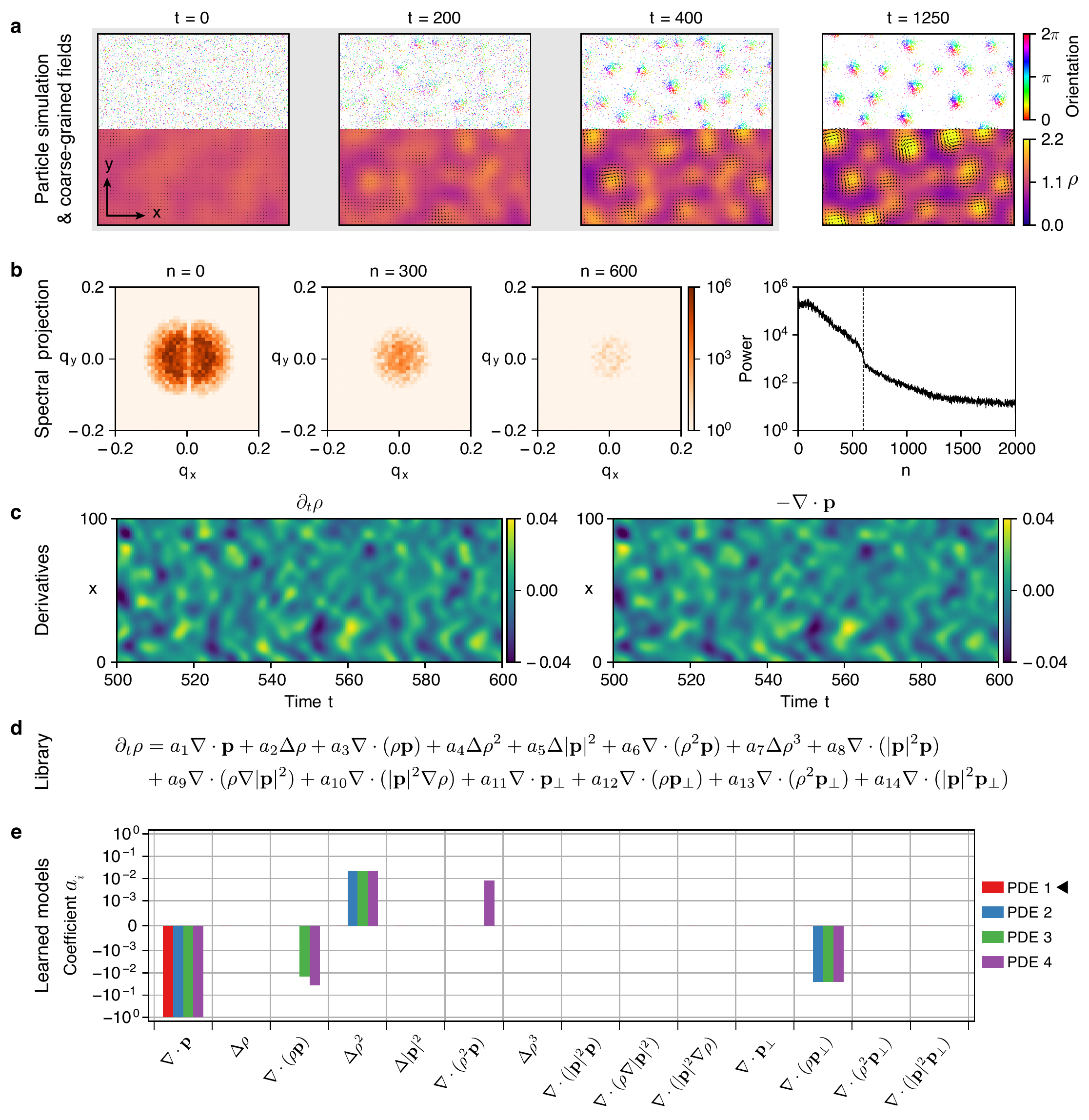}
\caption{Learning mass conservation dynamics. {\textbf{a},}~Top: Time evolution of positions and orientations of 12,000 particles following the dynamics in Eqs.~\eqref{eqn:EOM}. Bottom: Coarse-grained density $\rho$ (color code) and polarization field $\mathbf{p}$ (arrows). Starting from random initial conditions ($t=0$), a long-lived vortex pattern with well-defined handedness emerges ($t=1250$). Training data were randomly sampled from the time window $t\in[40, 400]$, enclosed within the gray box. Domain size: $100 \times 100$. {\textbf{b},}~Slices through the spatio-temporal power spectrum $S_{x;n,\mathbf{q}}=\vert\mathbf{e}_x\cdot\hat{\mathbf{p}}_{n, \mathbf{q}}\vert^2$ for different values of the Chebyshev polynomial order $n\in\{0,300,600\}$, corresponding to modes with increasing temporal frequencies. The rightmost panel depicts the total spatial spectral power $\sum_{\mathbf{q}}S_{x;n,\mathbf{q}}$ [see Eq.~(\ref{eq:Pproj})] of each Chebyshev mode~$n$. The slowly decaying long tail of fast modes indicates a regime in which fluctuations dominate over a smooth signal. The cut-off $n_0 = 600$ removes these modes, in line with the goal to learn a hydrodynamic model for the slow long-wavelength modes. {\textbf{c},}~Kymographs of the spectral derivatives $\partial_t \rho$ and $- \nabla \cdot \mathbf{p}$ at $y=50$, obtained from the spectrally truncated data. {\textbf{d},}~Mass conservation in the microscopic system restricts the physics-informed candidate library to terms that can be written as divergence of a vector field. {\textbf{e},}~Learned phenomenological coefficients $a_l$ of PDEs with increasing complexity (decreasing sparsity) (SI~Sec.~\ref{App:LearnedCoeffs}). PDE~1~($\blacktriangleleft$) is given by $\partial_t \rho = a_1 \nabla \cdot \mathbf{p}$ with $a_1 = -0.99$. As PDE~1 is the sparsest PDE that agrees well with analytic coarse-graining results~(Tab.~\ref{Tab:VortexModelCoeffs}), it is selected for the hydrodynamic model.}
\label{Fig:MassCons}
\end{figure*}

We found that, in the context of hydrodynamic model learning, the coarse-graining~\eqref{Eq:CoarseGrain} with a Gaussian kernel $K(\mathbf{x}) \propto \exp[- |\mathbf{x}|^2/(2 \sigma^2)]$ presents a useful preprocessing step that simplifies the use of fast transforms at later stages. \jd{The coarse-graining  scale~$\sigma$ determines the spatial resolution of the hydrodynamic theory. In practice, $\sigma$ must be chosen larger than the particles' mean-free path length or interaction scale, to ensure smoothness of the hydrodynamic fields but also smaller than the emergent collective structures. In accordance with these requirements, we fixed $\sigma = 5$ for the microscopic test data from Eqs.~\eqref{eqn:EOM} (Fig.~\ref{Fig:MassCons}a, SI~Fig.~\ref{Fig:SpecEntrVortex}). Interestingly, measuring the  spectral entropy as a function of $\sigma$ for both simulated and experimental data showed that  coarse-grained hydrodynamic fields typically maintain only about $1\%$ of the spectral  information contained in the fine-grained particle data~(SI~Sec.~\ref{App:InfCont}, SI ~Figs.~\ref{Fig:SpecEntrVortex} and \ref{Fig:SpecEntrQuincke}).}

\toclesslab\subsection{Spatiotemporal representation and differentiation}{Sec:Projection}

A central challenge in PDE learning is the computation of spatial and temporal derivatives of the coarse-grained fields. Our framework exploits that hydrodynamic models aim to capture the long-wavelength dynamics of the slow collective modes~\cite{Marchetti2013}. This fact allows us to project the coarse-grained fields on suitable basis functions that additionally enable sparse representations (high compression), fast transforms and efficient differentiation. Here, we work with representations of the form
\begin{subequations}
\begin{eqnarray}
\rho(t, \mathbf{x}) &= \sumsign{n, \mathbf{q}} \hat{\rho}_{n, \mathbf{q}} T_n(t) F_{\mathbf{q}}(\mathbf{x}),\\
\mathbf{p}(t, \mathbf{x}) &= \sumsign{n, \mathbf{q}} \hat{\mathbf{p}}_{n, \mathbf{q}} T_n(t) F_{\mathbf{q}}(\mathbf{x}),\label{eq:Pproj}
\end{eqnarray}
\label{Eq:BasisProjection}
\end{subequations}             
where $T_n(t)$ denotes a degree-$n$ Chebyshev polynomial of the first kind~\cite{Boyd2001, Mason2002}, $F_\mathbf{q} (\mathbf{x}) = \exp (2 \pi i\mathbf{q} \cdot \mathbf{x})$ is a Fourier mode with wave vector $\mathbf{q} = (q_x, q_y)^\top$, and $\hat{\rho}_{n, \mathbf{q}}$ and $\hat{\mathbf{p}}_{n, \mathbf{q}}$ are complex mode coefficients~(Fig.~\ref{Fig:MassCons}b and SI~Sec.~\ref{App:SpecRep}). Generally, the choice of the basis functions should be adapted to the spatiotemporal boundary conditions of the microscopic data~(Sec.~\ref{sec:AplExpData}). 

\par 

The spectral representation~\eqref{Eq:BasisProjection} enables the efficient and accurate computation of space and time derivatives~\cite{Bruno2012}. Preprocessing via spatial coarse-graining (Sec.~\ref{Sec:CoarseGraining}) ensures that the mode coefficients $\hat{\rho}_{n, \mathbf{q}}$ and $\hat{\mathbf{p}}_{n, \mathbf{q}}$ decay fast for $|\mathbf{q}|\gg 1/(2\pi\sigma)$ (Fig.~\ref{Fig:MassCons}b, left). \jd{If the asymptotic decay of the mode amplitudes with the temporal mode number $n$ is at least exponential then deterministic PDE descriptions are sufficient, whereas algebraically decaying temporal spectra indicate that stochastic PDEs may be required to capture essential aspects of the coarse-grained dynamics. For the simulated and experimental systems considered in this work, temporal spectra were found to decay exponentially (Fig.~\ref{Fig:MassCons}b) or super-exponentially (SI~Fig.~\ref{Fig:TempSpectraQuinckeFish}), suggesting the existence of deterministic PDE-based hydrodynamic models.  To infer such models from data}, we focus on the slow hydrodynamic modes and filter out the fast modes with~$n>n_0$ by keeping only the dominant Chebyshev terms in Eq.~\eqref{Eq:BasisProjection}. The cut-off value~$n_0$ can usually be directly inferred from a characteristic steep drop-off in the power spectrum of the data, which signals the transition to hydrodynamically irrelevant fast fluctuations~\cite{Aurentz2017}~(Fig.~\ref{Fig:MassCons}b, right). Choosing~$n_0$ according to this criterion yields accurate, spatiotemporally consistent derivatives as illustrated for the kymographs of the derivative fields $\partial_t\rho$ and $-\nabla\cdot\mathbf{p}$, which are essential to capture mass conservation. More generally, combining kernel-based and spectral coarse-graining also mitigates measurement noise, enabling a direct application to experimental data~(Sec.~\ref{sec:AplExpData}). 

\toclesslab\subsection{Inference of hydrodynamic equations}{Sec:LearningFramework}

To infer hydrodynamic models that are consistent with the coarse-grained projected fields~\eqref{Eq:BasisProjection}, we build on a recently proposed sparse regression framework~\cite{Brunton2016, Rudy2017a}. The specific aim is to determine sparse PDEs for the density and polarization dynamics of the form
\begin{subequations}
\begin{eqnarray}
\partial_t \rho &=& \sumsign{l} a_l \,C_l(\rho, \mathbf{p}), \label{Eqn:RhoPDELibrary}\\
\partial_t \mathbf{p} &=& \sumsign{l} b_l\, \boldsymbol{C}_l(\rho, \mathbf{p}).\label{Eqn:pPDELibrary}
\end{eqnarray}\label{Eqn:PDELibrary}
\end{subequations}
\am{Additional dynamic equations and libraries can be added to Eqs.~(\ref{Eqn:PDELibrary}) if, for example, higher-rank orientational order-parameters fields (such as a $\mathbf{Q}$-tensor field describing spatio-temporal nematic order; see SI Sec.~\ref{App:AnalyticCoarse2}) are dynamically relevant and can be extracted from microscopic data. For  self-propelled polar systems, the relaxation of higher-rank hydrodynamic fields is typically fast compared to the relaxation of the polar orientation field~\cite{Bertin2009}. In this case,  higher-rank tensorial fields are dynamically less relevant and can often be approximated by lower-rank fields and their derivatives through theoretically or empirically motivated closure relations~\cite{Toner1995,Farrell2012,Marchetti2013}. Accordingly, for the active particle data considered here, $\rho$ and $\mathbf{p}$ present a natural choice for the hydrodynamic  variables in a minimal mean-field description. This rationale is supported by generic analytic coarse-graining arguments (SI~Sec.~\ref{App:AnalyticCoarse}) which also suggest first-order-in-time dynamics as described by Eqs.~(\ref{Eqn:PDELibrary}).}

The candidate library terms $\left\{C_l(\rho, \mathbf{p})\right\}$ and $\left\{\boldsymbol{C}_l(\rho, \mathbf{p})\right\}$ are functions of the fields and their derivatives, which can be directly evaluated at various sample points using the spectral representation~\eqref{Eq:BasisProjection}. Equations~\eqref{Eqn:PDELibrary} thus define a linear system for the phenomenological coefficients $a_l$ and $b_l$, and the objective is to find sparse solutions such that the resulting hydrodynamic model recapitulates the collective particle dynamics. 

\par
Learned hydrodynamic models must respect the symmetries of the underlying microscopic dynamics. Prior knowledge of such symmetries can greatly accelerate the inference process by placing constraints on the model parameters $a_l$ and $b_l$. The learning ansatz~\eqref{Eqn:pPDELibrary} already assumes global rotational invariance by using identical coefficients $b_l$ for the $x$ and $y$ components of the polarization field equations. Generally, coordinate-independence of hydrodynamic models demands that the dynamical fields and the library functions $C_l$, $\boldsymbol{C}_l$, etc. have the correct scalar, vectorial or tensorial transformation properties. This fact imposes stringent constraints on permissible libraries, as do microscopic conservation laws.

\begin{figure*}
\includegraphics[width=\textwidth]{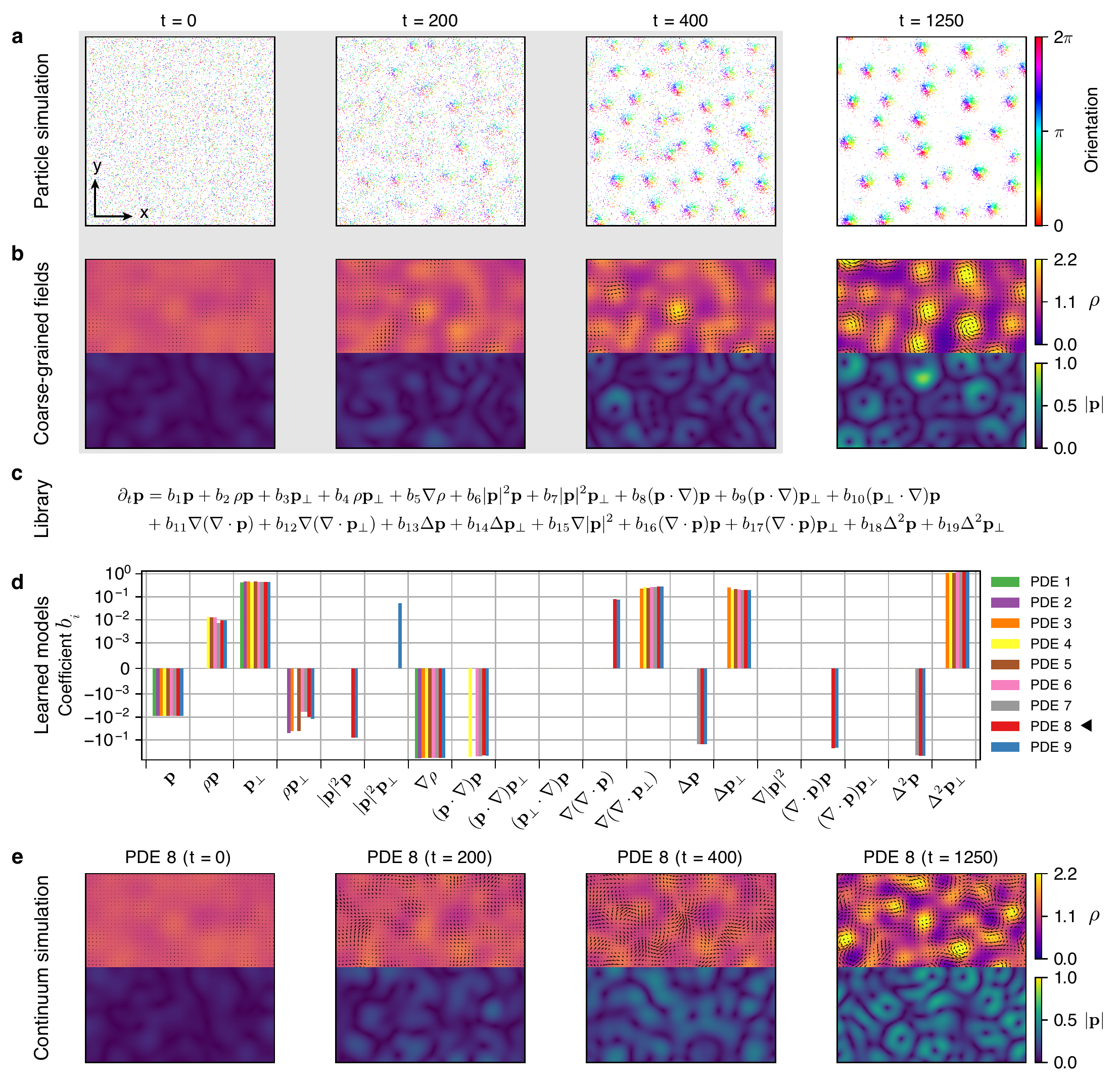}
\caption{Learning polarization dynamics. {\textbf{a},}~Same particle dynamics as in Fig.~\ref{Fig:MassCons}\textbf{a} for visual reference. {\textbf{b},}~Top: Coarse-grained density and polarization field as in Fig.~\ref{Fig:MassCons}a. Bottom: Magnitude $|\mathbf{p}|$ of the coarse-grained polarization field. Emerging vortices ($t=400,1250$) appear as ring-like patterns in $|\mathbf{p}|$. Training data were randomly sampled from the time window $t\in[40, 400]$, enclosed within the gray box. {\textbf{c},}~Physics-informed candidate library (with $b_1=-D_r$) including terms constructed from $\mathbf{p}_{\perp}=(-p_y,p_x)^\top$, which are allowed due to the chirality of the microscopic system. {\textbf{d},}~Learned phenomenological coefficients $b_l$ of PDEs with increasing complexity (SI~Sec.~\ref{App:LearnedCoeffs}). For all PDEs, learned coefficients of the linear terms $\mathbf{p}_\perp$ and $\nabla \rho$ compare well with analytic predictions~(Tab.~\ref{Tab:VortexModelCoeffs}, SI~Sec.~\ref{App:AnalyticCoarse2}). {\textbf{e},}~Simulation of the final hydrodynamic model (PDE~8 for the polarization dynamics and PDE~1 in Fig.~\ref{Fig:MassCons}e for the density dynamics). Starting from random initial conditions ($t=0$), long-lived vortex states emerge on a similar time scale, with similar spatial patterns, and with comparable density and polarization amplitudes as in the coarse-grained microscopic model data (\textbf{b}). Hydrodynamic models with PDEs sparser than PDE 8 do not form stable vortex patterns.}
\label{Fig:MomCons}
\end{figure*}

\toclesslab\subsubsection{Symmetries and conservation laws: Generating a physics-informed candidate library}{Sec:PhysicsInformedLib}

Whenever prior knowledge about (broken) symmetries and conservation laws is available, it should inform the candidate library construction to ensure that the PDE learning is performed within a properly constrained model space. \jd{A useful constraint that holds in many experimental active matter systems, as well as in the microscopic model~{[Eq.~\eqref{eqn:EOM}]} arises from particle number conservation. To impose a corresponding mass conservation in the learned hydrodynamic models, we can restrict the scalar library terms $C_l(\rho,\mathbf{p})$ in Eq.~(\ref{Eqn:RhoPDELibrary}) to expressions that can be written as the divergence of a vector field. In this case,  each term represents a different contribution to an overall mass flux and mass conservation holds by construction for any model that will be learned.} For the application considered in this work, we included fluxes up to first order in derivatives and third order in the fields (Fig.~\ref{Fig:MassCons}d). \am{If required, such an approach can easily be generalized to other conservation laws, which then require libraries to be constructed exclusively from divergences of suitable tensors.}
\par
\jd{The active particle model in Eq.~(\ref{eqn:EOM}) describes a chiral dynamical system with intrinsic microscopic rotation rates $\Omega_i \geq 0$. The space of valid hydrodynamic models therefore includes PDEs in which the mirror symmetry is explicitly broken. Formally, this implies the Levi-Civita symbol $\epsilon_{ij}$ can be used to generate a pseudo-vector \hbox{$\mathbf{p}_\perp:=\boldsymbol{\epsilon}^\top\cdot\mathbf{p} = (-p_y, p_x)^\top$} that has to be included in the construction of the candidate libraries $\left\{C_l(\rho, \mathbf{p})\right\}$ and~$\left\{\boldsymbol{C}_l(\rho, \mathbf{p})\right\}$.
 } The vectorial library $\left\{\boldsymbol{C}_l(\rho, \mathbf{p})\right\}$ for the chiral polarization dynamics, Eq.~\eqref{Eqn:pPDELibrary}, cannot be constrained further by symmetries or conservation laws. Mechanical substrate interactions with the environment as invoked by the microscopic model~\eqref{eqn:EOM} and present in many active matter experiments explicitly break Galilean invariance, leading to external forces and torques whose form is not known \textit{a priori}. We therefore included in Eq.~\eqref{Eqn:pPDELibrary} also vector fields that cannot be written as a divergence, such as $\mathbf{p}_{\perp}$, $\rho\mathbf{p}$ or $(\mathbf{p}\cdot\nabla)\mathbf{p}$, in our candidate library~$\left\{\boldsymbol{C}_l(\rho, \mathbf{p})\right\}$.

\par
In general, higher-order terms can be systematically constructed from the basic set of available fields and operators \hbox{$\mathcal{B}=\{\rho, \mathbf{p}, \mathbf{p}_{\perp}, \nabla\}$}. We illustrate the general procedure for an example library containing terms \am{up to linear order in $\rho$ and up to cubic order of the other terms in~$\mathcal{B}$}. The first step is to write the list of distinct rank-2 tensors 
\begin{equation}
\mathcal{S}=\left\{s\mathbb{I},\mathbf{p}\mathbf{p},\mathbf{p}\mathbf{p}_{\perp},\mathbf{p}_{\perp}\mathbf{p}_{\perp},\nabla\mathbf{p},\nabla\mathbf{p}_{\perp}\right\},\label{eq:termset}
\end{equation}
where $s\in\left\{1,\rho,\nabla\cdot\mathbf{p},\nabla\cdot\mathbf{p}_{\perp}\right\}$ represents one of the linearly independent scalars that can be formed from elements in~$\mathcal{B}$. From any tensor \smash{$\mathbf{\Sigma}\in\mathcal{S}$} and its transpose, we can then generate vectorial terms $\boldsymbol{C}_l$ by forming scalar products with the elements in $\mathcal{B}$. In particular, terms~$\nabla\cdot\mathbf{\Sigma}$ yield possible contributions from internal stresses and torques due to alignment interactions, while $\mathbf{\Sigma}\cdot\mathbf{p}$~and~$\mathbf{\Sigma}\cdot\mathbf{p}_{\perp}$ correspond to substrate-dependent interactions. \am{Note that we omitted $\boldsymbol{\epsilon}$ from the set $\mathcal{S}$, as it yields only one additional linearly independent term $\sim\nabla_\perp\rho$ that can be excluded for the microscopic dynamics in  Eq.~(\ref{eqn:drdt}) on the basis of generic  coarse-graining arguments~(see SI~Sec.~\ref{App:AnalyticCoarse2}).}

\par
For pattern-forming systems with emergent length scale selection, the library should be extended to include Swift-Hohenberg-type~\cite{Cross09} terms $\Delta^2 \mathbf{p}$, $\Delta^2 \mathbf{p}_\perp$, etc.~\cite{Wensink2012,James2018a}. Such terms can stabilize small-wavelength modes and, combined with $\Delta \mathbf{p}$ and $\Delta \mathbf{p}_\perp$, can give rise to patterns of well-defined length~\cite{Cross09}. \am{The final 19-term library with linearly independent terms~(SI~Sec.~\ref{App:LibLinDeps}) used to learn the polarization dynamics for the chiral particle model from Eq.~\eqref{eqn:EOM} is summarized in Fig.~\ref{Fig:MomCons}c.}

\toclesslab\subsubsection{Sparse model learning}{Sec:OptimSparseReg} 

To determine the hydrodynamic parameters~$a_l$~and~$b_l$ in Eqs.~(\ref{Eqn:PDELibrary}), we randomly sampled the coarse-grained fields $\rho(t, \mathbf{x})$ and $\mathbf{p}(t, \mathbf{x})$ and their derivatives at \hbox{$\sim10^6$} space-time points within a predetermined learning interval (SI~Sec.~\ref{App:Methods}). Generally, the success or failure of hydrodynamic model learning depends crucially on the choice of an appropriate space-time sampling interval. As a guiding principle, learning should be performed during the relaxation stage, when both time and space derivatives show the most substantial variation.
\par
Evaluating Eqs.~\eqref{Eqn:RhoPDELibrary} and~\eqref{Eqn:pPDELibrary} at all sample points yields
linear systems of the form $\mathbf{U}_t = \mathbf{\Theta}\boldsymbol{\xi}$, where the vector~$\mathbf{U}_t$ contains the time derivatives~(SI~Sec.~\ref{App:SparseReg}). The columns of the matrix~$\mathbf{\Theta}$ hold the numerical values of the library terms $C_l(\rho, \mathbf{p})$ and $\boldsymbol{C}_l(\rho, \mathbf{p})$ computed from the spectral representations~\eqref{Eq:BasisProjection}. 
The aim is to infer a parsimonious model so that the vector $\boldsymbol{\xi}$ containing the hydrodynamic parameters~$a_l$~or~$b_l$ is sparse. In this case, the corresponding PDE only contains a subset of the library terms, and we refer to the total number of terms in a PDE as its \textit{complexity}. 
\par
To estimate sparse parameters $\boldsymbol{\xi}$, we used the previously proposed sequentially thresholded least-squares (STLSQ) algorithm from SINDy~\cite{Brunton2016}. STLSQ first finds the least-squares estimate $\boldsymbol{\hat{\xi}} = \arg \min_{\boldsymbol{\xi}} ||\mathbf{U}_t - \mathbf{\Theta}\boldsymbol{\xi}||_2^2$. Subsequently, sparsity of $\boldsymbol{\hat{\xi}}$ is imposed by iteratively setting coefficients below a thresholding hyperparameter $\tau$ to zero. Adopting a stability selection approach \cite{Nicolai2010, Shah2013, Maddu2019, Maddu2021} in which~$\tau$ is systematically varied over a regularization path $ [\tau_{\textrm{max}}, \tau_{\textrm{min}}]$~(SI~Sec.~\ref{App:SparseReg}), we obtain candidate PDEs of increasing complexity~(Figs.~\ref{Fig:MassCons}e and \ref{Fig:MomCons}d) whose predictions need to be validated against the phenomenology of the input data. 

\toclesslab\subsubsection{Performance improvements and pitfalls}{Sec:PerfImprov}

Sparse regression-based learning becomes more efficient and robust if known symmetries or other available information can be used to reduce the number of undetermined parameters $a_l$ and $b_l$ in Eqs.~\eqref{Eqn:PDELibrary}. Equally helpful and important is prior knowledge of the relevant time and length scales. The coarse-grained field data need to be sampled across spatiotemporal scales that contain sufficient dynamical information; over-sampling in a steady-state typically prevents algorithms from learning terms relevant to the relaxation dynamics. Systems exhibiting slow diffusion time-scales can pose additional challenges. For example, generic analytic coarse-graining (SI~Sec.~\ref{App:AnalyticCoarse1}) shows that additive rotational noise as in Eq.~\eqref{eqn:dthetadt} implies the linear term $-D_r \mathbf{p}$ in the polarization dynamics~Eq.~\eqref{Eqn:pPDELibrary}. If the diffusive time scale $1/D_r$ approaches or exceeds the duration of the sampling time interval, then the learned PDEs may not properly capture the relaxation dynamics of the polarization field. From a practical perspective, this is not a prohibitive obstacle, as the rotational diffusion coefficient $D_r$ can be often measured independently from isolated single-particle trajectories~\cite{Edmond2012}. In this case, fixing $-D_r \mathbf{p}$ in~Eq.~\eqref{Eqn:pPDELibrary} and performing the regression over the remaining parameters produced satisfactory learning results (see Fig.~\ref{Fig:MomCons}, where $1/D_r\sim 100$ is comparable to the length of the learning interval $t\in[40, 400]$).

\toclesslab\subsection{Validation and discussion of learned models}{Sec:LearnedEqs}

The STLSQ algorithm with stability selection proposes PDEs of increasing complexity -- the final learning step is to identify the sparsest acceptable hydrodynamic model among these (Fig.~\ref{Fig:Schematic}). This can be achieved by simulating all the candidate PDEs (SI~Sec.~\ref{App:ContinuumSim}) and comparing their predictions against the original data and, if available, against analytic coarse-graining results (SI~Sec.~\ref{App:AnalyticCoarse}).

\par
For the microscopic particle model from Eq.~\eqref{eqn:EOM}, the sparsest learned PDE for the particle number density is \hbox{$\partial_t\rho = a_1\nabla \cdot \mathbf{p}$} ~(Fig.~\ref{Fig:MassCons}e); this mass conservation equation is also predicted by analytic coarse-graining (SI~Sec.~\ref{App:AnalyticCoarse}). The learned coefficient $a_1 = -0.99$ implies an effective number density flux $-a_1\mathbf{p}\approx\mathbf{p}$, which agrees very well with the analytic prediction $\langle v_i \rangle_p\mathbf{p}=\mathbf{p}$. Additional coefficients appearing in more complex models proposed by the algorithm are at least one order of magnitude smaller than~$a_1$~(Fig.~\ref{Fig:MassCons}e). Hence, as part of a hydrodynamic description of the microscopic system Eq.~\eqref{eqn:EOM}, we adopt the minimal density dynamics~$\partial_t\rho = a_1\nabla \cdot \mathbf{p}$ from now on. 

\par
The sparsest learned PDE for the dynamics of the polarization field $\mathbf{p}$ only contains three terms. However, together with the density dynamics, the resulting hydrodynamic models are either unstable or do not lead to the formation of vortex patterns. Our simulations showed that a certain level of complexity is required to reproduce the dynamics observed in the test data. In particular, there exists a unique sparsest model (PDE~8 in Fig.~\ref{Fig:MomCons}d) for which long-lived vortex states emerge from random initial conditions. The resulting hydrodynamic model exhibits density and polarization patterns quantitatively similar to those observed in the original particle system (Fig.~\ref{Fig:MomCons}a,b,e), which also form on a similar time scale. Furthermore, the learned coefficients of the linear terms $\sim\mathbf{p}_{\perp}$ and $\sim\nabla\rho$ agree well with the analytic predictions~(Tab.~\ref{Tab:VortexModelCoeffs}, SI~Sec.~\ref{App:AnalyticCoarse2}). \am{A direct comparison of temporal and spatial spectra from simulations of the learned hydrodynamic model with the coarse-grained original data shows close agreement between the characteristic length and time scales seen in each data set~(SI~Sec.~\ref{App:QuantComparison}, SI~Figs.~\ref{Fig:VortexSpectraComparison} and \ref{Fig:TimeSpectraComparison}). Furthermore, density profiles, vortex sizes, and the disordered nature of emergent vortex patterns are also  qualitatively and quantitatively similar between the coarse-grained particle data and the learned model~(SI~Fig.~\ref{Fig:VortexComparison}), confirming that the learned model captures key features of the collective hydrodynamic modes.}
\par
\am{The individual terms appearing in the learned hydrodynamic equations identify specific  physical mechanisms  that contribute to emergent pattern formation. The linear contributions are directly interpretable based on generic analytic coarse-graining arguments (SI~Sec.~\ref{App:AnalyticCoarse}): The term $b_1\mathbf{p}$ with $b_1<0$ corresponds to the lowest order mean-field contribution of rotational diffusion that suppresses orientational order at long times. The chiral  term $b_3\mathbf{p}_{\perp}$ with $b_3>0$ drives counter-clockwise rotations of the local polar field, since  $\partial_t\mathbf{p}=b_3\mathbf{p}_{\perp}$ is solved by the rotating vector field $\mathbf{p}=(\cos b_3t,\sin b_3t)$. This term represents the lowest-order chiral mean-field contribution to the dynamics and is a direct consequence of the active rotations $\sim\Omega_i$ of single particles in Eq.~(\ref{eqn:dthetadt}). The term $b_7\nabla\rho$ with $b_7<0$ comes from an effective extensile isotropic stress $\boldsymbol{\sigma}\sim-b_7\rho\mathbb{I}$ that arises entropically in systems with moving polar particles (SI~Sec.~\ref{App:AnalyticCoarse2}). The nonlinear $\rho \mathbf{p}$ and $|\mathbf{p}|^2 \mathbf{p}$ terms represent  density dependent polar alignment interactions, similar to ferromagnetic interactions in spin systems. Other higher-order and nonlinear terms can be identified as contributions from an effective closure relation, capturing the interplay between polar and nematic order in the particle system, or from effects of the microscopic parameter variability (a~detailed discussion is provided in SI~Sec.~\ref{App:PhysInterp}). }

As the learning only used coarse-grained field data in the time interval $t\in[40,400]$, simulation results for $t>400$ represent predictions of the learned hydrodynamic model (Fig.~\ref{Fig:MomCons}e). The close agreement between original data and the model simulations (Fig.~\ref{Fig:MomCons}b,e) shows that the inference framework has succeeded in learning a previously unknown hydrodynamic description for a chiral polar active particle system with broadly distributed microscopic parameters. 

\begin{table}[t]
\caption{Parameters of the hydrodynamic model learned for the microscopic dynamics in Eq.~(\ref{eqn:EOM}) and values predicted by analytic coarse-graining~(SI~Sec.~\ref{App:AnalyticCoarse2}). $\langle\cdot\rangle_p$ denotes averages over the distribution $p(v_i,\Omega_i)$ of particle velocities $v_i$ and rotation rates~$\Omega_i$~(SI~Sec.~\ref{App:PartSims}). \label{Tab:VortexModelCoeffs}}
\begin{ruledtabular} 
\begin{tabular}{P{2cm}|
>{\raggedleft\arraybackslash}p{0.6cm} @{{} = {}} >{\raggedright\arraybackslash}p{1.1cm}|
>{\raggedleft\arraybackslash}p{2.1cm} @{{} = {}} >{\raggedright\arraybackslash}p{0.7cm}}
Term
&\multicolumn{2}{c|}{Learned value} &\multicolumn{2}{c}{Analytic coarse-graining}\\
\hline 
\multicolumn{5}{l}{Density dynamics}\\
$a_1\nabla \cdot \mathbf{p}$ & $a_1$ &$-0.99$ &$-\langle v_i \rangle_p$ &$-1.00$\\
\hline
\multicolumn{5}{l}{Polarization dynamics}\\
$b_3\mathbf{p}_\perp$ &$b_3$ &\hphantom{$-$}$0.44$ &$\langle v_i \Omega_i \rangle_p/\langle v_i \rangle_p$ &\hphantom{$-$}$0.50$
 \\
$b_5\nabla \rho$ &$b_5$ &$-0.60$ &$-\frac{1}{2}\langle v_i^2 \rangle_p/\langle v_i \rangle_p$ &$-0.57$
\end{tabular}
\end{ruledtabular}
\end{table}

\begin{figure*}
\includegraphics[width=\textwidth]{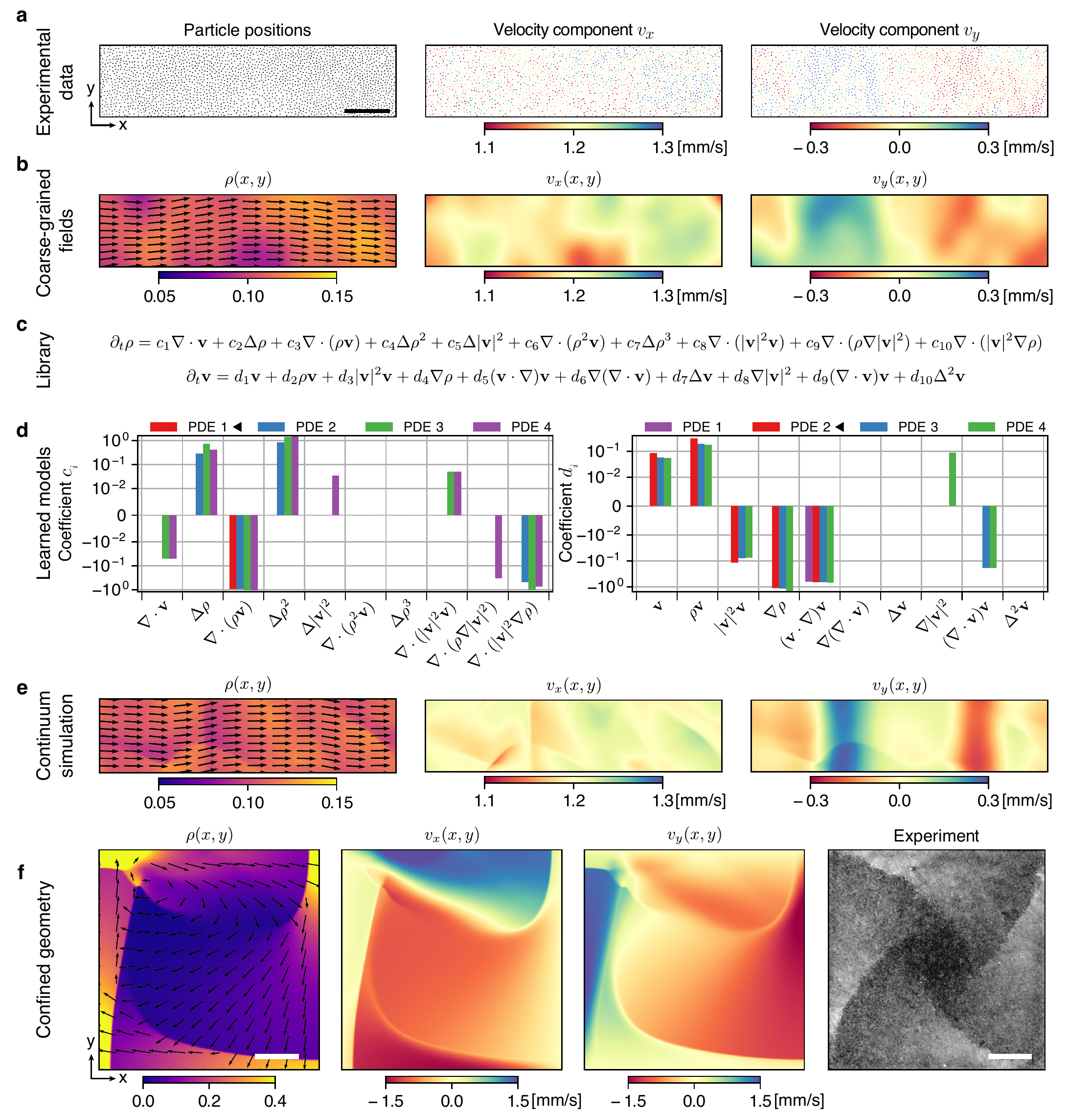}
\caption{Learning from active polar particle experiments. {\textbf{a},}~Snapshot of particle positions and velocity components of $\sim2,200$ spontaneously moving Quincke rollers in a microfluidic channel~\cite{Geyer2018a}. Scale bar, $200\, \si{\micro\meter}$. {\textbf{b},}~Coarse-grained density field $\rho(t, \mathbf{x})$, expressed as the fraction of area occupied by the rollers with diameter $D_c = 4.8\, \si{\micro\meter}$, and components $v_x(t, \mathbf{x})$ and $v_y(t, \mathbf{x})$ of the coarse-grained velocity field ($\sigma = 45\, \si{\micro \meter}$). $5\times 10^5$ randomly sampled data points from $\sim580$ such snapshots over a time duration of $1.4\,\si{\s}$ were used for the learning algorithm. {\textbf{c},}~Physics-informed candidate libraries for the density and velocity dynamics, $\{\bar{C}_l(\rho,\mathbf{v})\}$ and $\{\boldsymbol{\bar{C}}_l(\rho,\mathbf{v})\}$, respectively [Eq.~(\ref{Eqn:PDELexpibrary})]. These are the same libraries as shown in Figs.~\ref{Fig:MassCons}e and \ref{Fig:MomCons}d, but without the chiral terms and replacing $\mathbf{p}\rightarrow\mathbf{v}$.
{\textbf{d},}~Learned phenomenological coefficients $c_l$ and $d_l$ of the four sparsest PDEs for the density (left) and velocity (right) dynamics. The coefficients are non-dimensionalized with length scale $\sigma$ and time scale $\sigma/v_0$, where $v_0 = 1.2\si{\milli\meter\per\s}$ is the average roller speed.
 PDE~1 for density dynamics 
corresponds to $\partial_t \rho = c_3 \nabla \cdot (\rho \mathbf{v})$ with $c_3 \simeq -0.95$. PDE~2 for the velocity dynamics is shown in Eq.~\eqref{Eqn:QuinckeVelocity}. Learned coefficients compare well with the values reported in Ref.~\cite{Geyer2018a}~(Tab.~\ref{Tab:QuinckeCoeffs}). {\textbf{e},}~Simulation snapshot at $t=1.8$\,s of the learned hydrodynamic model (PDEs marked by $\blacktriangleleft$ in \textbf{d} in a doubly periodic domain. Spontaneous flow emerges from random initial conditions, and exhibits density and velocity fluctuations that show similar spatial patterns and amplitudes as seen in the experiments~(\textbf{a}). {\textbf{f},}~Simulation snapshots at $t=18.5$\,s of the same hydrodynamic model as in \textbf{\text{e}} on a square domain with reflective boundary conditions. The model predicts the emergence of a vortex-like flow permeated by density shock waves. This prediction agrees qualitatively with experimental observations (rightmost panel) of Quincke rollers in a $5 \si{\milli \meter} \times 5 \si{\milli \meter}$ confinement with average density $\rho_0 \approx 0.1$ (Image credits: Alexandre Morin, Delphine Geyer, and Denis Bartolo). Scale bars, $200 \, \si{\micro \meter}$ (simulation) and $1\,\si{\milli\meter}$ (experiment).}
\label{Fig:Quincke}
\end{figure*}

\toclesslab\section{Learning from experimental data}{sec:AplExpData} 

The inference framework can be readily applied to experimental data. We illustrate this by learning a quantitative hydrodynamic model directly from a video recorded in a recent study~\cite{Geyer2018a} of driven colloidal suspensions~(Fig.~\ref{Fig:Quincke}a). In these experiments, an electro-hydrodynamic instability enables micron-sized particles to self-propel with speeds up to a few millimeters per second across a surface. The rich collective dynamics of these so-called Quincke rollers~\cite{Bricard2013,Geyer2018a} provides a striking experimental realization of self-organization in active polar particle systems~\cite{Vicsek1995,Toner2005,Marchetti2013}.
 
\par
\toclesslab\subsection{Coarse-graining and spectral representation of experimental data}{sec:QuinckeCoarseGraining}

To gather dynamic particle data from experiments, we extracted particle positions $\mathbf{x}_i(t)$ from the Supplementary Movie~S2 in~Ref.~\cite{Geyer2018a}, with particle velocities $\mathbf{v}_i(t)=\frac{d}{dt}\mathbf{x}_i$ replacing the particle orientations~$\mathbf{p}_i(t)$ from before. This data set captures a weakly compressible suspension of Quincke rollers in a part of a racetrack-shaped channel~(Fig.~\ref{Fig:Quincke}a). We then applied the kernel coarse-graining [Eqs.~(\ref{Eq:CoarseGrain}), $\sigma=45 \si{\micro\meter}$, \am{see SI~Fig.~\ref{Fig:SpecEntrQuincke}}] to obtain the density field $\rho$ and the velocity field \hbox{$\mathbf{v}=\mathbf{p}/\rho$}. Accounting for the non-periodicity of the data, $\rho$ and $\mathbf{v}$ were projected on a Chebyshev polynomial basis [Eq.~\eqref{Eq:BasisProjection}] in time and space~(Fig.~\ref{Fig:Quincke}b). Filtering out non-hydrodynamic fast modes with temporal mode numbers $n>n_0$, we found that the final learning results were robust for a large range of cut-off modes~$n_0$~(SI~Sec.~\ref{App:LearnedCoeffs}). 

\toclesslab\subsection{Physics-informed library}{Sec:QuinckePhysLib}

The goal is to learn a hydrodynamic model of the form 
\begin{subequations}
\begin{eqnarray}
\partial_t \rho &=& \sumsign{l} c_l \,\bar{C}_l(\rho, \mathbf{v}), \label{Eqn:RhoexpPDELibrary}\\
\partial_t \mathbf{v} &=& \sumsign{l} d_l\, \boldsymbol{\bar{C}}_l(\rho, \mathbf{v}),\label{Eqn:vexpPDELibrary}
\end{eqnarray}\label{Eqn:PDELexpibrary}
\end{subequations}
where $\bar{C}_l(\rho, \mathbf{v})$ and $\boldsymbol{\bar{C}}_l(\rho, \mathbf{v})$ denote library terms with coefficients $c_l$ and $d_l$, respectively. The experimental Quincke roller system shares several key features with the particle model in Eq.~\eqref{eqn:EOM}, so the construction of the candidate libraries \smash{$\left\{\bar{C}_l(\rho, \mathbf{v})\right\}$} and \smash{$\left\{\boldsymbol{\bar{C}}_l(\rho, \mathbf{v})\right\}$} follows similar principles~(Fig.~\ref{Fig:Quincke}c). Conservation of particle number implies that $\bar{C}_l$ can be written as divergences of vector fields. However, rollers do not explicitly break mirror symmetry, so chiral terms can be dropped from the {$\left\{\boldsymbol{\bar{C}}_l(\rho, \mathbf{v})\right\}$} library, leaving the candidate terms shown in Fig.~\ref{Fig:Quincke}c. 

\toclesslab\subsection{Learned hydrodynamic equations and validation}{Sec:LearnedEqsQuincke}

The sparse regression algorithm proposed a hierarchy of hydrodynamic models with increasing complexity (Fig.~\ref{Fig:Quincke}d). The sparsest learned model that recapitulates the experimental observations is given by
\begin{subequations}
\begin{eqnarray}
 \partial_t \rho &=& c_3 \nabla \cdot (\rho \mathbf{v}),\label{Eqn:QuinckeDensity}\\
    \partial_t \mathbf{v} &=& d_1 \mathbf{v} + d_2 \rho \mathbf{v} + d_3 \vert \mathbf{v} \vert^2 \mathbf{v} + d_4 \nabla \rho + d_5 (\mathbf{v} \cdot \nabla) \mathbf{v}.
    \phantom{\hspace{0.5cm}} 
    \label{Eqn:QuinckeVelocity}
\end{eqnarray}
\label{Eqn:QuinckeModel}
\end{subequations}
Notably, Eqs.~\eqref{Eqn:QuinckeModel} contain all the relevant terms to describe the propagation of underdamped sound waves, a counter-intuitive, but characteristic feature of overdamped active polar particle systems~\cite{Geyer2018a}. 
\begin{table}[!bt]
\caption{Parameters of the learned hydrodynamic model for the Quincke roller system are close to values expected from analytic coarse-graining (*) and reported in Ref.~\cite{Geyer2018a} for experiments performed at mean area fraction $\rho_0\approx0.11$. \label{Tab:QuinckeCoeffs}}
\begin{ruledtabular}
\begin{tabular}{P{1.75cm}|r@{{} = {}}l|P{2.7cm}}
Term & \multicolumn{2}{c|}{Learned values} & Ref. \cite{Geyer2018a}\\
\hline
\multicolumn{3}{l}{Density dynamics}&\\
$c_3\nabla \cdot (\rho \mathbf{v})$ &
$c_3$&$-0.95$ &$-1.0$*\\
\hline
\multicolumn{3}{l}{Velocity dynamics}&\\
\begin{tabular}{c}$(d_1+d_2\rho)\mathbf{v}$\\$+ d_3|\mathbf{v}|^2\mathbf{v}$ \end{tabular}& 
$\sqrt{\frac{d_1+d_2 \rho_0}{-d_3}}$
&\hphantom{$-$}1.21\,mm/s &$1.20$\,mm/s\\
$d_4\nabla \rho$ & $d_4$ &$-1.62$\,mm$^2$/s$^2$ & $-5.0\pm$
$2.0$\,mm$^2$/s$^2$\\
$d_5(\mathbf{v}\cdot\nabla)\mathbf{v}$ &$d_5$ &$-0.67$ &$-0.7\pm0.1$ 
\end{tabular}  
\end{ruledtabular}
\end{table}

\begin{figure}[!b]
    \centering
    \includegraphics[width=0.87\linewidth]{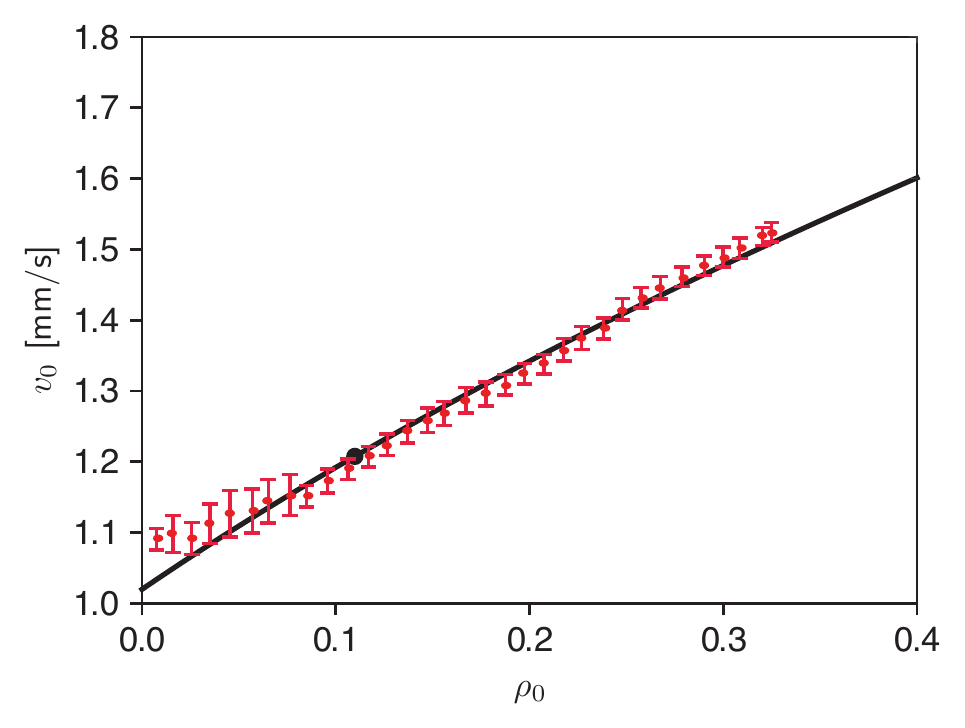}  \vspace{-0.4cm}
    \caption{The learned model accurately predicts collective Quincke roller speeds $v_0$ at different average area fractions~$\rho_0$. Although Eq.~\eqref{Eqn:QuinckeModel} was learned from a single experiment (Supplementary Movie~S2 in Ref.~\cite{Geyer2018a}) at fixed average area fraction $\rho_0=0.11$ (filled black circle), the model prediction $v_0(\rho_0) = \sqrt{-(d_1 + d_2 \rho_0)/d_3}$ (solid line) with inferred parameters $d_1,d_2,d_3$ (SI Tab.~\ref{Tab:QuinckeMomCoeffs}), agrees well the experimentally measured speed values (red symbols) reported in Supplementary~Fig.~4 of Ref.~\cite{Geyer2018a}.
   }
    \label{Fig:QuinckeVel}
\end{figure}
\par
Although the finite experimental observation window and imperfect particle tracking was expected to limit the accuracy of the learned models, the learned coefficient values agree well with corresponding parameters estimated in Ref.~\cite{Geyer2018a} by fitting a linearized Toner-Tu model to the experimental data~(Tab.~\ref{Tab:QuinckeCoeffs}). The coefficient $c_3 \simeq -0.95$ in the mass conservation equation is close to the theoretically expected value $-1$. The learned coefficient $d_4$ in the velocity Eq.~\eqref{Eqn:QuinckeVelocity} is of similar magnitude but slightly less negative than the dispersion-based estimate in Ref.~\cite{Geyer2018a}. The learned coefficients $d_1$, $d_2$, and $d_3$ (Tab.~\ref{Tab:QuinckeMomCoeffs}), to our knowledge, had not been determined previously. Despite being inferred from a single video, these parameters yield a remarkably accurate prediction \hbox{$v_0(\rho_0)=\sqrt{-(d_1+d_2\rho_0)/d_3}$} for the typical roller speed as a function of the area fraction $\rho_0$ (Supplementary Fig.~4 in Ref~\cite{Geyer2018a} and Fig.~\ref{Fig:QuinckeVel}). Similarly, the learned coefficient~$d_5$ of the nonlinear advective term $\sim(\mathbf{v} \cdot \nabla)\mathbf{v}$, is in excellent agreement with the value reported in Ref~\cite{Geyer2018a}. Interestingly, $d_5\ne-1$ reveals the broken Galilean invariance~\cite{Toner1995,Marchetti2013} due to fluid-mediated roller-substrate interaction, a key physical aspect of the experimental system that is robustly discovered by the hydrodynamic model learning framework.

To validate the learned hydrodynamic model, we simulated Eqs.~\eqref{Eqn:QuinckeModel} on a periodic domain comparable to the experimental observation window (Fig.~\ref{Fig:Quincke}e, SI~Sec.~\ref{App:ContinuumSim}). Starting from random initial conditions, spontaneously flowing states emerge, even though the spontaneous onset of particle flow is not a part of the experimental data from which the model was learned. The emergent density and flow patterns are quantitatively similar to the experimentally observed ones. In particular, the learned model predicts the formation of transverse velocity bands as seen in the experiments~(Fig.~\ref{Fig:Quincke}b,e). 

\toclesslab\subsection{Predicting collective roller dynamics in confinement}{sec:RollConf}

Useful models can make predictions for a variety of experimental conditions. At minimum, if a learned hydrodynamic model captures the most relevant physics of an active system, then it should remain valid in different geometries and boundary conditions. To confirm this for the Quincke system, we simulated Eqs.~\eqref{Eqn:QuinckeModel} on a square domain using no-flux and shear-free boundary conditions (SI~Sec.~\ref{App:ContinuumSim}). Starting from random initial conditions, our learned model predicts the formation of a vortex-like flow, permeated by four interwoven density shock waves, which arise from reflections at the boundary~(Fig.~\ref{Fig:Quincke}f, left). Remarkably, this behavior has indeed been observed in experiments~\cite{Bricard2013}, in which Quincke rollers were confined within a square domain~(Fig.~\ref{Fig:Quincke}f, right). These results demonstrate the practical potential of automated model learning for complex active matter systems.

\toclesslab\section{Discussion \& conclusions}{Sec:Discussion}

Leveraging spectral representations of field observables, we have presented a PDE learning framework that robustly identifies quantitative hydrodynamic models for the self-organized dynamics of active matter systems. \am{To illustrate its broad practical potential and applicability, we demonstrated the automated inference of interpretable  hydrodynamic models from microscopic simulation data as well as from experimental video data for active and living systems (SI~Sec.~\ref{App:Fish}). The underlying  computational framework   complements modern machine learning approaches, including model-free methods~\cite{Pathak2018, Brunton2020} and others that leverage \textit{a priori} known model structure to predict complex dynamics~\cite{Raissi2019, Zhang2020, Wallin2020,Shankar2020}, infer specific model parameters~\cite{Colen2021} or hidden fields~\cite{Raissi2020}, partially replace PDE models with suitably trained neural networks~\cite{Bar-Sinai2019, Rackauckas2020}, or use them for dimensionality reduction~\cite{Linot2020, Linot2021}.  Inferring sparse  hydrodynamic models from coarse-grained active matter data also complements analytic coarse-graining techniques \cite{Bertin2009,Farrell2012,Marchetti2013,Liebchen2016,Chate2020}, which generally} require \textit{ad hoc} moment closures to truncate infinite hierarchies of coupled mode equations (SI~Sec.~\ref{App:AnalyticCoarse}). Such closures typically neglect correlations and rely on approximations that may not be valid in interacting active matter systems. Automated learning of hydrodynamic equations yields data-informed closure relations, while simultaneously providing quantitative measurements of phenomenological coefficients (viscosities, elastic moduli, etc.) from video data~\cite{Colen2021}.  

\par 
Successful model learning requires both good data and a good library. Good data need to sample all dynamically relevant length and time scales~\cite{Champion2018}. A good library is large enough to include all hydrodynamically relevant terms and small enough to enable robust sparse regression~\cite{Maddu2021}. Since the number of possible terms increases combinatorially with the number of fields and differential operators, library construction should be guided by prior knowledge of global, local, and explicitly broken symmetries. Such physics-informed libraries ensure properly constrained model search spaces, promising more robust and efficient sparse regression. Equally important is the use of suitable spectral field representations -- without these an accurate evaluation of the library terms seems nearly impossible even for very-high quality data.
\par

In view of the above successful applications, we expect that the computational framework presented here can be directly applied to a wide variety of passive and active matter systems; \markblue{for example,  SI~Sec.~\ref{App:Fish} demonstrates  automated hydrodynamic model inference for the collective dynamics of sunbleak fish~\cite{walter2021trex}}. In parallel, there is vast potential for future enhancements by combining recent advances in sparse regression~\cite{Maddu2019, Zheng2019} and weak formulations~\cite{Reinbold2020} with statistical information criteria~\cite{Mangan2017} and cross-validation~\cite{Hastie2001} for model selection. \markblue{Furthermore, an extension to three dimensions is conceptually and computationally straightforward: Kernel-based coarse-graining, spectral data representation, the implementation of conservation laws through suitable restrictions of library terms, and the sparse regression scheme all extend naturally to higher dimensions in a parallelizable manner. Given the rapid progress in experimental imaging and tracking techniques~\cite{Hartmann2019,Shah2019,Power2017,Stelzer2015,walter2021trex}}, 
we anticipate that many previously intractable physical and biological systems will soon find interpretable quantitative continuum descriptions that may reveal novel ordering and self-organization principles.

\toclesslab\section{Acknowledgments}{Sec:Acknow}
We thank Keaton Burns for helpful advice on the continuum simulations, Henrik Ronellenfitsch for insightful discussions about learning methodologies, and the MIT SuperCloud~\cite{Reuther2018} for providing us access to HPC resources.
\jd{We thank Tristan Walter and Iain Couzin for sharing and explaining the sunbleak data.} This work was supported by a MathWorks Engineering Fellowship (R.S.), a Graduate Student Appreciation Fellowship from the MIT Mathematics Department (B.S.), \markblue{a National Science Foundation Mathematical Sciences Postdoctoral Research Fellowship (DMS-2002103, G.P.T.C.),} a Longterm Fellowship from the European Molecular Biology Organization (ALTF~528-2019, A.M.), a Postdoctoral Research Fellowship from the Deutsche Forschungsgemeinschaft (Project~431144836, A.M.), a Complex Systems Scholar Award from the James S. McDonnell Foundation (J.D.) and the Robert E. Collins Distinguished Scholarship Fund (J.D.). 

\vspace{0.25cm}
\noindent
\textbf{Data availability.}
The data sets generated during and/or analyzed during the current study are available from the corresponding author on reasonable request.
\vspace{0.25cm}\\
\textbf{Code availability.}
All codes generated during and/or used during the current study are available from the corresponding author on reasonable request.  

\onecolumngrid
\newpage
\centerline{\large\textbf{Supplementary Information}}
\setcounter{page}{1}
\appendix
\renewcommand\thefigure{S\arabic{figure}}    
\setcounter{figure}{0}
\renewcommand\thetable{S\,\Roman{table}}    
\setcounter{table}{0}

\tableofcontents

\section{Methods}
\label{App:Methods}

\subsection{Particle simulations} \label{App:PartSims}
\begin{figure}[b]
    \centering
    \includegraphics[width=0.5\linewidth]{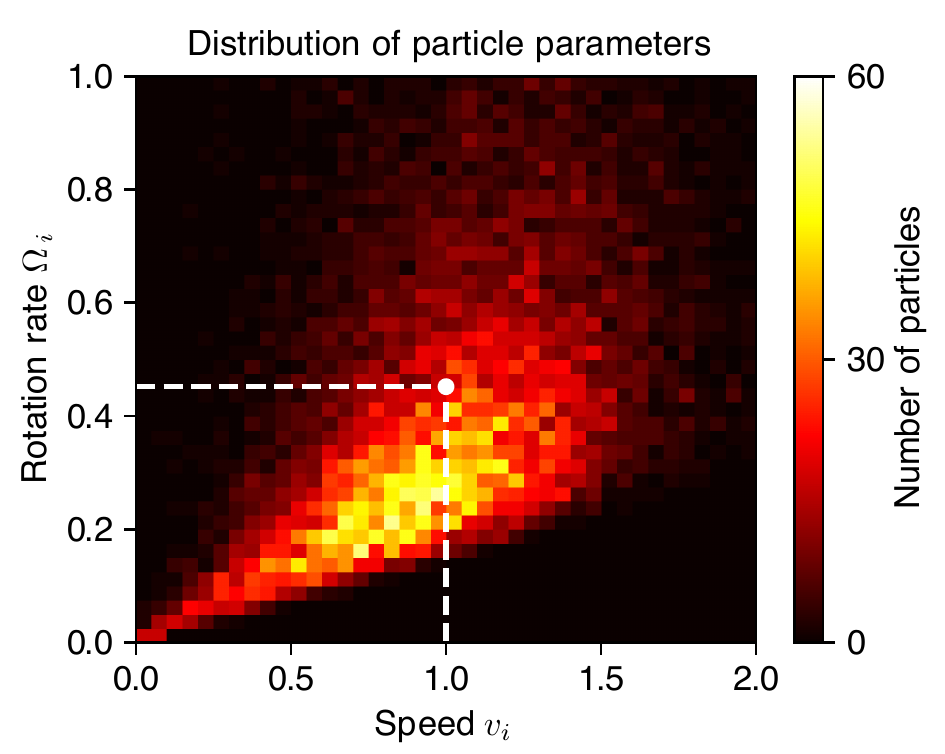} 
    \caption{Distribution of particle speeds~$v_i$ and rotation rates~$\Omega_i$ used to simulate Eqs.~\eqref{eqn:EOM}. Those parameters were drawn from a heuristic distribution $p(v_i,\Omega_i)$ that is explained in more detail in SI~Sec.~\ref{App:PartSims}. The white marker and dashed lines depict the mean velocity $\langle v_i\rangle_p\simeq1$ and $\langle\Omega_i\rangle_p\simeq0.45$.}
    \label{Fig:ParamDist}
\end{figure}

The microscopic model in Eqs.~(\ref{eqn:EOM}) has been previously studied for fixed particle velocities $v_i=v_0$ and rotation frequencies $\Omega_i=\Omega_0$. In this scenario, particles form small clusters of aligned particles and each cluster orbits on a circle of radius $\sim v_0/\Omega_0$~\cite{Liebchen2017}. To generate the microscopic test data used in Sec.~\ref{Sec:LearningFramework}, we considered instead a heuristic distribution $p(v_i,\Omega_i)$ for which particles spontaneously organize into proper vortices~(Fig.~\ref{Fig:MassCons}a, top). It is convenient to define and draw from this distribution using propagation speeds~$v_i$ and the curvature radii~$R_i=v_i/\Omega_i$ of a particle's noise-free trajectory as independent variables. In particular, we considered \hbox{$\tilde{p}(v_i,R_i)\sim G(v_i;\mu_v,\sigma_v)G(R_i;\mu_R,\sigma_R)$}, where $G(x;\mu_x,\sigma_x)$ represents a Gaussian normal distribution with mean~$\mu_x$ and standard deviation~$\sigma_x$. $\tilde{p}(v_i,R_i)$ then defines $p(v_i,\Omega_i)$ implicitly through the relation \hbox{$v_i=\Omega_iR_i$}. In units of the characteristic scales -- mean velocity $\langle v_i\rangle_p$ and interaction radius $R$ -- the particle properties $v_i$ and $\Omega_i=v_i/R_i$ used for simulating Eqs.~(\ref{eqn:EOM})~(Fig.~\ref{Fig:ParamDist}) were drawn from $\tilde{p}(v_i,R_i)$ with $\mu_v=1$ ($\langle v_i\rangle_{\tilde{p}}=\langle v_i\rangle_p = 1$), $\sigma_v=0.4$, $\mu_R=2.2$ and $\sigma_R=1.7$. From these samples, we finally removed all particles with $\Omega_i>1.4$. 
 
For simulations of the microscopic model in Eqs.~(\ref{eqn:EOM}), we set \markblue{$g \simeq 0.018$} and $D_r \simeq 0.009$ ($\Rightarrow D_r\ll\langle \Omega_i \rangle_p,\ D_r\ll\langle v_i \rangle_p$) and initially placed particles randomly distributed and oriented on a domain of size 100$\times$100 (in units of the interaction radius). Equations~(\ref{eqn:EOM}) were then numerically integrated for particles interacting within the interaction radius $R=1$ using the Euler-Maruyama method with a time step of $dt \simeq 0.0176$. For the subsequent coarse-graining, the data were saved at time intervals of~$\Delta t \simeq 0.44$.  

\subsection{Kernel coarse-graining with periodic and non-periodic boundaries} \label{App:CG_nearB}
To coarse-grain the discrete microscopic data through Eqs.~\eqref{Eq:CoarseGrain}, we used a 2D Gaussian kernel 
\begin{equation}
K(\mathbf{x}) = (2\pi\sigma^2)^{-1}\exp{(-\vert\mathbf{x}\vert^2/2\sigma^2)}. 
\end{equation}
Periodicity of the coarse-grained fields for the microscopic test data~(Sec.~\ref{Sec:CoarseGraining}) was ensured by placing ghost particles periodically around the domain.
\par

Coarse-graining in non-periodic domains (Sec.~\ref{sec:AplExpData}) was performed by truncating and renormalizing the kernel. This was achieved by defining the integral over the non-periodic domain $\mathcal{X}$ to be $N(\mathbf{x}) = \int_\mathcal{X}d^2 \mathbf{x}'\, K(\mathbf{x}' - \mathbf{x})$, and then replacing $K[\mathbf{x}-\mathbf{x}_i(t)]$ with $K[\mathbf{x}-\mathbf{x}_i(t)]/N(\mathbf{x})$ in Eqs.~\eqref{Eq:CoarseGrain}. 
This renormalization ensured that the coarse-grained density $\rho(t, \mathbf{x})$ integrated to the total particle number and strongly reduced artefacts near the boundary. 

\subsection{Spectral representation}
\label{App:SpecRep}
The coarse-grained hydrodynamic fields [Eq.~\eqref{Eq:CoarseGrain}] were evaluated at $[N_t, N_x, N_y]$ uniformly spaced grid points in the respective directions. The resulting discrete data were projected onto the spectral basis functions~[Eq.~\eqref{Eq:BasisProjection}] using multidimensional discrete cosine and Fourier transforms provided by the FFTW library~\cite{Matteo2005}, with an efficient time complexity of $O(N\log(N))$, where $N = N_t N_x N_y$. For the Chebyshev transforms, the data were interpolated onto the required Chebyshev extrema grid using spline functions of degree 5.

\subsection{Sparse regression} 
\label{App:SparseReg}
To perform sparse regression using the sequentially thresholded least squares (STLSQ) algorithm~\cite{Brunton2016}, we used the same parameters when working with data from the test microscopic model~(Sec.~\ref{Sec:MicroModel}) as well as for the experiments using Quincke rollers~(Sec.~\ref{sec:AplExpData}) \am{and sunbleak fish~(SI~Sec.~\ref{App:Fish})}. \am{As the sparse regression approach used to infer hydrodynamic equations only requires the evaluation of input data within the data domain, it is independent of the system's boundary conditions.} Details of the concrete steps of the learning framework are provided in the following. 
\par
\textbf{Construction of linear systems}: To construct the linear system $\mathbf{U}_t = \mathbf{\Theta}\boldsymbol{\xi}$, we randomly sampled the coarse-grained fields at $N_d=5\times10^5$ time-space points. The explicit form of the linear systems constructed for Eqs.~\eqref{Eqn:PDELibrary} was given by
\begin{subequations}
\label{eq:LinearSysAll}
\begin{align}
\underbrace{
\left[ \begin{array}{c}
\vdots \\ \partial_t \rho \\ \vdots
\end{array} \right]}_{\mathbf{U}_t (N_d \times 1)}
&=
\underbrace{
\left[ \begin{array}{ccc}
\vdots & & \vdots  \\
\nabla \cdot \mathbf{p} &\cdots & \nabla \cdot (\rho \mathbf{p}_\perp) \\
\vdots & & \vdots
\end{array} \right]}_{\mathbf{\Theta}(N_d \times r)}
\underbrace{
\left[\begin{array}{c}
a_1\\a_2\\\vdots \\a_r
\end{array}\right]}_{\mathbf{\xi}(r\times1)},  \label{Eqn:LinearSysrho}\\
\underbrace{
\left[\begin{array}{c}
\partial_t p_x \\
\vdots\\\hline\\ \partial_t p_y\\\vdots
\end{array}\right]}_{\mathbf{U}_t (2N_d \times 1)}
&=
\underbrace{\left[ \begin{array}{ccc}
(\nabla \rho)_x & \cdots & ((\mathbf{p} \cdot \nabla) \mathbf{p})_x  \\
\vdots & & \vdots  \\
\hline\\
(\nabla \rho)_y &\cdots & ((\mathbf{p} \cdot \nabla) \mathbf{p})_y \\
\vdots & & \vdots 
\end{array}\right]}_{\mathbf{\Theta} (2N_d \times m)}
\underbrace{
\left[\begin{array}{c}
b_1\\b_2\\\vdots \\b_m
\end{array}\right]}_{\boldsymbol{\xi} (m \times 1)}.
\label{Eqn:LinearSysp}
\end{align}
\end{subequations}
Here, the subscripts denote components of the vectors, and $r, m$ are the total number of library terms in each equation. The vertical dots denote the respective terms evaluated at different time-space $(t, \mathbf{x})$ locations.  The linear system in Eq.~\eqref{Eqn:LinearSysp} was generated by stacking data for the $x$- and $y$-components of the time-derivatives and the library terms. Such a construction enforced the same coefficients for both the components of the polarization equation, ensuring rotational invariance (coordinate-independence) of the learned PDE. \markblue{For the Quincke roller system (Sec.~\ref{sec:AplExpData}, main text) and the sunbleak system (SI~Sec.~\ref{App:Fish}) linear systems analog to Eqs.~(\ref{eq:LinearSysAll}) were constructed for $N_d=5\times10^{5}$ sampling points each, with term libraries described in the corresponding sections.}

\textbf{Pre-processing}: Since the thresholding hyperparameter $\tau$ in STLSQ is agnostic to the scales of the library terms, as a pre-processing step, we performed transformations so that columns of the data matrix $\mathbf{\Theta}$ had zero mean and unit variance, and the time-derivative vector $\mathbf{U}_t$ had zero mean. \am{Column standardization has been widely applied in equation discovery approaches~\cite{Schaeffer2017, Rudy2017a, Maddu2019}, where it can mitigate numerical resolution limitations and may easily be extended to regularized regression techniques, such as ridge regression or LASSO.}

\textbf{Stability selection}~\cite{Maddu2019}: With equal spacing on a $\log_{10}$ scale, we chose $40$ values for $\tau$ over the regularization path $[\tau_{\textrm{max}}, \epsilon \tau_{\textrm{max}}]$. The value of $\tau_{\textrm{max}}$ was chosen so that all the terms get thresholded out and $\epsilon$ was set to $10^{-2}$. 
For every $\tau$, the data were split into $200$ sub-samples each with $50\%$ randomly selected data points. Every library term was assigned an importance score as the fraction of sub-samples in which it was learned by STLSQ; in general, this importance score was larger for smaller values of~$\tau$. Along the regularization path, unique combinations of terms that had an importance score larger than~$0.6$ were considered and their coefficients were refitted to the full data without normalization. This procedure resulted in a small number of PDEs of increasing complexity~(Figs.~\ref{Fig:MassCons}e, \ref{Fig:MomCons}d, and \ref{Fig:Quincke}d).

\subsection{Linear dependencies of the library terms}
\label{App:LibLinDeps}
The procedure outlined in Sec.~\ref{Sec:PhysicsInformedLib} leads to a few library terms for the polarization dynamics [Eq.~\eqref{Eqn:pPDELibrary}] that are linearly dependent on each other. For completeness, we provide here a list of non-trivial identities that can be used to eliminate these dependencies: 
\begin{subequations}
\begin{align}
\frac{1}{2}\nabla|\mathbf{p}|^2&=(\mathbf{p}\cdot\nabla)\mathbf{p}+(\nabla\cdot\mathbf{p}_{\perp}) \mathbf{p}_{\perp},\label{Eqn:Iden1}\\
(\nabla \cdot \mathbf{p}) \mathbf{p}_\perp  &= (\nabla \mathbf{p}) \cdot \mathbf{p}_\perp + (\mathbf{p} \cdot \nabla)\mathbf{p}_\perp,  \label{Eqn:Iden2}\\
(\mathbf{p} \cdot \nabla)\mathbf{p}_\perp &+ (\mathbf{p}_\perp \cdot \nabla)\mathbf{p} \nonumber \\&= (\nabla \cdot \mathbf{p})\mathbf{p}_\perp  +  (\nabla \cdot \mathbf{p}_\perp)\mathbf{p}.
\end{align}
\end{subequations}
In Eq.~\eqref{Eqn:Iden2}, we follow the convention, $[(\nabla \mathbf{a})\cdot\mathbf{b}]_{i} = b_j\partial_i a_j$, with $i=x,y$ and repeated indices indicating summation. One may set $\mathbf{p} \rightarrow \mathbf{p}_\perp$ and \hbox{$\mathbf{p}_\perp \rightarrow - \mathbf{p}$} in Eqs.~\eqref{Eqn:Iden1} and \eqref{Eqn:Iden2} to obtain two additional identities. 

To ensure in general that no linear dependencies remain after a library of terms has been constructed, it should be checked that the columns the data matrix $\mathbf{\Theta}$ are linearly independent, for example by using a singular value decomposition of $\mathbf{\Theta}$.

\subsection{\am{Numerical simulations of PDEs and boundary conditions}}
\label{App:ContinuumSim}
Continuum simulations were performed using the spectral PDE solver Dedalus~\cite{Burns2020} with four-step Runge-Kutta time stepping scheme~RK443. For simulation of the PDEs learned from the microscopic test data~(Fig.~\ref{Fig:MomCons}e) we used $256 \times 256$ Fourier modes in a doubly periodic domain with time step $4\times10^{-3}$. To facilitate a comparison, simulations shown in Fig.~\ref{Fig:MomCons}e were initialized using the initial density and polarization field of the coarse-grained particle data. It was verified that similar vortex patterns also form from fully random initial conditions.
\par
For the doubly periodic simulation of Eqs.~\eqref{Eqn:QuinckeModel}~(Fig.~\ref{Fig:Quincke}e), we used $1024 \times 1024$ Fourier modes and time step $10^{-4} \si{\s}$. The initial conditions were random with mean density $0.11$, and mean horizontal and vertical velocities, \hbox{$\langle v_x\rangle=0.1\,\si{\milli\meter\per\s}$} and $\langle v_y\rangle=0$, respectively. 
\par
The simulation presented in Fig.~\ref{Fig:Quincke}f was performed on a confined square domain using the Sine/Cosine basis functions with $1024 \times 1024$ modes and time step $10^{-4}\si{\s}$. The basis combinations in the $(x,y)$ directions were chosen to be ($\cos$, $\cos$) for density $\rho$, ($\sin$, $\cos$) for $v_x$ and ($\cos$, $\sin$) for $v_y$. These imply that normal density gradients, normal velocities and all remaining shear rates $\partial_xv_y$ and $\partial_yv_x$ vanish at the domain boundaries. The simulations were initialized with random perturbations around a mean density of $0.11$. \markblue{Simulations of models learned for the confined motion of sunbleak fish (see SI~Sec.~\ref{App:Fish}) were run with the same boundary conditions and basis functions with $256\times256$ modes and time step $10^{-3}$ s.}
\par
Since the model in Eqs.~\eqref{Eqn:QuinckeModel} learned for the Quincke roller dynamics generates density shock waves, we added numerical diffusivities of $10^{-4}\si{\milli\meter^2\per\s}$~(Fig.~\ref{Fig:Quincke}e) and $10^{-3}\si{\milli\meter^2\per\s}$~(Fig.~\ref{Fig:Quincke}f) in both density and velocity equation to avoid \markblue{Gibbs} ringing.  

\section{Analytic coarse-graining of the particle model}
\label{App:AnalyticCoarse}
We describe in this section two approaches to analytically determine mean-field approximations of the microscopic model~{[Eqs.~(\ref{eqn:EOM})]} and compare their predictions and methodology with our learning approach. In general, analytic coarse-graining can $(i)$~provide guidance for developing a physics-informed learning library, $(ii)$~allows discussing our PDE learning framework as a tool to effectively infer moment closure relations, and $(iii)$~predict the dependency of PDE coefficients of linear terms on distributions of microscopic parameters, which can be used to validate learned hydrodynamic models. \am{However, to produce  interpretable mean-field equations, analytic coarse-graining procedures  typically have to
\begin{itemize}[leftmargin=7cm]    
    \item[(i)] factorize pair-correlations;   
    \item[(ii)] impose closure relations; and 
    \item[(iii)] neglect microscopic parameter variability.
\end{itemize}
Such approximations particularly affect non-linear terms and terms with higher-order derivatives in the resulting coarse-grained equations. It is, therefore, instructive to directly compare the coefficients from analytic approximations against the results obtained by hydrodynamic model learning, where the latter effectively infers a suitable expansion of pair-correlations, as well as a closure relation directly from the underlying~data and naturally integrates microscopic parameter variability.

\subsection{Dynamic equation of the one-particle probability density}
\label{App:AnalyticCoarse1}
\am{We first describe a common analytic coarse-graining procedure that is often applied to microscopic models like Eq.~(\ref{eqn:EOM}) \textit{for constant and homogeneous parameters} $v_i=v_0$ and $\Omega_i=\Omega_0$.} Specifically, this approach aims to find an approximate dynamic equation for the one-particle probability density~\cite{Dean1996,Bertin2009,Farrell2012,Marchetti2013,Liebchen2016,Liebchen2017} 
\begin{equation}
    f(t,\theta,\mathbf{x})=\sum_{i=1}^{N}\langle\delta\left(\theta-\theta_i(t)\right)\delta\left(\mathbf{x}-\mathbf{x}_i\right)\rangle,
\end{equation}
where $\langle\cdot\rangle$ denotes a Gaussian white noise average. To find such a dynamic equation, this approach proceeds in two steps: First, a dynamic equation for the angular moments of $f(t,\theta,\mathbf{x})$ is derived and second, to truncate the infinite hierarchy of dynamic equations, a closure relation is imposed.}

\am{
\newpage
\subsubsection{Angular moment expansion}}
Neglecting multiplicative noise and factorizing pair-correlations gives rise to a nonlinear integro-differential equation~\cite{Farrell2012,Liebchen2017} that can be transformed into an infinite hierarchy of coupled PDEs for the angular moments $f_n(t,\mathbf{x})$ defined by
\begin{equation}
    f_n(t,\mathbf{x}) = \int_0^{2\pi}d\theta f(t,\theta,\mathbf{x})e^{in\theta}.
\end{equation}
For the microscopic model~{[Eqs.~(\ref{eqn:EOM})]} and equal microscopic parameters, $v_i=v_0$ and $\Omega_i=\Omega_0$ for all particles, this procedure leads to~\cite{Farrell2012,Liebchen2017}


\begin{equation}
  \partial_t  f_n+\frac{v_0}{2}\left[\partial_x\left(f_{n+1}+f_{n-1}\right)-i\partial_y\left(f_{n+1}-f_{n-1}\right)\right] =n(i\Omega_0-D_r n)f_n+\frac{gn\pi}{2}\left(f_{n-1}f_1-f_{n+1}f_{-1}\right). 
  \label{eq:moddyn}
\end{equation}

Each complex angular moment $f_n$ can be identified as a mean-field variable that represents different orientational order parameters encoded by the probability density $f(t,\theta,\mathbf{x})$~\cite{Marchetti2013}. In particular, $f_0$ represents the particle number density $\rho$ and $f_1=:p_x+ip_y$ represents the polarization density $\mathbf{p}=(p_x,p_y)^{\top}$. These fields correspond to the coarse-graining information in Eqs.~(\ref{Eq:CoarseGrain}) that our learning framework extracts explicitly from given microscopic data. For $n=0$ and $n=1$, we can therefore write Eqs.~(\ref{eq:moddyn}) as
\begin{subequations}\label{eq:dyneqCG1}
\begin{align}
\partial_t\rho+v_0\nabla\cdot\mathbf{p}&=0,\\
\partial_t\mathbf{p}+\frac{v_0}{2}(\nabla \rho + \nabla \cdot \mathbf{Q})&=\Omega_0\mathbf{p}_{\perp}-D_r\mathbf{p}+\frac{g\pi}{2}\left(\rho\mathbb{I}-\mathbf{Q}\right)\cdot\mathbf{p}\label{eq:dyneqp},   \end{align}
\end{subequations}
which also shows the coupling to the next higher mode \hbox{$f_2=:Q_{xx}+iQ_{xy}$}, corresponding to the independent degrees of freedom of a nematic tensor. The chiral term $\sim\Omega_0\mathbf{p}_{\perp}$ with \hbox{$\mathbf{p}_{\perp}=(-p_y,p_x)^\top$} breaks the mirror symmetry. Terms constructed from $\mathbf{p}_{\perp}$ are therefore generally allowed in chiral systems and consequently included into the library in Eqs.~(\ref{Eqn:PDELibrary}). 

\am{\subsubsection{\am{Closure relation and comparison with learned models}} \label{App:closerel}}
The final step that is key to analytically close the infinite hierarchy of Eqs.~(\ref{eq:moddyn}) requires the introduction of moment closure assumptions~\cite{Bertin2009,Farrell2012,Geyer2018a}. Depending on the structure of the mode coupling, the resulting closure relation allows to express the nearest coupled modes with \hbox{$|n|=k$} in terms of modes \hbox{$|n|<k$} and neglects the remaining modes. For example, in the case of Eq.~(\ref{eq:dyneqCG1}), a moment closure assumption must provide an expression~$\mathbf{Q}(\rho,\mathbf{p})$~\cite{Farrell2012,Geyer2018a}. \am{As modes with higher mode numbers $n$ are increasingly suppressed by rotational noise, which can be seen by the prefactor~$-n^2D_r$ in Eq.~(\ref{eq:moddyn}), a common closure assumption is $\partial_t\mathbf{Q}\approx0$~\cite{Liebchen2016,Liebchen2017,Marchetti2013}. Equation~(\ref{eq:moddyn}) for $n=2$ then implies the desired closure relation $\mathbf{Q}(\rho,\mathbf{p})$ that can be used in Eq.~(\ref{eq:dyneqp}) and leads to
\begin{align}
\partial_t\mathbf{p}&=-\frac{v_0}{2}\nabla\rho+\frac{\bar{g}\rho}{2}\mathbf{p}+\Omega\mathbf{p}_{\perp}-D_r\mathbf{p}\nonumber\\
&+\frac{1}{8}\frac{D_r}{\Omega_0^2+4D_r^2}\left[2v_0^2\Delta\mathbf{p}+\bar{g}v_0\left(5\nabla|\mathbf{p}|^2-6\mathbf{p}\cdot\nabla\mathbf{p}-10\mathbf{p}\nabla\cdot\mathbf{p}\right)
-4\bar{g}^2\mathbf{p}^2\mathbf{p}\right]\nonumber\\
&+\frac{1}{8}\frac{\Omega_0}{\Omega_0^2+4D_r^2}\left[v_0^2\Delta\mathbf{p}_{\perp}-\bar{g}v_0\left(3\mathbf{p}\cdot\nabla\mathbf{p}_{\perp}+5\mathbf{p}_{\perp}\cdot\nabla\mathbf{p}\right)-2\bar{g}^2\mathbf{p}^2\mathbf{p}_{\perp}\right],\label{eq:dtpgen}
\end{align}
where we defined $\bar{g}=\pi g$. Equation~(\ref{eq:dtpgen}) is equivalent to the result given in Ref.~\cite{Liebchen2017}. The coefficients in Eq.~(\ref{eq:dtpgen}) are listed and computed in Tab.~\ref{Tab:FullCGcomparison}, where we used $\Omega_0=\langle v_i \Omega_i \rangle_p/\langle v_i \rangle_p=0.5$ and $v_0=\langle v_i^2\rangle_p/\langle v_i \rangle_p=1.14$. These values for $\Omega_0$ and $v_0$ are suggested by an analytic kernel coarse-graining~(see SI~Sec.~\ref{App:AnalyticCoarse2}) of the microscopic model with a distribution $p(v_i,\Omega_i)$ of kinetic particle parameters, as considered in this work~(see SI~Sec.~\ref{App:PartSims}). 

As discussed in the main text (see Sec.~\ref{Sec:LearnedEqs} and Tab.~\ref{Tab:VortexModelCoeffs}), coefficients associated with terms linear in the fields and derivatives inferred from the model learning agree well with analytic predictions. However, Tab.~\ref{Tab:FullCGcomparison} shows that coefficients associated with terms that are non-linear in the fields or derivatives can substantially differ between these two approaches (e.g. $b_5$, $b_8$ or $b_{13}$ in Tab.~\ref{Tab:FullCGcomparison}) or they do not even appear in the analytically coarse-grained dynamics~(e.g. $b_4$, $b_{11}$~or~$b_{12}$ in Tab.~\ref{Tab:FullCGcomparison}). At the same time, the analytically coarse-grained values of these non-linear terms are most strongly affected by the various approximations listed in the introduction of SI~Sec.~\ref{App:AnalyticCoarse1}. Our learning framework on the other hand does not invoke such approximations but instead takes a different route by inferring -- directly from the data -- an effective closure relation that best explains the observed dynamics and thereby accounts for non-trivial effects from correlations and microscopic parameter variability that are not captured by standard analytic coarse-graining methods.\\

\begin{table*}
\am{\caption{Comparison of analytic coarse-graining (CG) results with learned coefficients suggest a limited validity of common analytic coarse-graining approximations in systems with microscopic parameter variability and mesoscopic pattern formation. Analytic coefficients (CG~coefficient) have been obtained by coarse-graining the model in Eq.~(\ref{eqn:EOM}) for constant microscopic parameters $v_i=v_0$ and $\Omega_i=\Omega_0$ [see SI~Sec.~\ref{App:AnalyticCoarse1} and Eq.~(\ref{eq:dtpgen})] using a common closure relation~\cite{Liebchen2017}. Models were learned (PDE~1, PDE~4 and PDE~8 shown here;  see Tab.~\ref{Tab:VortexMomCoeffs}for the complete list) from coarse-grained fields~(see Sec.~\ref{Sec:CoarseGraining} and Fig.~\ref{Fig:MomCons}) of the microscopic dynamics in Eq.~(\ref{eqn:EOM}) with a distribution of parameters $v_i$ and $\Omega_i$ (see SI~Sec.~\ref{App:PartSims} and Fig.~\ref{Fig:ParamDist}). Analytic coefficients were calculated (CG value) using $\Omega_0=\langle v_i \Omega_i \rangle_p/\langle v_i \rangle_p=0.5$, $v_0=\langle v_i^2\rangle_p/\langle v_i \rangle_p=1.14$, $D_r=0.009$, $\bar{g}=\pi g=1.1$ and $c_0=(8\Omega_0^2+32D_r^2)^{-1}$. While the analytic coefficients for linear terms agree well with the learned models (see Sec.~\ref{Sec:LearnedEqs} and Tab.~\ref{Tab:VortexModelCoeffs}), coefficients of higher order terms in the fields and derivatives can differ significantly (e.g. $b_5$, $b_8$ or $b_{13}$) or are not predicted by the analytic coarse-graining to contribute to the mean-field dynamics~(e.g. $b_4$, $b_{11}$~or~$b_{12}$).\label{Tab:FullCGcomparison}}
\begin{ruledtabular}
\begin{tabular}{l|ccccc}
Term&CG coefficient & CG value & PDE~1& PDE~4 & PDE~8$\blacktriangleleft$\\
\hline$b_{1}$$\mathbf{p}$& $-D_r$ &-0.009 & -0.009 & -0.009 & -0.009\\
$b_{2}$$\rho \mathbf{p}$& $\bar{g}/2$ & 0.0283 & -- & 0.013  &\hphantom{-}0.009\\
$b_{3}$$\mathbf{p}_\perp$& $\Omega_0$ & 0.5 & 0.414 & 0.428 &\hphantom{-}0.440\\
$b_{4}$$\rho \mathbf{p}_\perp$& -- & -- & -- & -- &-0.010\\
$b_{5}$$|\mathbf{p}|^2 \mathbf{p}$& $-4c_0\bar{g}^2D_r$ & $-6\times10^{-5}$ & -- & --&-0.080\\
$b_{6}$$|\mathbf{p}|^2 \mathbf{p}_\perp$& $-2c_0\bar{g}^2\Omega_0$ & -0.0016 & --& -- & --\\
$b_{7}$$\nabla\rho$& $-v_0/2$& -0.57 & -0.638 & -595 & -0.595\\
$b_{8}$$(\mathbf{p} \cdot \nabla) \mathbf{p}$& $-6c_0\bar{g}v_0D_r$& -0.0017 & -- & -0.536 &-0.463\\
$b_{9}$$(\mathbf{p} \cdot \nabla) \mathbf{p}_\perp$& $-3c_0\bar{g}v_0\Omega_0$ & -0.0483 & -- & -- &--\\
$b_{10}$$(\mathbf{p}_\perp \cdot \nabla) \mathbf{p}$& $-5c_0\bar{g}v_0\Omega_0$ & -0.0805 &-- & -- &--\\
$b_{11}$$\nabla(\nabla \cdot \mathbf{p})$ & -- & -- & -- &-- &\hphantom{-}0.078\\
$b_{12}$$\nabla(\nabla \cdot \mathbf{p}_\perp)$& -- & -- & -- & 0.265 &\hphantom{-}0.277\\
$b_{13}$$\Delta \mathbf{p}$& $2c_0v_0^2D_r$ & 0.0117 & -- &-- &-0.155\\
$b_{14}$$\Delta \mathbf{p}_\perp$& $c_0v_0^2\Omega_0$ & 0.3245 & -- & 0.202 & \hphantom{-}0.196\\
$b_{15}$$\nabla \vert \mathbf{p}\vert^2$& $5c_0\bar{g}v_0D_r$ & 0.0014 & -- & -- &--\\
$b_{16}$$(\nabla \cdot \mathbf{p})\mathbf{p}$& $-10c_0\bar{g}v_0D_r$ & -0.0029 & -- &-- &-0.225\\
$b_{17}$$(\nabla \cdot \mathbf{p})\mathbf{p}_\perp$& -- & -- &-- & -- &--\\
$b_{18}$$\Delta^2 \mathbf{p}$& -- & -- & -- & -- & -0.483\\
$b_{19}$$\Delta^2 \mathbf{p}_\perp$& -- & -- & -- & 1.197 & \hphantom{-}1.235\\
    \end{tabular}
    \end{ruledtabular}}
\end{table*}
}

\subsection{Dynamic equations from conventional kernel coarse-graining}
\label{App:AnalyticCoarse2}
While the previous approach provides a clear coarse-graining strategy to find a closed set of PDEs from a system of stochastic ODEs with homogeneous microscopic parameters, it is more challenging to understand how the phenomenological coefficients will depend on the distribution $p(v_i,\Omega_i)$ of microscopic kinetic parameters described in SI~Sec.~\ref{App:PartSims}. We therefore consider an alternative strategy, for which we write Eqs.~(\ref{eqn:EOM}) as
\begin{subequations}
\begin{eqnarray}
\frac{d\mathbf{x}_i}{dt} & = & v_i \mathbf{p}_i, \label{eqn:drdt_cf} \\
\frac{d\mathbf{p}_i}{dt} & = & \Omega_i\boldsymbol{\epsilon}\cdot\mathbf{p}_{i} + \mathbf{F}_i, \label{eqn:dthetadt_cf}
\end{eqnarray} \label{eqn:EOM_cf}
\end{subequations}
where $\boldsymbol{\epsilon}\cdot\mathbf{p}_{i}=\mathbf{p}_{i,\perp}=(-\sin\theta_i,\cos\theta_i)^\top$, and $\mathbf{F}_i$ contains forces from interactions and rotational diffusion. Taking directly the time derivative of the coarse-graining prescription in Eq.~(\ref{eq:densCG}) and using Eq.~(\ref{eqn:drdt_cf}), we find
\begin{equation}\label{eq:ContEqcg}
\partial_t\rho(t,\mathbf{x})+\nabla\cdot\mathbf{J}(t,\mathbf{x})=0,
\end{equation}
where we have defined a flux
\begin{equation}\label{eq:J}
\mathbf{J}(t,\mathbf{x})=
\sum_i K\left[\mathbf{x}-\mathbf{x}_i(t)\right]\,v_i\mathbf{p}_i(t).
\end{equation}
Using this definition and Eq.~(\ref{eqn:dthetadt_cf}), we find a dynamic equation for $\mathbf{J}$ of the form
\begin{equation}\label{eq:FluxEqcg}
\partial_t\mathbf{J}(t,\mathbf{x})+\nabla\cdot\boldsymbol{\sigma}(t,\mathbf{x})=\mathbf{T}(t,\mathbf{x})+\boldsymbol{\Phi}(t,\mathbf{x}).
\end{equation}
Here, we have defined the tensor and vector fields
\begin{subequations}\label{eq:cgfjorn_a}
\begin{align}
\boldsymbol{\sigma}(t,\mathbf{x})&=
\sum_i
K\left[\mathbf{x}-\mathbf{x}_i(t)\right]\,
v_i^2\mathbf{p}_i(t)\mathbf{p}_i(t),\\
\mathbf{T}(t,\mathbf{x})&=
\boldsymbol{\epsilon}\cdot\sum_i
K\left[\mathbf{x}-\mathbf{x}_i(t)\right]\,
v_i\Omega_i\mathbf{p}_i(t),\\
\boldsymbol{\Phi}(t,\mathbf{x})&=
\sum_i
K\left[\mathbf{x}-\mathbf{x}_i(t)\right]\,
v_i\mathbf{F}_i(t).
\end{align}
\end{subequations}
Averaging the fields in Eq.~(\ref{eq:J}) and Eqs.~(\ref{eq:cgfjorn_a}) over the particle parameter distribution $p(v_i,\Omega_i)$ yields
\begin{subequations}\label{eq:cgfjorn}
\begin{eqnarray}
\langle\mathbf{J}(t,\mathbf{x})\rangle_p&=&
\left\langle \sum_iK\left[\mathbf{x}-\mathbf{x}_i(t)\right]v_i\mathbf{p}_i(t)\right\rangle_p,\\
\langle\boldsymbol{\sigma}(t,\mathbf{x})\rangle_p&=&
\left\langle\sum_i
K\left[\mathbf{x}-\mathbf{x}_i(t)\right]\,
v_i^2\,\mathbf{p}_i(t)\mathbf{p}_i(t)\right\rangle_p,\phantom{\hspace{1cm}}\\
\langle\mathbf{T}(t,\mathbf{x})\rangle_p&=&
\left\langle\boldsymbol{\epsilon}\cdot\sum_i
K\left[\mathbf{x}-\mathbf{x}_i(t)\right]v_i\Omega_i\,\mathbf{p}_i(t)\right\rangle_p,\\
\langle\boldsymbol{\Phi}(t,\mathbf{x})\rangle_p&=&
\left\langle\sum_i
K\left[\mathbf{x}-\mathbf{x}_i(t)\right]\,
v_i\mathbf{F}_i(t)\right\rangle_p.
\end{eqnarray}
\end{subequations}
We then adopt a moment factorization approximation
\begin{subequations}\label{eq:cgfjorn_avg}
\begin{align}
\langle\mathbf{J}(t,\mathbf{x})\rangle_p&\simeq\langle v_i\rangle_p\mathbf{p},\\
\langle\boldsymbol{\sigma}(t,\mathbf{x})\rangle_p&\simeq
\frac{1}{2}\langle v_i^2\rangle_p\left(\rho\mathbb{I}+\mathbf{Q}\right),\label{eq:nemmom}\\
\langle\mathbf{T}(t,\mathbf{x})\rangle_p&\simeq
\langle v_i\Omega_i\rangle_p\mathbf{p}_{\perp},
\end{align}
\end{subequations}
where we used the definition of the particle number density in Eq.~(\ref{eq:densCG}), the polarization density in Eq.~(\ref{eq:pCG}), and \hbox{$\vert\mathbf{p}_i\vert^2=1$}. Additionally, we have defined in Eq.~(\ref{eq:nemmom}) a nematic moment of the form
\begin{equation}\label{eq:NemTens}
\mathbf{Q}=\sum_i K\left[\mathbf{x}-\mathbf{x}_i(t)\right]\left[2\mathbf{p}_i(t)\mathbf{p}_i(t)-\mathbb{I}\right].
\end{equation}
Averaging Eqs.~(\ref{eq:ContEqcg}) and (\ref{eq:FluxEqcg}) over the microscopic parameter distributions and using Eqs.~(\ref{eq:cgfjorn_avg}), we obtain
\begin{subequations}\label{eq:CoeffComp}
\begin{align}
\partial_t\rho+\langle v_i\rangle_p\nabla\cdot\mathbf{p}&=0,\\
\partial_t\mathbf{p}+\frac{\langle v_i^2\rangle_p}{2\langle v_i\rangle_p}\left(\nabla\rho+\nabla\cdot\mathbf{Q}\right)&=\frac{\langle v_i\Omega_i\rangle_p}{\langle v_i\rangle_p}\mathbf{p}_{\perp}+\langle v_i\rangle_p^{-1}\langle\boldsymbol{\Phi}\rangle_p.\label{eq:dtp_kcg}
\end{align}
\end{subequations}
From this, we can read off predictions about the coefficients we expect to find from the learning framework for the terms $\nabla\cdot\mathbf{p}$, $\nabla\rho$ and $\mathbf{p}_{\perp}$~(Tab.~\ref{Tab:VortexModelCoeffs}).

\am{Following the same analytic coarse-graining strategy, but starting from a more general microscopic position dynamics
\begin{equation}\label{eq:pscal}
\frac{d\mathbf{x}_i}{dt} = v_i \mathbf{p}_i+v'\boldsymbol{\epsilon}\cdot\mathbf{p}_i,
\end{equation}
yields an additional term $\sim v'\,\nabla_\perp\rho$ in the polar dynamics SI Eq.~(\ref{eq:dtp_kcg}). It is therefore a simple, non-interacting term -- for which analytic coarse-graining reliably predicts coefficients (see Tab.~\ref{Tab:FullCGcomparison}) -- that would result from a chiral propagation of particles. However, such a chiral propagation is absent in the microscopic model used in our work [see Eq.~(\ref{eqn:drdt}) in the main text where $v'_i=0$], such that the term $\nabla_\perp\rho$ was omitted from the library, which is equivalent to omitting the Levi-Civita tensor $\boldsymbol{\epsilon}$ from the set $\mathcal{S}$ given in Eq.~(\ref{eq:termset}).}

\am{\subsection{\am{Physical interpretation of higher-order terms and relation to microscopic particle properties}}\label{App:PhysInterp}
The analytic coarse-graining results in Tab.~\ref{Tab:FullCGcomparison} are based on the simplifying assumption of constant microscopic parameters $v_i=v_0$ and $\Omega_i=\Omega_0$ and invoke the various other approximations described in SI~Sec.~\ref{App:AnalyticCoarse1}. Nevertheless, a more detailed comparison with learning results is also instructive for terms that are non-linear in the fields or derivatives. In the following, we discuss this comparison for three different groups of terms.

\begin{itemize}
\item Terms that are present in the learned models and also expected by the analytic coarse-graining 
$$
\{b_5|\mathbf{p}^2|\mathbf{p}, b_8(\mathbf{p}\cdot\nabla)\mathbf{p}, b_{13}\Delta\mathbf{p}, b_{13}\Delta\mathbf{p}_{\perp}, b_{16}(\nabla\cdot\mathbf{p})\mathbf{p}\}
$$ 
can be understood as dynamic coupling of the polar order parameter~$\mathbf{p}$ to the nematic order~$\mathbf{Q}$ that naturally arise from symmetry arguments [see terms $\nabla\cdot\mathbf{Q}$ and $\mathbf{Q}\cdot\mathbf{p}$ in Eq.~(\ref{eq:dyneqp})]. In particular, we can consider a general expansion of nematic tensors constructed in terms of the polar order parameter field
\begin{equation}
\mathbf{Q}=\nu_1(\mathbf{p}\mathbf{p}-\mathbb{I}/2)+\nu_2(\nabla\mathbf{p}+\nabla\mathbf{p}^{\top}-\nabla\cdot\mathbf{p}\mathbb{I})+\nu_3(\nabla\mathbf{p}_{\perp}+\nabla\mathbf{p}_{\perp}^{\top}-\nabla\cdot\mathbf{p}_{\perp}\mathbb{I})+...,
\end{equation}
which is similar to defining a closure relation, where the coefficients $\nu_1,\nu_2,\nu_3,...$ are here inferred via a data-driven approach.

\item A second group of non-linear terms can be defined as those that the learning approach identifies as relevant contributions, but which would not be predicted by the analytic coarse-graining for constant microscopic parameters 
$$
\{
b_4\rho\mathbf{p}_{\perp}, b_{11}\nabla(\nabla\cdot\mathbf{p}), b_{12}\nabla(\nabla\cdot\mathbf{p}), b_{18}\Delta^2\mathbf{p}, b_{19}\Delta^2\mathbf{p}
\}.
$$
Here, the term $b_4\rho\mathbf{p}_{\perp}$ ($b_4<0$) describes a density-dependent reduction of average particle rotations through collective effects, which (mildly) counteracts the term $b_3\mathbf{p}_{\perp}$ ($b_3>0$) related to single particle rotations. Such a contribution is most likely a consequence of the skewed distribution of single particle rotation rates $\Omega_i$ (Fig.~\ref{Fig:ParamDist}): A larger amount of particles has rotation frequencies $\Omega_i<\Omega_0$, such that the average rotation frequency of the finite subsample of particles present in a given vortex (i.e. regions with high density) tends to be reduced as compared to the mean value~$\Omega_0$ of the overall distribution. The terms $b_{11}\nabla(\nabla\cdot\mathbf{p})$ and $b_{12}\nabla(\nabla\cdot\mathbf{p})$ with $b_{11},b_{12}>0$ correspond to non-standard diffusive terms~\cite{Toner1995} that can generally exist in any flocking-type model and in our case arise through the combination of rotational diffusion and microscopic parameter variability. The terms $b_{18}\Delta^2\mathbf{p}$ and $b_{19}\Delta^2\mathbf{p}$ with $b_{18},b_{19}>0$ correspond to Swift-Hohenberg-type terms, which typically appear in systems that exhibit pattern formation on mesoscopic length scales~\cite{Cross09}. In our case, the relevant length scales are determined by the microscopic parameters as $v_0/\Omega_0$, corresponding to the approximate radius of vortices, as well as by the finite interaction range among particles, which was also recently shown to give rise to Swift-Hohenberg-type operators on the mean-field level~\cite{mietke2021}.

\item Lastly, we identify a third group of non-linear terms 
$$
\{b_6|\mathbf{p}^2|\mathbf{p}_{\perp},  b_9(\mathbf{p}\cdot\nabla)\mathbf{p}_{\perp}, 
b_{10}(\mathbf{p}_{\perp}\cdot\nabla)\mathbf{p},
b_{15}\nabla|\mathbf{p}|^2\}
$$ 
as those that are expected by the analytic coarse-graining of a system with constant microscopic parameters, but which do not appear in the most parsimonious model that recapitulates the dynamics (PDE~8 in Tabs.~\ref{Tab:FullCGcomparison} and~\ref{Tab:VortexMomCoeffs}). However, these terms will be consecutively added to the learned models, when looking for those with higher complexity~{(Tab.~\ref{Tab:VortexMomCoeffs2})}, but their effect on the overall dynamics and pattern formation is minute~(Fig.~\ref{Fig:VortexSpectraComparison}, mid- and bottom-row). 
\end{itemize}
}
\begin{table*}
\caption{Parameters $a_l$ of the density dynamics PDE~(Fig.~\ref{Fig:MassCons}d) learned from simulations the microscopic active particle system in Eq.~\eqref{eqn:EOM}. The sparsest model ($\blacktriangleleft$) agrees well with the analytic coarse-graining prediction~(Tab.~\ref{Tab:VortexModelCoeffs}).  \label{Tab:VortexMassCoeffs}}
\begin{ruledtabular}
\begin{tabular}{l|cccc}
Term&PDE 1$\blacktriangleleft$&PDE 2&PDE 3&PDE 4\\\hline$a_{1}$$\nabla \cdot \mathbf{p}$&-0.991&-0.991&-0.972&-0.957\\$a_{2}$$\Delta \rho$&--&--&--&--\\$a_{3}$$\nabla \cdot (\rho \mathbf{p})$&--&--&-0.015&-0.037\\$a_{4}$$\Delta \rho^2$&--&\hphantom{-}0.022&\hphantom{-}0.022&\hphantom{-}0.022\\$a_{5}$$\Delta \vert \mathbf{p} \vert^2$&--&--&--&--\\$a_{6}$$\nabla \cdot (\rho^2 \mathbf{p})$&--&--&--&\hphantom{-}0.008\\$a_{7}$$\Delta \rho^3$&--&--&--&--\\$a_{8}$$\nabla \cdot (\vert \mathbf{p}\vert^2 \mathbf{p})$&--&--&--&--\\$a_{9}$$\nabla \cdot (\rho \nabla \vert \mathbf{p}\vert^2)$&--&--&--&--\\$a_{10}$$\nabla \cdot (\vert \mathbf{p}\vert^2 \nabla\rho)$&--&--&--&--\\$a_{11}$$\nabla \cdot \mathbf{p}_\perp$&--&--&--&--\\$a_{12}$$\nabla \cdot (\rho \mathbf{p}_\perp)$&--&-0.026&-0.026&-0.026\\$a_{13}$$\nabla \cdot (\rho^2 \mathbf{p}_\perp)$&--&--&--&--\\$a_{14}$$\nabla \cdot (\vert \mathbf{p}\vert^2 \mathbf{p}_\perp )$&--&--&--&--\\
    \end{tabular}
    \end{ruledtabular}
\end{table*}

\begin{table*}
\caption{Parameters $b_l$ of the nine sparsest PDEs for the polarization dynamics~(Fig.~\ref{Fig:MomCons}c), learned from simulations of the microscopic system in Eq.~\eqref{eqn:EOM}. PDE 8 ($\blacktriangleleft$) reproduces the characteristic vortex dynamics as in the microscopic simulations~(Fig.~\ref{Fig:MomCons}a,b,e) and the coefficients of the linear terms compare well with analytic coarse-graining predictions~(Tab.~\ref{Tab:VortexModelCoeffs}). Four additional PDEs with more terms are shown in Tab.~\ref{Tab:VortexMomCoeffs2}.  \label{Tab:VortexMomCoeffs}}
\begin{ruledtabular}
\begin{tabular}{l|ccccccccc}
Term&PDE 1&PDE 2&PDE 3&PDE 4&PDE 5&PDE 6&PDE 7&PDE 8$\blacktriangleleft$&PDE 9\\\hline$b_{1}$$\mathbf{p}$&-0.009&-0.009&-0.009&-0.009&-0.009&-0.009&-0.009&-0.009&-0.009\\$b_{2}$$\rho \mathbf{p}$&--&--&--&\hphantom{-}0.013&\hphantom{-}0.013&\hphantom{-}0.013&\hphantom{-}0.007&\hphantom{-}0.009&\hphantom{-}0.009\\$b_{3}$$\mathbf{p}_\perp$&\hphantom{-}0.414&\hphantom{-}0.476&\hphantom{-}0.477&\hphantom{-}0.428&\hphantom{-}0.478&\hphantom{-}0.436&\hphantom{-}0.436&\hphantom{-}0.440&\hphantom{-}0.441\\$b_{4}$$\rho \mathbf{p}_\perp$&--&-0.050&-0.040&--&-0.040&-0.006&-0.006&-0.010&-0.012\\$b_{5}$$|\mathbf{p}|^2 \mathbf{p}$&--&--&--&--&--&--&--&-0.080&-0.080\\$b_{6}$$|\mathbf{p}|^2 \mathbf{p}_\perp$&--&--&--&--&--&--&--&--&\hphantom{-}0.054\\$b_{7}$$\nabla \rho$&-0.638&-0.637&-0.600&-0.595&-0.601&-0.596&-0.596&-0.595&-0.595\\$b_{8}$$(\mathbf{p} \cdot \nabla) \mathbf{p}$&--&--&--&-0.536&--&-0.510&-0.510&-0.463&-0.479\\$b_{9}$$(\mathbf{p} \cdot \nabla) \mathbf{p}_\perp$&--&--&--&--&--&--&--&--&--\\$b_{10}$$(\mathbf{p}_\perp \cdot \nabla) \mathbf{p}$&--&--&--&--&--&--&--&--&--\\$b_{11}$$\nabla(\nabla \cdot \mathbf{p})$&--&--&--&--&--&--&--&\hphantom{-}0.078&\hphantom{-}0.077\\$b_{12}$$\nabla(\nabla \cdot \mathbf{p}_\perp)$&--&--&\hphantom{-}0.225&\hphantom{-}0.265&\hphantom{-}0.248&\hphantom{-}0.265&\hphantom{-}0.270&\hphantom{-}0.277&\hphantom{-}0.277\\$b_{13}$$\Delta \mathbf{p}$&--&--&--&--&--&--&-0.151&-0.155&-0.156\\$b_{14}$$\Delta \mathbf{p}_\perp$&--&--&\hphantom{-}0.252&\hphantom{-}0.202&\hphantom{-}0.222&\hphantom{-}0.203&\hphantom{-}0.198&\hphantom{-}0.196&\hphantom{-}0.197\\$b_{15}$$\nabla \vert \mathbf{p}\vert^2$&--&--&--&--&--&--&--&--&--\\$b_{16}$$(\nabla \cdot \mathbf{p})\mathbf{p}$&--&--&--&--&--&--&--&-0.225&-0.213\\$b_{17}$$(\nabla \cdot \mathbf{p})\mathbf{p}_\perp$&--&--&--&--&--&--&--&--&--\\$b_{18}$$\Delta^2 \mathbf{p}$&--&--&--&--&--&--&-0.475&-0.483&-0.484\\$b_{19}$$\Delta^2 \mathbf{p}_\perp$&--&--&\hphantom{-}1.100&\hphantom{-}1.197&\hphantom{-}1.085&\hphantom{-}1.212&\hphantom{-}1.215&\hphantom{-}1.235&\hphantom{-}1.243\\
    \end{tabular}
    \end{ruledtabular}
\end{table*}

\begin{table*}
\markblue{\caption{Parameters $b_l$ of PDEs 10-13 for the polarization dynamics~(Fig.~\ref{Fig:MomCons}c), learned from simulations of the microscopic system in Eq.~\eqref{eqn:EOM}. These are learned in addition to the PDEs mentioned in Tab.~\ref{Tab:VortexMomCoeffs}. 
\label{Tab:VortexMomCoeffs2}}}
\begin{ruledtabular}
\markblue{
\begin{tabular}{l|cccc}
Term&PDE 10&PDE 11&PDE 12&PDE 13\\\hline$b_{1}$$\mathbf{p}$&-0.009&-0.009&-0.009&-0.009\\$b_{2}$$\rho \mathbf{p}$&\hphantom{-}0.010&\hphantom{-}0.010&\hphantom{-}0.010&\hphantom{-}0.010\\$b_{3}$$\mathbf{p}_\perp$&\hphantom{-}0.442&\hphantom{-}0.442&\hphantom{-}0.442&\hphantom{-}0.448\\$b_{4}$$\rho \mathbf{p}_\perp$&-0.012&-0.012&-0.012&-0.017\\$b_{5}$$|\mathbf{p}|^2 \mathbf{p}$&-0.079&-0.060&-0.066&-0.065\\$b_{6}$$|\mathbf{p}|^2 \mathbf{p}_\perp$&\hphantom{-}0.055&\hphantom{-}0.055&\hphantom{-}0.055&\hphantom{-}0.065\\$b_{7}$$\nabla \rho$&-0.595&-0.595&-0.595&-0.594\\$b_{8}$$(\mathbf{p} \cdot \nabla) \mathbf{p}$&-0.479&-0.479&-0.480&-0.461\\$b_{9}$$(\mathbf{p} \cdot \nabla) \mathbf{p}_\perp$&--&\hphantom{-}0.057&\hphantom{-}0.058&\hphantom{-}0.058\\$b_{10}$$(\mathbf{p}_\perp \cdot \nabla) \mathbf{p}$&--&--&\hphantom{-}0.054&\hphantom{-}0.054\\$b_{11}$$\nabla(\nabla \cdot \mathbf{p})$&\hphantom{-}0.078&\hphantom{-}0.078&\hphantom{-}0.076&\hphantom{-}0.076\\$b_{12}$$\nabla(\nabla \cdot \mathbf{p}_\perp)$&\hphantom{-}0.277&\hphantom{-}0.277&\hphantom{-}0.277&\hphantom{-}0.278\\$b_{13}$$\Delta \mathbf{p}$&-0.153&-0.150&-0.138&-0.138\\$b_{14}$$\Delta \mathbf{p}_\perp$&\hphantom{-}0.197&\hphantom{-}0.197&\hphantom{-}0.197&\hphantom{-}0.195\\$b_{15}$$\nabla \vert \mathbf{p}\vert^2$&--&--&--&-0.023\\$b_{16}$$(\nabla \cdot \mathbf{p})\mathbf{p}$&-0.213&-0.215&-0.218&-0.202\\$b_{17}$$(\nabla \cdot \mathbf{p})\mathbf{p}_\perp$&-0.117&-0.151&-0.171&-0.171\\$b_{18}$$\Delta^2 \mathbf{p}$&-0.489&-0.454&-0.403&-0.403\\$b_{19}$$\Delta^2 \mathbf{p}_\perp$&\hphantom{-}1.244&\hphantom{-}1.244&\hphantom{-}1.245&\hphantom{-}1.236\\    \end{tabular}
}
    \end{ruledtabular}
\end{table*}

\begin{table*}
\caption{Parameters $c_l$ of the PDE for the density dynamics~(Fig.~\ref{Fig:Quincke}c) learned from experimental data for self-propelled Quincke rollers~(Supplementary Movie S2 of Ref.~\cite{Geyer2018a}). The dimensions of the coefficients are such that $[\mathbf{v}]=$mm/s and $[\rho]=1$, where the density $\rho$ represents the area fraction of rollers of diameter $D_c=4.8\,\si{\micro\meter}$. The four sparsest PDEs are shown corresponding to the cut-off $n_0 \in \{50, 100\}$ above which the temporal Chebyshev modes in Eq.~\eqref{Eq:BasisProjection} are set to zero to ignore high frequencies. The sparsest PDEs ($\blacktriangleleft$) have coefficients close to each other and agree well with the mass conservation equation obtained from analytic coarse-graining~(Ref.~\cite{Geyer2018a}). \label{Tab:QuinckeMassCoeffs}}
\begin{ruledtabular}
\begin{tabular}{lr|cccc|cccc}
&&\multicolumn{4}{c|}{$n_0=50$}&\multicolumn{4}{c}{$n_0 = 100$}\\
Term& Unit&PDE 1$\blacktriangleleft$&PDE 2&PDE 3&PDE 4&PDE 1$\blacktriangleleft$&PDE 2&PDE 3&PDE 4\\\hline$c_{1}$$\nabla \cdot \mathbf{v}$&--&--&--&-0.052&-0.052&--&-0.051&-0.051&-0.055\\$c_{2}$$\Delta \rho$&mm$^{2}$ s$^{-1}$&--&\hphantom{-}0.016&\hphantom{-}0.040&\hphantom{-}0.023&--&\hphantom{-}0.055&\hphantom{-}0.040&\hphantom{-}0.041\\$c_{3}$$\nabla \cdot (\rho \mathbf{v})$&--&-0.950&-0.950&-1.068&-1.067&-0.945&-1.057&-1.054&-0.985\\$c_{4}$$\Delta \rho^2$&mm$^{2}$ s$^{-1}$&--&\hphantom{-}0.047&\hphantom{-}0.080&\hphantom{-}0.081&--&-0.076&-0.051&-0.062\\$c_{5}$$\Delta \vert \mathbf{v} \vert^2$&s&--&--&--&\hphantom{-}0.001&--&--&\hphantom{-}0.001&\hphantom{-}0.001\\$c_{6}$$\nabla \cdot (\rho^2 \mathbf{v})$&--&--&--&--&--&--&--&--&-0.313\\$c_{7}$$\Delta \rho^3$&mm$^{2}$ s$^{-1}$&--&--&--&--&--&\hphantom{-}0.427&\hphantom{-}0.341&\hphantom{-}0.366\\$c_{8}$$\nabla \cdot (\vert \mathbf{v}\vert^2 \mathbf{v})$&mm$^{-2}$ s$^{2}$&--&--&\hphantom{-}0.035&\hphantom{-}0.035&--&\hphantom{-}0.034&\hphantom{-}0.034&\hphantom{-}0.034\\$c_{9}$$\nabla \cdot (\rho \nabla \vert \mathbf{v}\vert^2)$&s&--&--&--&-0.013&--&--&-0.008&-0.007\\$c_{10}$$\nabla \cdot (\vert \mathbf{v}\vert^2 \nabla\rho)$&s&--&-0.018&-0.039&-0.028&--&-0.036&-0.028&-0.027\\
    \end{tabular}
    \end{ruledtabular}
\end{table*}

\section{Parameters \am{and parameter robustness} of learned models} \label{App:LearnedCoeffs}
The parameters of the PDEs learned from simulations of the active polar particle model in Eq.~\eqref{eqn:EOM} are summarized in Tab.~\ref{Tab:VortexMassCoeffs} (density dynamics) and Tab.~\ref{Tab:VortexMomCoeffs} (polarization dynamics). 
For the experimental Quincke roller system~\cite{Geyer2018a}, the learned hydrodynamic model parameters are given in Tab.~\ref{Tab:QuinckeMassCoeffs} (density dynamics) and Tab.~\ref{Tab:QuinckeMomCoeffs} (velocity dynamics).

\begin{figure*}
    \centering
    \includegraphics[width=0.9\linewidth]{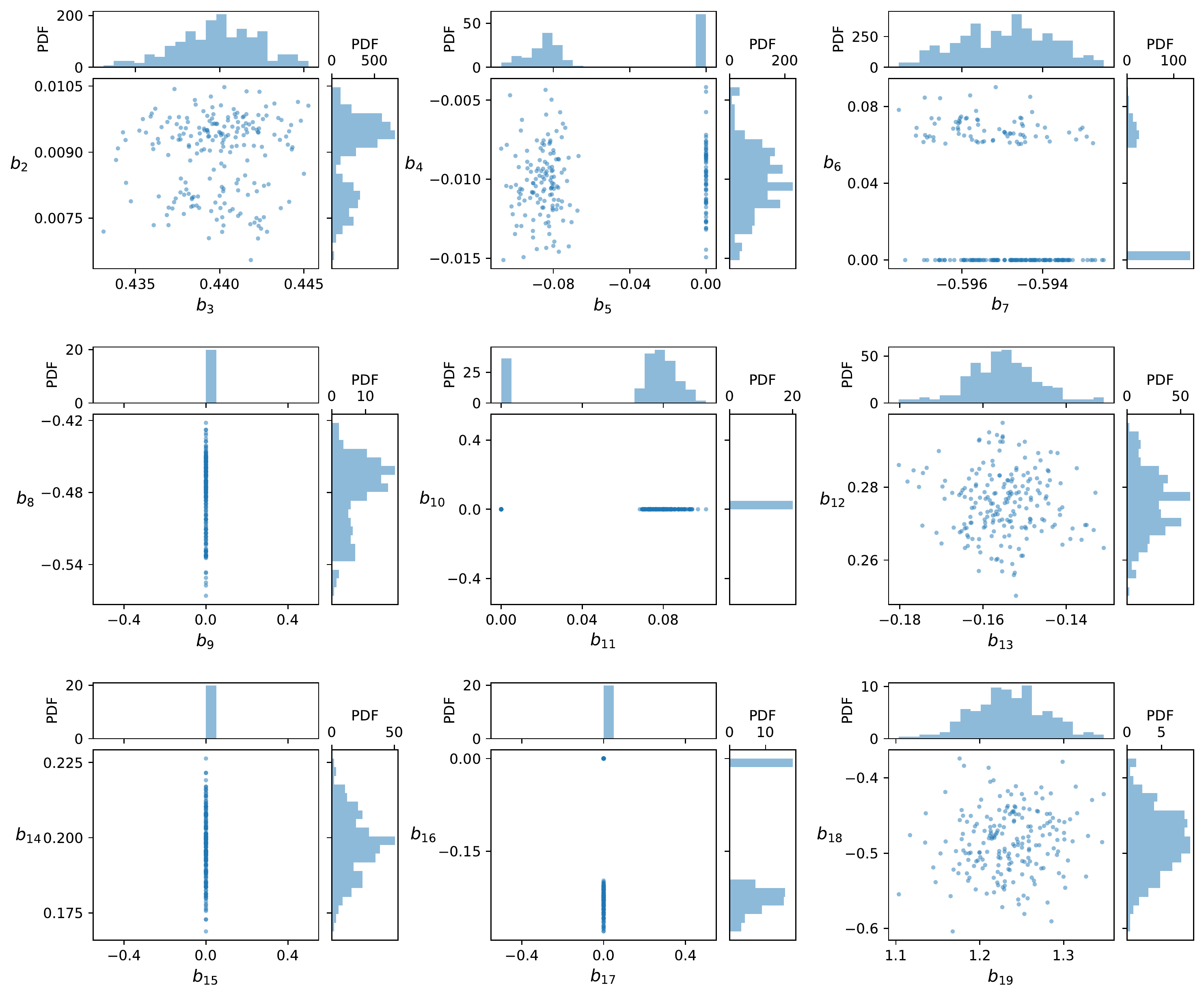}
    \markblue{
    \caption{Parameters $b_l$ of the polarization dynamics~{(Eq.~(\ref{Eqn:pPDELibrary}), Fig.~\ref{Fig:MomCons}c)} obtained by applying the STLSQ algorithm on 200 data sub-samples with 50\,\% randomly selected data points~{(see SI~Sec.~\ref{App:SparseReg})}. The units of the coefficients are the same as in~{Tab.~\ref{Tab:VortexMomCoeffs}}. This plot is generated for the thresholding parameter $\tau = 1.34 \times 10^{-4}$, at which PDE~8~{(Tab.~\ref{Tab:VortexMomCoeffs})} is found in more than $60$\,\% of the sub-samples. Histograms indicate the marginal probability density functions (PDFs) of the corresponding coefficients. \label{Fig:CoeffsThresholded}}
    }
\end{figure*}

\begin{figure*}
    \centering
    \includegraphics[width=0.9\linewidth]{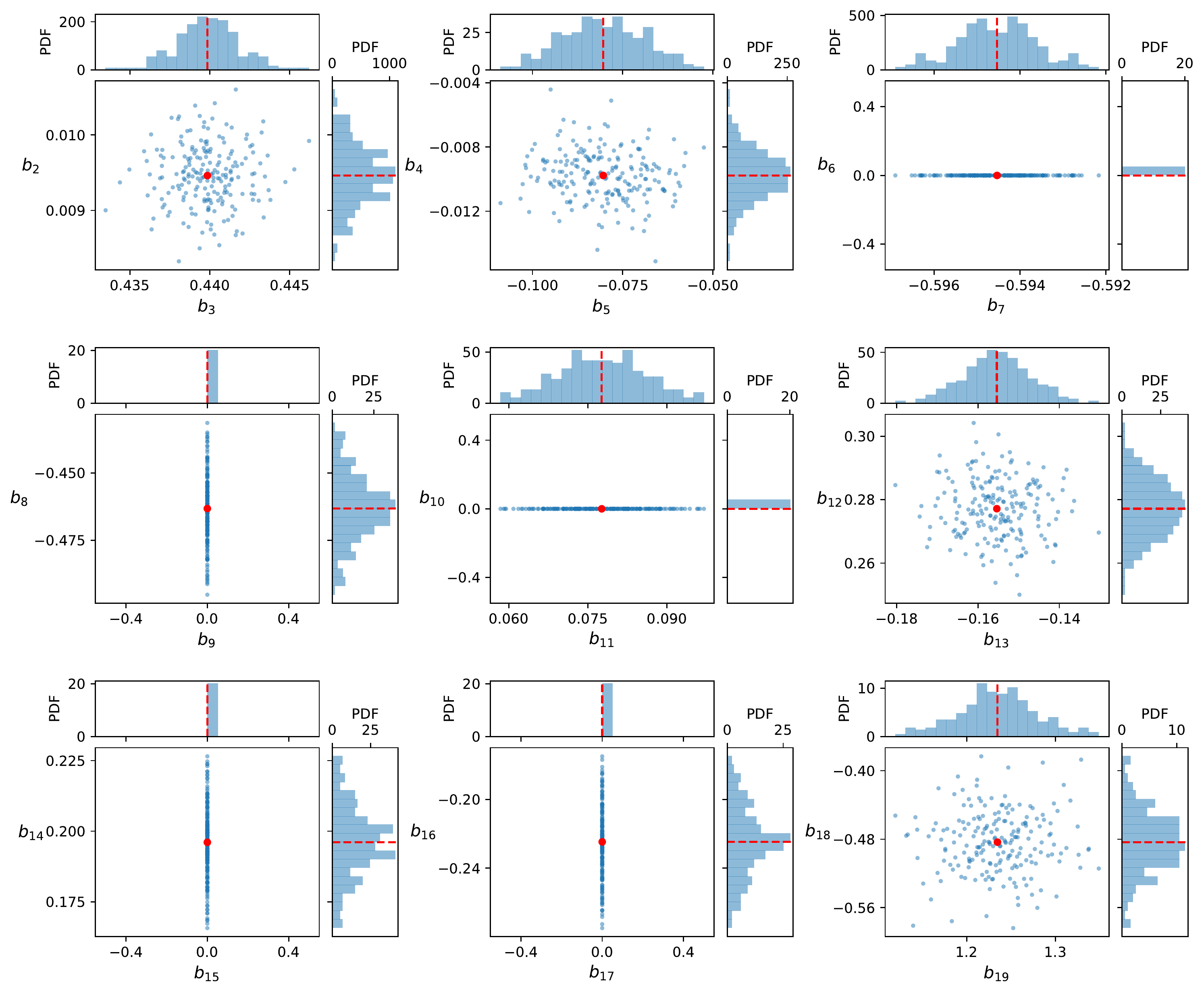}
    \markblue{
    \caption{Parameters $b_l$ of the polarization dynamics~{(Eq.~(\ref{Eqn:pPDELibrary}), Fig.~\ref{Fig:MomCons}c)}  obtained from the least-squares method by setting the thresholded coefficients in Fig.~\ref{Fig:CoeffsThresholded} to zero. The units of the coefficients are the same as in~{Tab.~\ref{Tab:VortexMomCoeffs}}. A total of 200 points (blue) are presented from 200 sub-samples with $50$\% randomly chosen data points. The red points indicate the fitted coefficients on the entire data set, which are the same as for PDE 8 in Tab.~\ref{Tab:VortexMomCoeffs}. The histograms indicate the marginal probability density functions (PDFs) of the corresponding coefficients, which can be used to find uncertainties for the coefficients. The fitted values and standard deviations of the coefficients are (mean\,$\pm$\,standard deviation): $b_{2}=0.009\pm0.0004$, $b_{3}=0.440\pm0.0020$, $b_{4}=-0.010\pm0.0016$, $b_{5}=-0.080\pm0.0111$, $b_{7}=-0.595\pm0.0009$, $b_{8}=-0.463\pm0.0124$, $b_{11}=0.078\pm0.0083$, $b_{12}=0.277\pm0.0092$, $b_{13}=-0.155\pm0.0082$, $b_{14}=0.196\pm0.0125$, $b_{16}=-0.225\pm0.0204$, $b_{18}=-0.483\pm0.0360$, $b_{19}=1.235\pm0.0434$. \label{Fig:CoeffsBootstrapped}}
    }
\end{figure*}

\begin{figure*}
    \centering
    \includegraphics[width=0.8\linewidth]{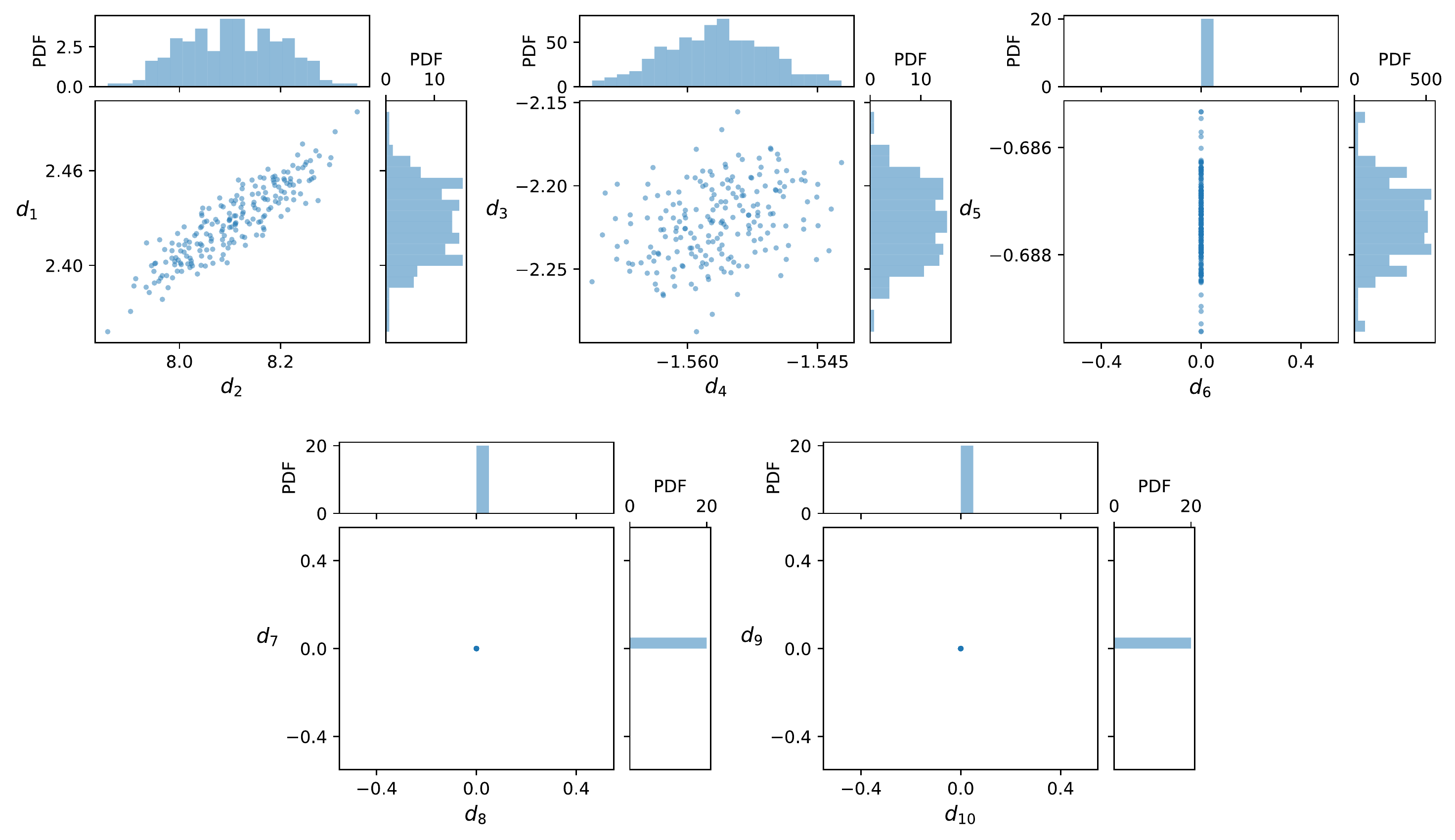}
    \markblue{
    \caption{Parameters $d_l$ of the velocity dynamics~{(Eq.~(\ref{Eqn:vexpPDELibrary}), Fig.~\ref{Fig:Quincke}c)}  obtained by applying the STLSQ algorithm on 200 data sub-samples with 50\% randomly selected data points~{(see SI~Sec.~\ref{App:SparseReg})}. The units of the coefficients are the same as in~{Tab.~\ref{Tab:QuinckeMomCoeffs}}. This plot is generated for the thresholding parameter $\tau = 9.59 \times 10^{-2}$, at which PDE~2~{(Tab.~\ref{Tab:QuinckeMomCoeffs} and $n_0 = 50$)} is found in more than $60$\% of the sub-samples. The histograms indicate the marginal probability density functions (PDFs) of the corresponding coefficients. \label{Fig:CoeffsThresholdedQuincke} }
    }
\end{figure*}

\begin{figure*}
    \centering
    \includegraphics[width=0.8\linewidth]{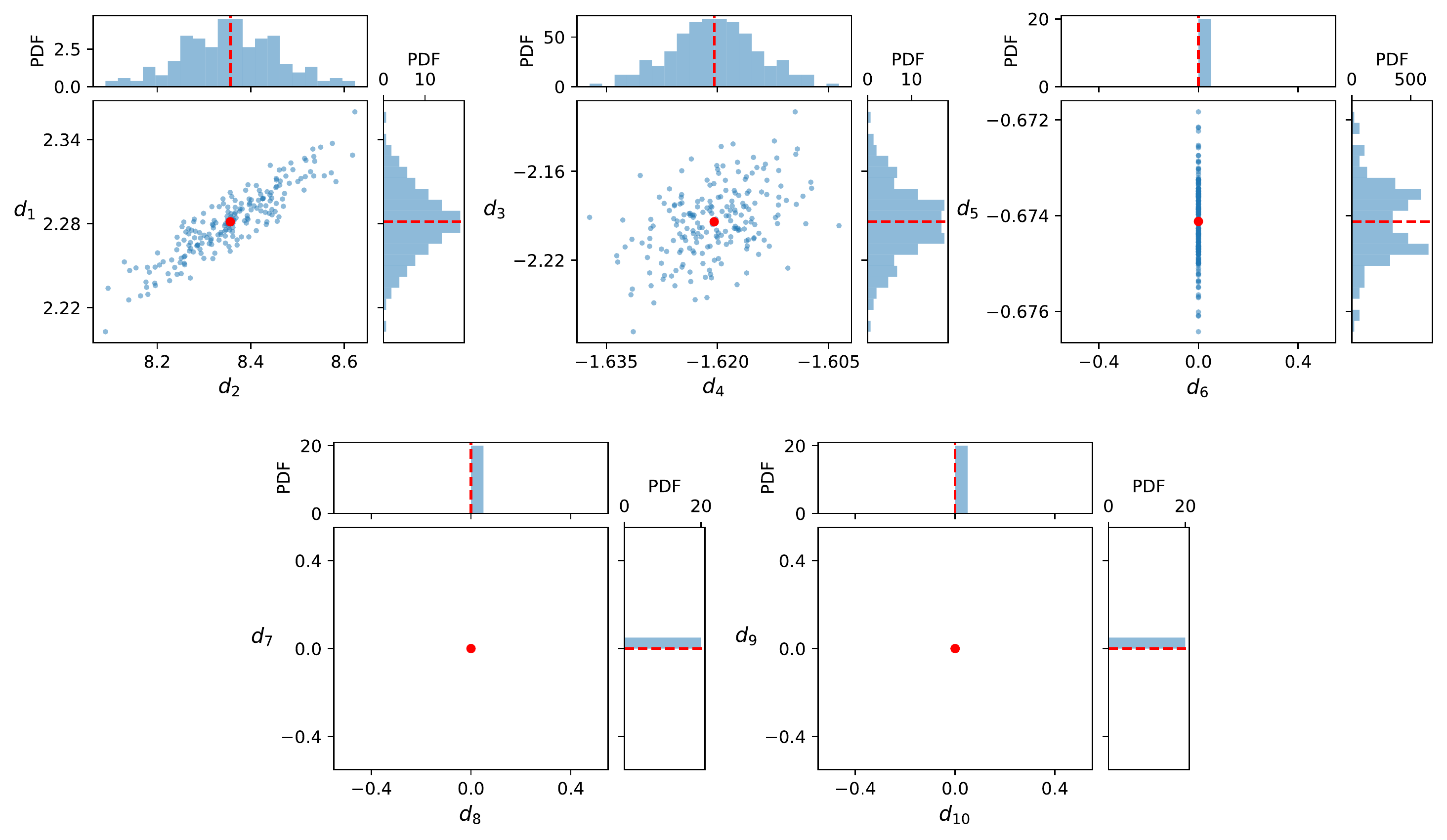}
    \markblue{
  \caption{Parameters $d_l$ of the velocity dynamics~{(Eq.~(\ref{Eqn:vexpPDELibrary}), Fig.~\ref{Fig:Quincke}c)} obtained from the least-squares method by setting the thresholded coefficients in Fig.~\ref{Fig:CoeffsThresholdedQuincke} to zero. The units of the coefficients are the same as in~{Tab.~\ref{Tab:QuinckeMomCoeffs}}. A total of 200 points (blue) are presented from 200 sub-samples with $50$\% randomly chosen data points. The red points indicate the fitted coefficients on the entire data set, which are the same as for PDE~2 in Tab.~\ref{Tab:QuinckeMomCoeffs}. The histograms indicate the marginal probability density functions (PDFs) of the corresponding coefficients. The fitted values and the standard deviations of the coefficients are~(mean\,$\pm$\,standard deviation): $d_{1}=2.281\pm0.0242$, $d_{2}=8.356\pm0.1027$, $d_{3}=-2.194\pm0.0239$, $d_{4}=-1.620\pm0.0059$, $d_{5}=-0.674\pm0.0008$. \label{Fig:CoeffsBootstrappedQuincke}}
  }
\end{figure*}

\markblue{The robustness of the sparse regression through STLSQ~{(SI~Sec.~\ref{App:SparseReg})} is demonstrated in Fig.~{\ref{Fig:CoeffsThresholded}} for the identification of the polarization dynamics~{[Eq.~(\ref{Eqn:pPDELibrary})]}. As specified in the stability selection procedure (SI~Sec.~\ref{App:SparseReg}), selected terms have non-zero coefficients in at least 60\,\% of the subsamples. To further quantify the uncertainty in the values of the coefficients, we performed a bootstrapping procedure by performing least-squares regression only on the terms identified. This leads to empirical probability density functions whose standard deviations can be used to quantify uncertainties in the coefficients~{(Fig.~\ref{Fig:CoeffsBootstrapped})}. Similar results for the velocity dynamics for the Quincke roller system~{[Eq.~\eqref{Eqn:vexpPDELibrary}]} are presented in Figs.~{\ref{Fig:CoeffsThresholdedQuincke}} and~{\ref{Fig:CoeffsBootstrappedQuincke}}.} 

\begin{table*}[h]
\caption{Parameters $d_l$ of the PDE for the velocity dynamics~(Fig.~\ref{Fig:Quincke}c) learned from experimental data for self-propelled Quincke rollers~(Supplementary Movie S2 of Ref.~\cite{Geyer2018a}). The dimensions of the coefficients are such that $[\mathbf{v}]=$mm/s and $[\rho]=1$, where the density $\rho$ represents the area fraction of rollers of diameter $D_c=4.8\,\si{\micro\meter}$. The four sparsest PDEs are shown corresponding to the cut-off $n_0 \in \{50, 100\}$ above which the temporal Chebyshev modes in Eq.~\eqref{Eq:BasisProjection} are set to zero to ignore high frequencies. The sparsest PDEs which reproduce the experimental observations ($\blacktriangleleft$) have coefficients that are close to each other for different values of $n_0$, and they agree well with corresponding values reported in Ref.~\cite{Geyer2018a}~(Tab.~\ref{Tab:QuinckeCoeffs}). \label{Tab:QuinckeMomCoeffs}}
\begin{ruledtabular}
\begin{tabular}{lr|cccc|cccc}
&&\multicolumn{4}{c|}{$n_0=50$}&\multicolumn{4}{c}{$n_0 = 100$}\\
Term& Unit&PDE 1&PDE 2$\blacktriangleleft$&PDE 3&PDE 4&PDE 1&PDE 2$\blacktriangleleft$&PDE 3&PDE 4\\\hline$d_{1}$$\mathbf{v}$&s$^{-1}$&--&\hphantom{-}2.281&\hphantom{-}1.524&\hphantom{-}1.491&--&\hphantom{-}1.825&\hphantom{-}1.252&\hphantom{-}1.122\\$d_{2}$$\rho \mathbf{v}$&s$^{-1}$&--&\hphantom{-}8.356&\hphantom{-}5.156&\hphantom{-}4.745&--&\hphantom{-}6.135&\hphantom{-}3.143&\hphantom{-}3.083\\$d_{3}$$|\mathbf{v}|^2 \mathbf{v}$&mm$^{-2}$ s&--&-2.194&-1.436&-1.382&--&-1.710&-1.095&-0.999\\$d_{4}$$\nabla \rho$&mm$^2$ s$^{-2}$&--&-1.620&-1.711&-2.074&--&-1.689&-2.438&-2.430\\$d_{5}$$(\mathbf{v}\cdot \nabla) \mathbf{v}$&--&-0.639&-0.674&-0.678&-0.679&-0.662&-0.696&-0.702&-0.702\\$d_{6}$$\nabla (\nabla \cdot \mathbf{v})$&mm$^{2}$ s$^{-1}$&--&--&--&--&--&--&--&--\\$d_{7}$$\Delta \mathbf{v}$&mm$^{2}$ s$^{-1}$&--&--&--&--&--&--&--&\hphantom{-}0.002\\$d_{8}$$\nabla (\vert \mathbf{v}\vert^2)$&--&--&--&--&\hphantom{-}0.090&--&--&\hphantom{-}0.169&\hphantom{-}0.168\\$d_{9}$$(\nabla \cdot \mathbf{v}) \mathbf{v}$&--&--&--&-0.189&-0.190&--&--&-0.178&-0.179\\$d_{10}$$\Delta^2 \mathbf{v}$&mm$^{4}$ s$^{-1}$&--&--&--&--&--&--&--&\hphantom{-}0.000\\
    \end{tabular}
    \end{ruledtabular}
\end{table*}

\begin{figure}[t]
    \centering
    \includegraphics[width=0.9\linewidth]{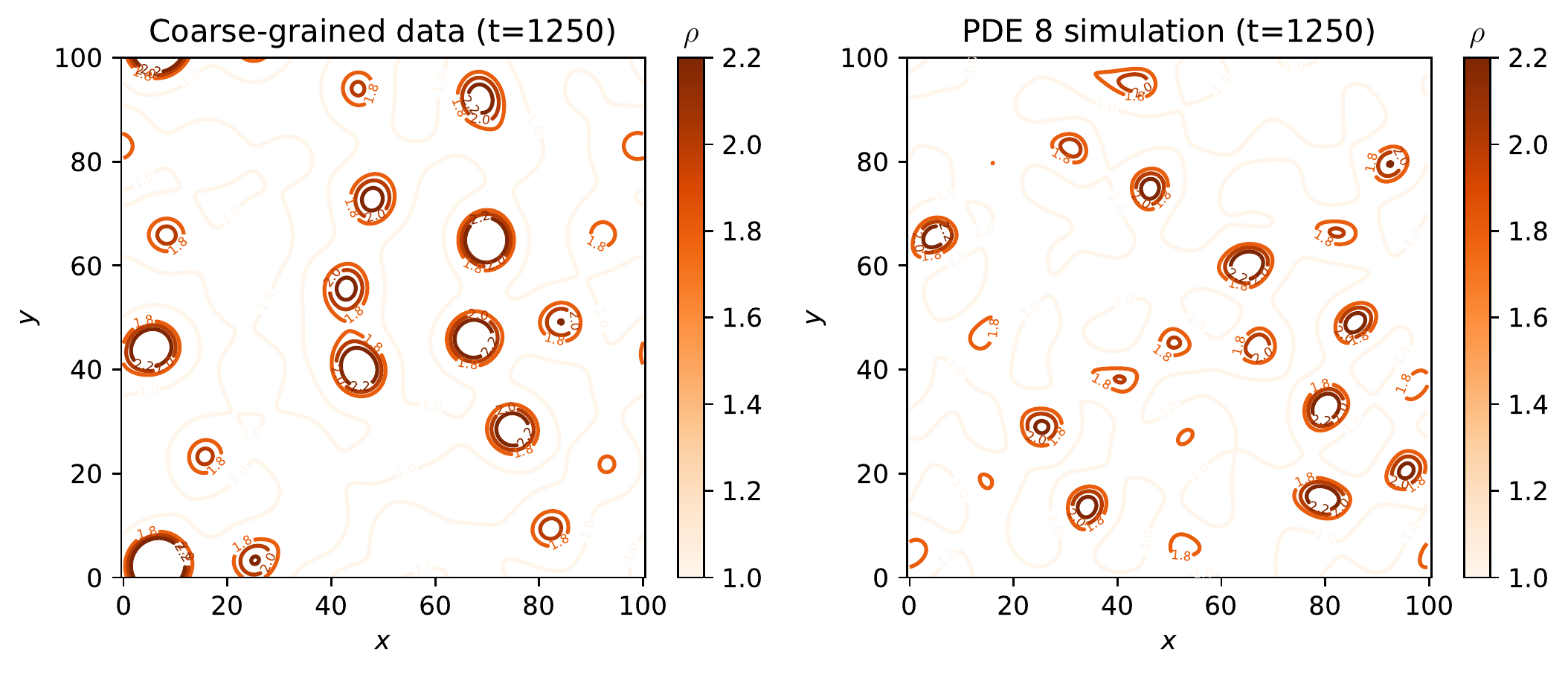}
    \markblue{
    \caption{Comparison of the contour plots of the density field $\rho(t, \mathbf{x})$ for the coarse-grained data (left) and the PDE simulation (right) corresponding to the chiral active Brownian model. Contour levels [1.0, 1.8, 2.0, 2.2] are the same in both plots. The contours distinctly isolate the most prominent vortices, which can be identified by the closely spaced red contours [1.8, 2.0, 2.2]. The representation shows that the number, density profiles, sizes, and disordered nature of vortices emerging in the learned model are very similar to the structures seen in the coarse-grained data.\label{Fig:VortexComparison}}
    }
\end{figure}

\begin{figure}
\centering
\includegraphics[width=0.8\textwidth]{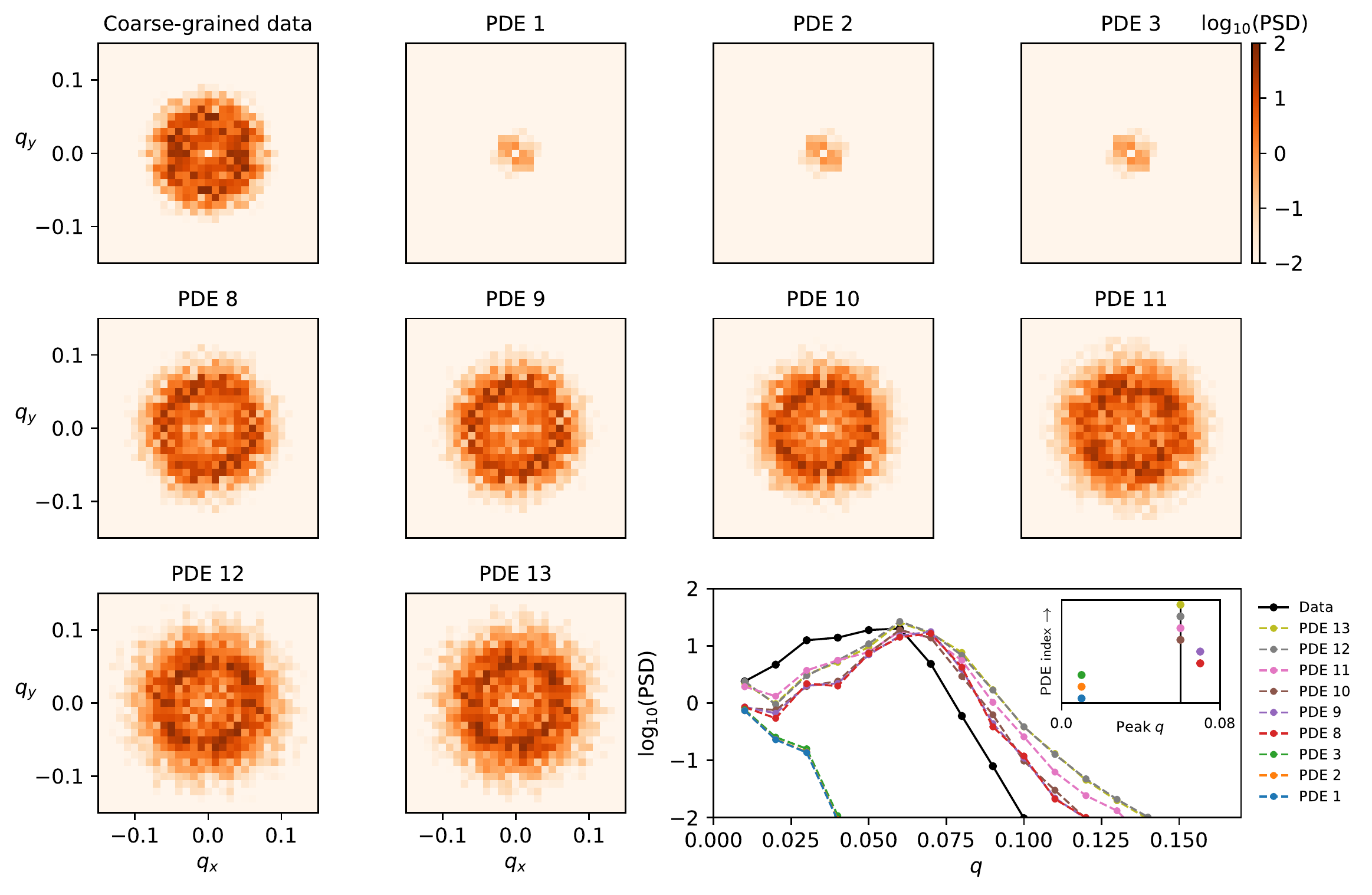}
\markblue{
\caption{Spatial power spectral density $S_{\rho; \mathbf{x}}(t, \mathbf{q}) = \mathcal{A}^{-1} \left|\int d^2 \mathbf{x} \, \rho(t, \mathbf{x})\exp(2\pi i \mathbf{q}\cdot\mathbf{x})\right|^2$ averaged over $100$ time points around $t=1250$ between the coarse-grained data (top left) and simulations of the learned polarization PDEs (other color plots). Bottom right: azimuthal average of the power spectrum. For PDE 8 and above, the peak of the spectrum compares well with that of the data. Inset: wavenumber $q$ corresponding to the peak of the spectrum for each PDE. This peak gets close to the peak of the data (black line) as the PDE index gets larger, that is, the PDE incorporates more terms. To improve further the agreement in the tails of spectra, additional higher-order derivatives~\cite{Slomka2017} must be accounted for in the library.\label{Fig:VortexSpectraComparison}}
}
\end{figure}

\begin{figure}
\centering
\includegraphics[width=0.5\textwidth]{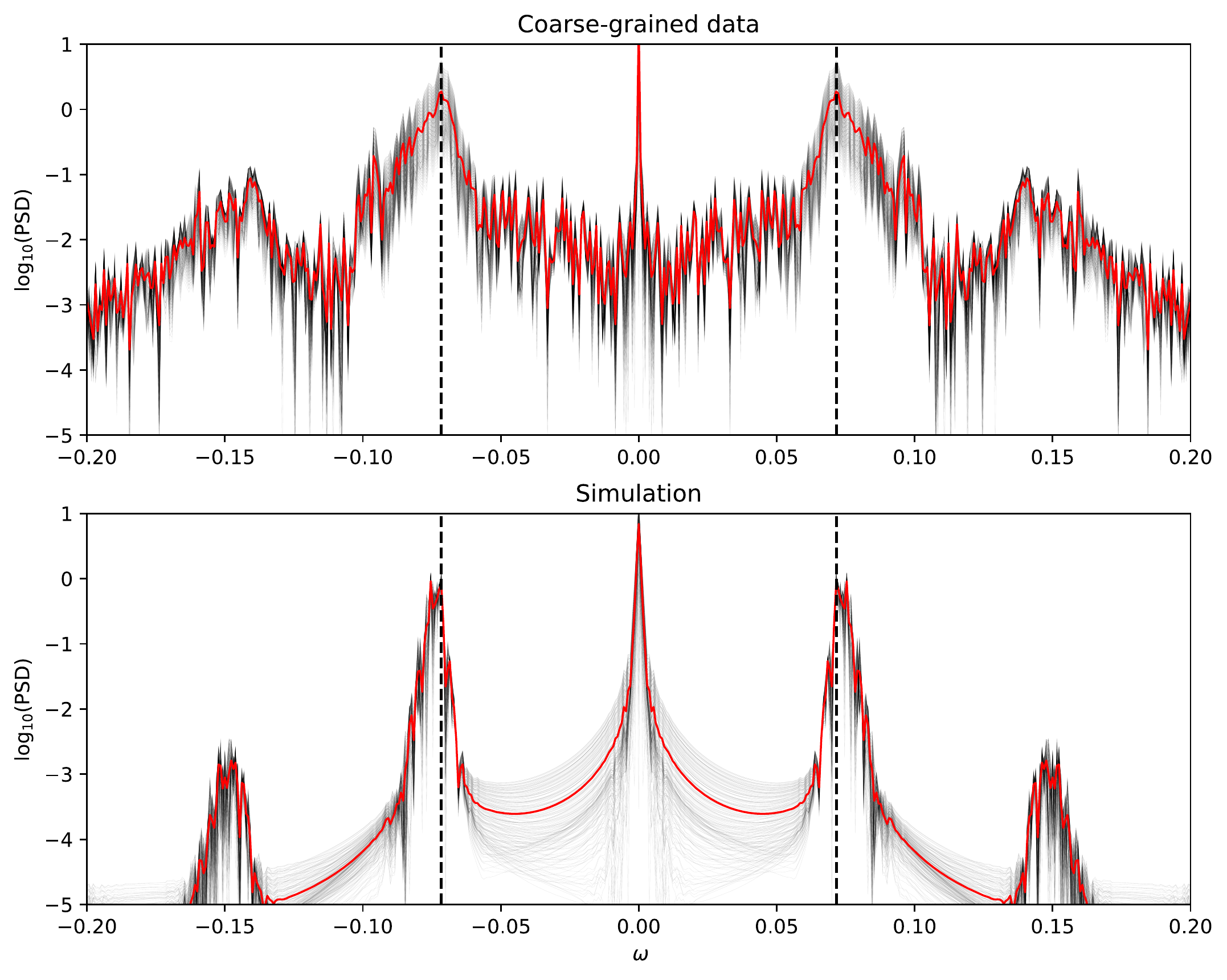}
\markblue{
\caption{Temporal power spectral density, $S_{p_x; t} (\omega, \mathbf{x}) = T^{-1} \left|\int dt \, p_x(t, \mathbf{x})\exp(2\pi i \omega t)\right|^2$ computed at grid points in the spatial window $x\in [45, 50]$, $y \in [45, 50]$ (black lines) for the time window $t\in[0, 1700]$. The red lines show the average over the spatial locations. The non-zero peaks in the spectrum for the coarse-grained data (top) and the simulation of PDE 8 for the polarization equation (Tab.~\ref{Tab:VortexMomCoeffs}, bottom) compare well with $\omega = \langle \Omega_i \rangle_p/2 \pi$ (black dotted lines). This implies that the simulation captures the bulk temporal dynamics in the input coarse-grained data at the average rotation frequency of the particles~{(see Fig.~\ref{Fig:ParamDist})}. The secondary peaks that result from nonlinearities also compare well between the simulations and the data. Note that the noise background in the coarse-grained particle data is several orders of magnitude smaller than the power in the characteristic peak frequencies. \label{Fig:TimeSpectraComparison}}
}
\end{figure}

\clearpage
\markblue{
\section{\am{Quantitative comparison of the particle model data and continuum simulations}}
\label{App:QuantComparison}
To facilitate a direct visual comparison between vortex patterns observed in the coarse-grained particle data and in simulations of the learned model~{(Fig.~\ref{Fig:MomCons})} the corresponding density fields are in SI~Fig.~{\ref{Fig:VortexComparison} depicted as a level set representation. Key pattern characteristics, such as the number, density profiles, sizes, and disordered nature of emerging vortices, show very good agreement between patterns seen in the coarse-grained data and in the simulated model.}\\ 

\noindent To further quantify this similarity between vortex patterns, we show in Fig.~\ref{Fig:VortexSpectraComparison} the spatial power spectral density, 
\begin{equation}\label{eq:SPSD}
S_{\rho; \mathbf{x}}(t, \mathbf{q}) = \mathcal{A}^{-1} \left|\int d^2 \mathbf{x} \, \rho(t, \mathbf{x})\exp(2\pi i \mathbf{q}\cdot\mathbf{x})\right|^2    
\end{equation} 
of the long-lived vortex states at $t=1250$ in the coarse-grained data and simulations. Here,  $\mathcal{A}$ is the domain area. The simulation data correspond to PDE~1 for the density equation~{(Tab.~\ref{Tab:VortexMassCoeffs})}, and PDEs~1-3 and~8-13 for the polarization equation~{(Tab.~\ref{Tab:VortexMomCoeffs}; PDEs~4-7 are found to be numerically unstable)}.  The comparison shows that PDE~8 is the sparsest model that captures the dominant flow length scale indicated by the shape of the spectra as well as close agreement with the peak. As more terms are incorporated in the PDE (that is, the PDE index gets larger), the peak of the inferred spectrum moves closer to the maximum seen in the data.\\ 

\noindent  In Fig.~\ref{Fig:TimeSpectraComparison}, we plot the temporal power spectral density
\begin{equation}\label{eq:TPSD}
S_{p_x; t} (\omega, \mathbf{x}) = T^{-1} \left|\int dt \, p_x(t, \mathbf{x})\exp(2\pi i \omega t)\right|^2   \end{equation}
over a time window $[0, T]$ and averaged over a spatial window. The spectra for the coarse-grained data and simulation of PDE 8 for the momentum equation~{(Tab.~\ref{Tab:VortexMomCoeffs})} show similar overall shape with four distinct peaks at non-zero frequencies. The first set of peaks occur at $2 \pi \omega = \pm \langle \Omega_i \rangle_p$ (the dotted lines), which is the average rotation frequency of the particles. This shows that the learned model reproduces the bulk temporal dynamics as seen in the input coarse-grained data.
}

\am{
\section{\am{Information content of coarse-grained data}}\label{App:InfCont}
To quantify the information loss due to coarse-graining as a function of the coarse-graining length scale $\sigma$, we use spectral entropy~\cite{Zhang2008,Pan2009} as a measure of the information content that remains in the coarse-grained fields. Specifically, we define the spectral entropy as
\begin{equation}\label{eq:SpecEntr}
H(\sigma)=-\sum_{\mathbf{q}}\hat{S}_{\rho;\textbf{x}}^{(\sigma)}(t,\mathbf{q})\log_2\hat{S}^{(\sigma)}_{\rho;\textbf{x}}(t,\mathbf{q}),
\end{equation}
where the normalized spatial power spectral density $\hat{S}_{\rho;\textbf{x}}^{(\sigma)}$ is defined as \begin{equation}\label{eq:NSPSD}
\hat{S}_{\rho; \mathbf{x}}(t, \mathbf{q}) = S_{\rho; \mathbf{x}}(t, \mathbf{q}) \left(\int d^2 \mathbf{q} \, S_{\rho; \mathbf{x}}(t, \mathbf{q})\right)^{-1},
\end{equation} 
with $S_{\rho; \mathbf{x}}(t, \mathbf{q})$ the spatial power spectral density defined in Eq.~(\ref{eq:SPSD}).  The additional index $\sigma$ indicates the Gaussian kernel smoothing width (\lq coarse-graining length scale\rq) with which the underlying density field $\rho(t,\textbf{x})$ was computed from the raw particle data.
For our analysis, we rescale the spectral entropy $H$ given in Eq.~(\ref{eq:SpecEntr}) by the spectral entropy of the raw particle data, yielding a normalized spectral entropy between $0$ and $1$ (see Figs.~\ref{Fig:SpecEntrVortex}~and~\ref{Fig:SpecEntrQuincke}).

\begin{figure}
    \centering
    \includegraphics[width=0.8\linewidth]{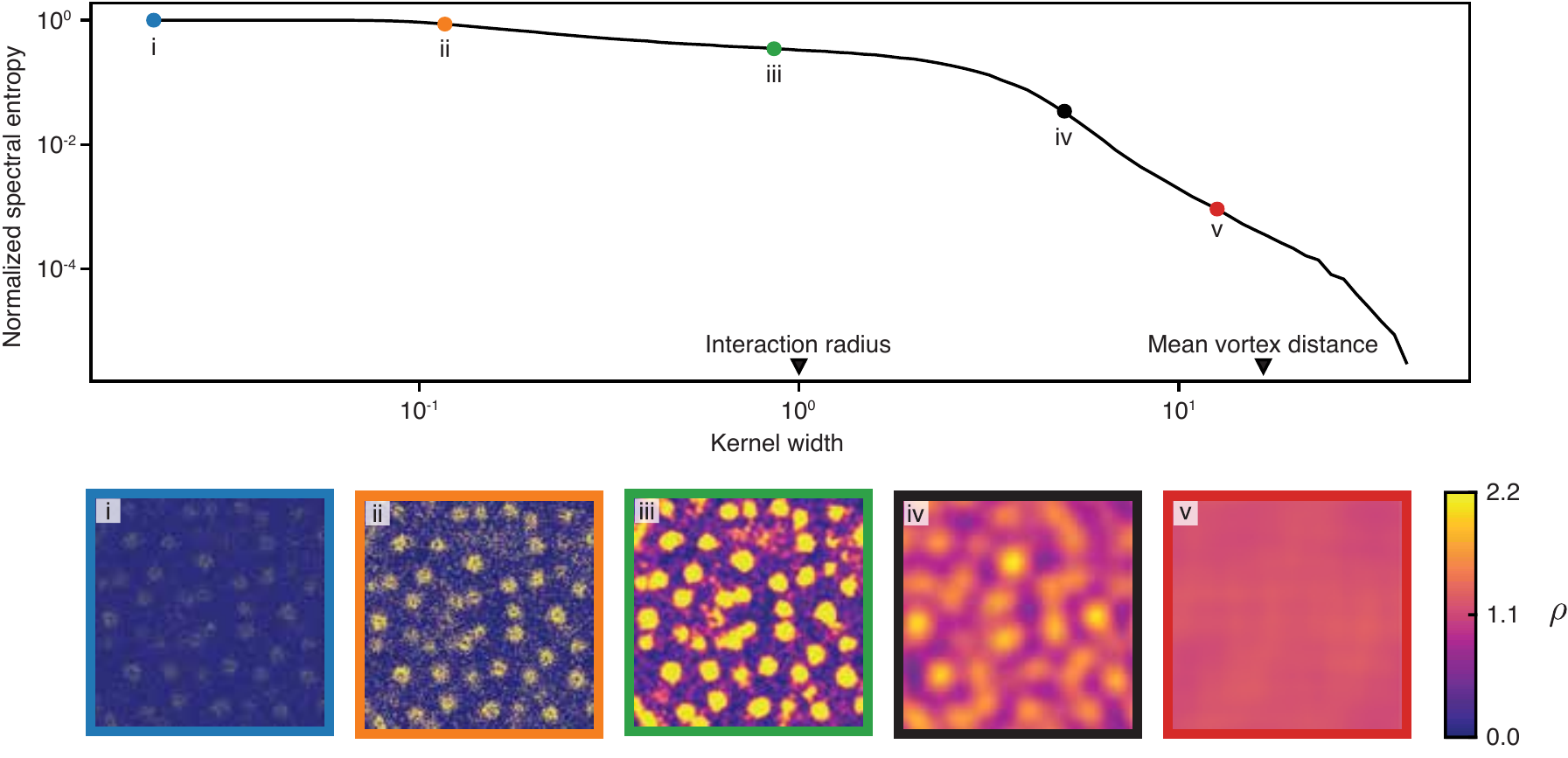}
    \markblue{
    \caption{
    Normalized spectral entropy, Eq.~(\ref{eq:SpecEntr}), as a function of the Gaussian kernel width~$\sigma$ (see SI~Sec.~\ref{App:CG_nearB}) for the chiral particle model data  (top) quantifies the fraction of information that remains in Fourier space after coarse graining. Representative snapshots of coarse-grained fields are shown in the bottom panels. Characteristic  scales in units of particle-particle interaction distance:  Median vortex distance $\sim 17$ (obtained from a Delaunay triangulation of density peaks), box size 100.
    (i,~$\sigma=0.02$):~Raw image before coarse graining. (i,~$\sigma=0.02$)--(ii,~$\sigma = 0.12$):~The discrete nature of the particle data remains present in the data, leading to little information loss. (ii,~$\sigma = 0.12$)--(iii,~$\sigma = 0.86$):~As the coarse-graining scale approaches the interaction length scale, $\sigma\rightarrow 1$,  coarse-grained data starts losing single particle information and vortices become more prominent. (iii,~$\sigma = 0.86$)--(iv,~$\sigma = 5$):~Vortices start to be smoothed out as $\sigma$ exceeds the particle interaction distance and vortex size. Data from (iv,~$\sigma=5$) was used for inferring a  continuum model from the chiral-particle simulation date; this choice of the coarse-graining scale  ensures that the hydrodynamic fields are sufficiently smooth while still containing sufficient information about density fluctuations and vortex patterns. (v, $\sigma = 12.6$):~When the kernel width $\sigma$ approaches the typical vortex-vortex distance, coarse-graining results in a constant homogeneous density and all spatially heterogeneous information is lost. 
    \label{Fig:SpecEntrVortex}}
    }
\end{figure}

\begin{figure}
    \centering
    \includegraphics[width=0.8\linewidth]{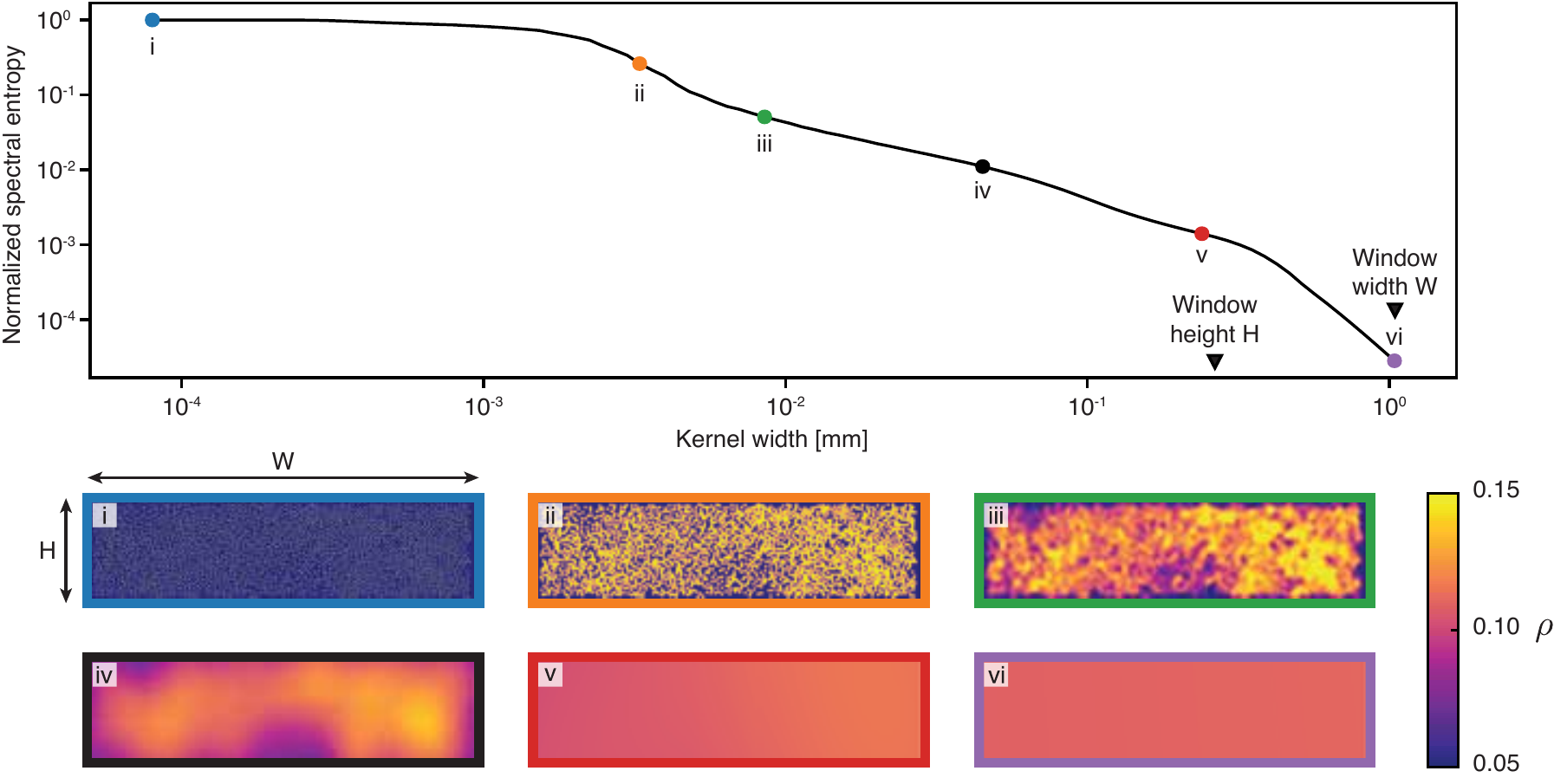}
    \markblue{
    \caption{
    Normalized spectral entropy, Eq.~(\ref{eq:SpecEntr}), as a function of the Gaussian kernel width~$\sigma$ (see SI~Sec.~\ref{App:CG_nearB}) for the Quincke roller system (top) quantifies the fraction of information that remains in Fourier space after coarse-graining. Representative snapshots of coarse-grained fields are shown in the bottom panels. Characteristic length scales: Roller diameter~$4.8\,\mu$m, mean roller-roller centroid distance $\sim 11\mu$m,  window height~$H=0.286$\,mm, window width~$W=1.146$\,mm.
    (i,~$\sigma=10^{-5}$\,mm):~Raw image before coarse graining. (ii,~$\sigma = 0.0033$\,mm)--(iii,~$\sigma = 0.0085$mm):~Single rollers are increasingly smoothed out, leading to an initial decrease in information. (iii,~$\sigma = 0.0085$\,mm)--(iv,~$\sigma=0.045$\,mm):~Large-scale density fluctuations become increasingly smoothed out by the coarse graining. Data from (iv,~$\sigma=0.045$\,mm) was used for model learning from experimental Quincke roller data, providing a compromise between sufficiently smooth data and well-resolved details of density fluctuation in both spatial directions. (v,~$\sigma = 0.24$\,mm):~As the coarse-graining scale $\sigma$ becomes comparable to the window height $H$,  density fluctuations in the vertical direction have been smoothed out leading to an effectively one-dimensional density pattern that varies only along the horizontal direction. (vi,~$\sigma = 1.04$\,mm):~As $\sigma$ becomes comparable to the window width $W$, all density variations disappear and the coarse-graining yields a constant homogeneous density. 
    \label{Fig:SpecEntrQuincke}
    }
    }
\end{figure}
}

\begin{figure}
    \centering
    \includegraphics[width=\linewidth]{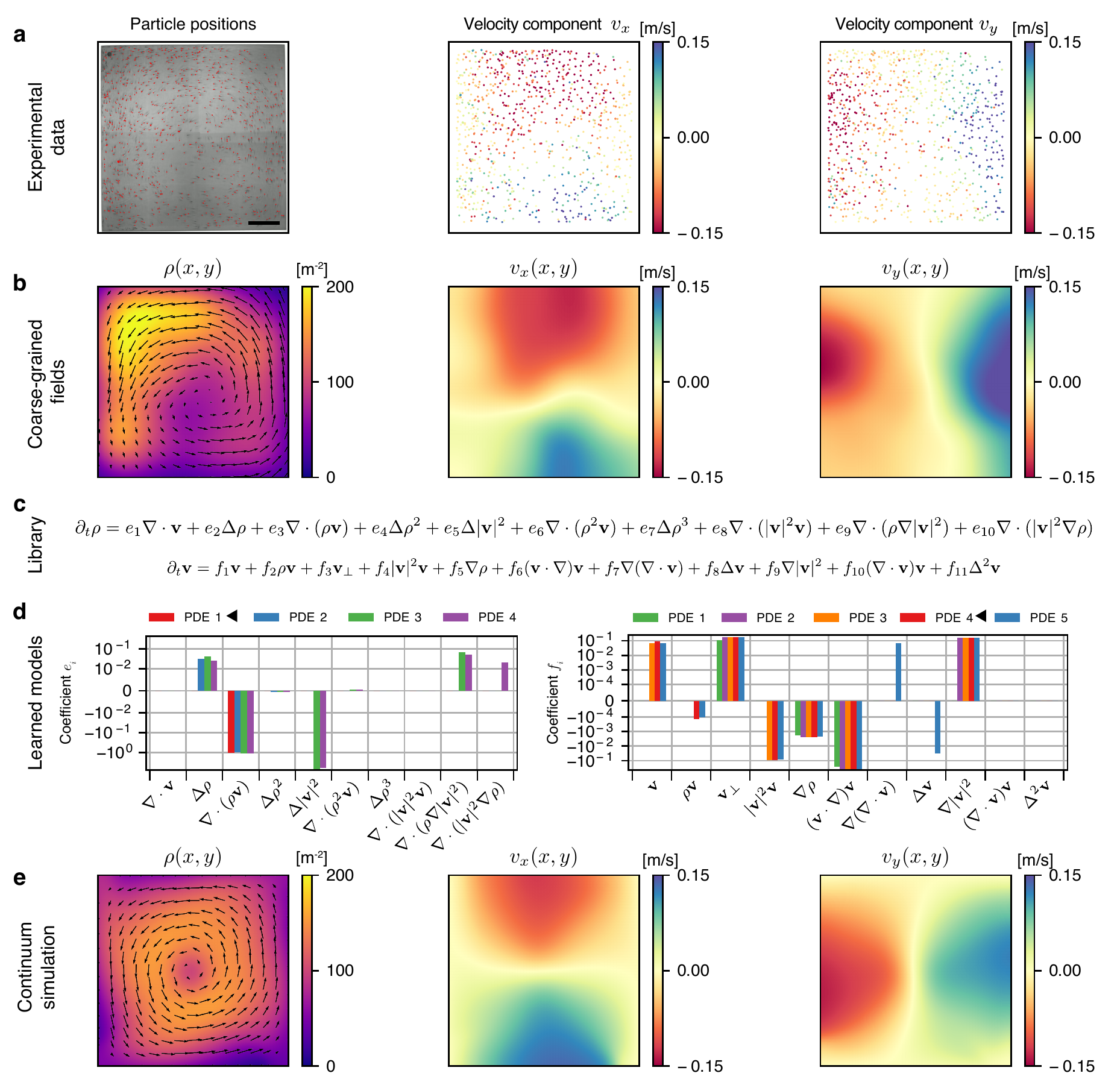}
    \markblue{
    \caption{Learning hydrodynamic equations for the collective motion of sunbleak fish. \textbf{a}, Identified locations and velocities of approximately 1024 fish from experiments in a quasi-two-dimensional tank~\cite{walter2021trex}. The fish are stimulated with a rotating pattern projected at the base of the tank, due to which they exhibit a counter-clockwise motion. Scale bar, $0.5$ m. \textbf{b}, Coarse-grained density $\rho(x,y)$ and velocity components, $v_x(x,y)$ and $v_y(x,y)$, at a representative time point. The coarse-graining width $\sigma$ is $0.3$m. \textbf{c}, Physics-informed libraries for the density and velocity dynamics. These are the same libraries as in Fig.~{\ref{Fig:Quincke}\textbf{c}} along with a $\mathbf{v}_\perp$ term in the velocity equation to take into account the external rotating stimulus provided to the fish by the rotating pattern. \textbf{d}, Learned phenomenological coefficients $e_{l}$ and $f_l$ of the four sparsest PDEs
    for the density (left) and velocity (right) dynamics. The coefficients are non-dimensionalized with length scale $\sigma$ and time scale $\sigma/v_0$, where $v_0 = 0.13$ m/s  is the average speed of the fish. \textbf{e}, Simulation snapshot at $t=500$ s of the learned hydrodynamic model (PDEs marked by $\blacktriangleleft$ in \textbf{d}) in a square domain with reflective boundary conditions. Starting from random initial conditions, this is the sparsest model for which spontaneous flow emerges and the flow settles into an anti-clockwise vortex as seen in the coarse-grained data. Furthermore, the magnitudes of the velocity components agree well with the coarse-grained data in \textbf{b}. \label{Fig:FishLearning}}
    }
\end{figure}

\begin{figure}
    \centering
    \includegraphics[width=0.8\textwidth]{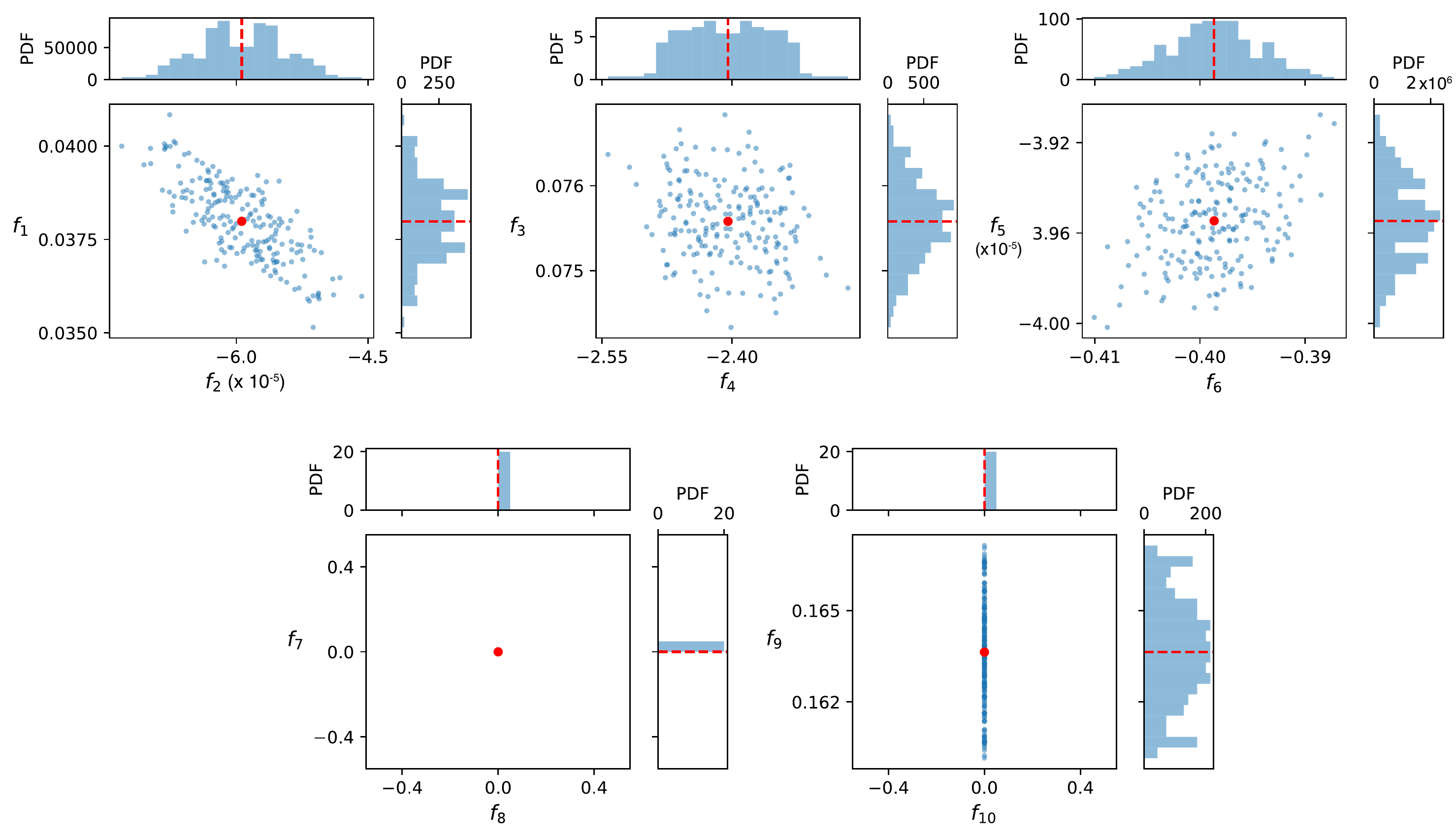}
    \markblue{\caption{Parameters $f_l$ of the velocity dynamics~{(Fig.~\ref{Fig:FishLearning}c)} that are obtained from the least-squares method by setting the thresholded coefficients in PDE 4~{(Tab.~\ref{Tab:FishMomCoeffs})} to zero. The units of the coefficients are the same as in~{Tab.~\ref{Tab:FishMomCoeffs}}. A total of 200 points (blue) are presented from 200 sub-samples with $50$\% randomly chosen data points. The red points indicate the fitted coefficients on the entire data set, which are the same as for PDE~4 in Tab.~\ref{Tab:FishMomCoeffs}. The histograms indicate the marginal probability density functions (PDFs) of the corresponding coefficients. The coefficient $f_{11}$ (not shown) is thresholded out and set to zero. The fitted values and the standard deviations of the coefficients that remain in the PDE are~\hbox{(mean $\pm$ standard deviation):} $f_{1} = (3.80 \pm 0.10)\times 10^{-2}$,  $f_{2} = (-5.94 \pm 0.50)\times 10 ^{-5}$,  $f_{3} = (7.56 \pm 0.05)\times 10^{-2}$,  $f_{4} = -2.40 \pm 0.05$,  $f_{5} = (-3.95 \pm 0.02)\times 10^{-5}$,  $f_{6} = (-3.99 \pm 0.04)\times 10^{-1}$,  $f_{9} = (1.64 \pm 0.02)\times 10^{-1}$.
    \label{Fig:CoeffsBootstrapedFish}
    }}
\end{figure}

\clearpage
\markblue{
\section{\am{Learning hydrodynamic equations for the collective motion of sunbleak fish}}
\label{App:Fish}
To demonstrate the generalizability of the presented learning framework to other active matter systems, we considered experimental data for the collective motion of sunbleak fish in Ref.~\cite{walter2021trex}. These experiments were done in a quasi-two-dimensional tank in which the motion of $\sim$1024 fish was influenced by a rotating dotted pattern that was projected on the bottom of the tank. This results in the fish swimming in an anti-clockwise motion~{(Fig.~{\ref{Fig:FishLearning}})} that was recorded using overhead cameras.

\subsection{\am{Tracking of swimming trajectories}}
To track the fish motion between two consecutive video frames $I_n$ and $I_{n+1}$, we first detected feature points in $I_n$ using the \texttt{detectSURFFeatures} function in the MATLAB Computer Vision Toolbox, which implements the Speeded Up Robust Features (SURF) algorithm~\cite{bay2008speeded}. The parameters \texttt{MetricThreshold} (strongest feature threshold) and \texttt{NumScaleLevels} (number of scale levels per octave) in the SURF algorithm were set to be 1200 and 4 respectively. After getting the SURF features in $I_n$, we tracked all corresponding feature point pairs in the two frames $I_n$ and $I_{n+1}$ using the \texttt{vision.PointTracker} function in the MATLAB Computer Vision Toolbox, which implements the Kanade-Lucas-Tomasi (KLT) tracking algorithm~\cite{lucas1981iterative,tomasi1991detection}. The parameter \texttt{MaxBidirectionalError} (forward-backward error threshold) in the KLT algorithm was set to be 1. To further remove the effect of outliers on the subsequent computation, we discarded all feature point pairs with displacement being outside of 2 standard deviations of the mean displacement.

\subsection{\am{Data coarse-graining and model learning}}
Starting from the particle data, we applied the same procedure as for the Quincke roller data~{(Sec.~\ref{sec:AplExpData} and Fig.~\ref{Fig:Quincke})}. We considered the particle positions $\mathbf{x}_i(t)$ and velocities $\mathbf{v}_{i}(t)$~{(Fig.~\ref{Fig:FishLearning}a)} and applied kernel coarse-graining~{(Eqs~\eqref{Eq:CoarseGrain}; $\sigma = 0.3$\,m)} to obtain the density field $\rho$ and velocity field $\mathbf{v} = \mathbf{p}/\rho$~{(Fig.~\ref{Fig:FishLearning}b)}. These data were then projected onto Chebyshev basis functions in space and time, and a temporal mode cut-off $n>n_0$ was imposed to retain the hydrodynamically relevant time scales. Our goal was to learn equations of the form in Eq.~\eqref{Eqn:QuinckeModel} and similar physics-informed libraries as in Sec.~\ref{Sec:QuinckePhysLib} were considered. The rotating pattern stimulus was accounted for by including an additional term $\mathbf{v}_\perp = \boldsymbol{\epsilon}^\top \cdot \mathbf{v} = (-v_y, v_x)^\top$ in the velocity equation. The complete libraries are shown in Fig.~\ref{Fig:FishLearning}c. Application of the STLSQ algorithm along with the stability selection procedure~{(SI~Sec.~\ref{App:SparseReg})} resulted in hydrodynamic models that are summarized in Fig.~\ref{Fig:FishLearning} and Tabs.~\ref{Tab:FishDensityCoeffs} and \ref{Tab:FishMomCoeffs}. The sparsest density equation is $\partial_t \rho= e_3 \nabla \cdot (\rho \mathbf{v})$ with $e_3 \simeq -1$, which is the expected continuity equation for the system. Along with this density equation, we simulated all the PDEs for the velocity equation starting from random initial conditions. These simulations were performed in a closed square domain with reflective boundary conditions to mimic the experimental tank with side walls~{(see SI Sec.~\ref{App:ContinuumSim})}. We found that PDE 4~{(marked by $\blacktriangleleft$ in Fig.~\ref{Fig:FishLearning})} was the sparsest velocity equation that resulted in a spontaneously formed anti-clockwise vortex as seen in the coarse-grained data~{(Fig.~\ref{Fig:FishLearning})}. Furthermore, 100\% of the subsamples at the corresponding thresholding parameter $\tau$ result in the same terms as in PDE 4. The uncertainties in the parameters of this PDE are quantified by a bootstrapping procedure in Fig.~\ref{Fig:CoeffsBootstrapedFish}.
}

\begin{table*}[hb]
\markblue{\caption{Phenomenological parameters $e_{l}$ for the density equation learned from the experimental data for the collective motion of sunbleak fish~{(Fig.~\ref{Fig:FishLearning})}. The dimensions of the coefficients are such that $[\mathbf{v}] =$ m/s and $[\rho] = $ m$^{-2}$. \label{Tab:FishDensityCoeffs}}}
\begin{ruledtabular}
\markblue{
\begin{tabular}{lr|cccc}
Term& Unit&PDE 1$\blacktriangleleft$&PDE 2&PDE 3&PDE 4\\\hline$e_{1}$$\nabla \cdot \mathbf{v}$&m$^{-2}$&--&--&--&--\\$e_{2}$$\Delta \rho$&m$^{2}$ s$^{-1}$&--&\hphantom{-}1.26$\times$10$^{-3}$&\hphantom{-}1.69$\times$10$^{-3}$&\hphantom{-}9.87$\times$10$^{-4}$\\$e_{3}$$\nabla \cdot (\rho \mathbf{v})$&--&-1.02$\times$10$^{0}$&-1.01$\times$10$^{0}$&-1.13$\times$10$^{0}$&-1.14$\times$10$^{0}$\\$e_{4}$$\Delta \rho^2$&m$^{4}$ s$^{-1}$&--&-1.57$\times$10$^{-5}$&-1.81$\times$10$^{-5}$&-1.66$\times$10$^{-5}$\\$e_{5}$$\Delta \vert \mathbf{v} \vert^2$&m$^{-2}$ s&--&--&-1.68$\times$10$^{1}$&-1.37$\times$10$^{1}$\\$e_{6}$$\nabla \cdot (\rho^2 \mathbf{v})$&m$^2$&--&--&\hphantom{-}5.96$\times$10$^{-4}$&\hphantom{-}6.20$\times$10$^{-4}$\\$e_{7}$$\Delta \rho^3$&m$^{6}$ s$^{-1}$&--&\hphantom{-}3.77$\times$10$^{-8}$&\hphantom{-}4.10$\times$10$^{-8}$&\hphantom{-}3.79$\times$10$^{-8}$\\$e_{8}$$\nabla \cdot (\vert \mathbf{v}\vert^2 \mathbf{v})$&m$^{-4}$ s$^{2}$&--&--&--&--\\$e_{9}$$\nabla \cdot (\rho \nabla \vert \mathbf{v}\vert^2)$&s&--&--&\hphantom{-}1.62$\times$10$^{-1}$&\hphantom{-}1.23$\times$10$^{-1}$\\$e_{10}$$\nabla \cdot (\vert \mathbf{v}\vert^2 \nabla\rho)$&s&--&--&--&\hphantom{-}4.91$\times$10$^{-2}$\\
    \end{tabular}
    }
    \end{ruledtabular}
\end{table*}

\begin{table*}[hb]
\markblue{
\caption{Phenomenological parameters $f_{l}$ for the velocity equation learned from the experimental data for the collective motion of sunbleak fish~{(Fig.~\ref{Fig:FishLearning})}. Simulations the of velocity dynamics PDE~4~{($\blacktriangleleft$)} along with density dynamics PDE~1 in Tab.~\ref{Tab:FishDensityCoeffs} show the spontaneous formation of an anti-clockwise vortex~{(Fig.~\ref{Fig:FishLearning}e)}, recapitulating the pattern observed in the input data~{(Fig.~\ref{Fig:FishLearning}b)}. The dimensions of the coefficients are such that $[\mathbf{v}] =$ m/s and $[\rho] = $ m$^{-2}$. \label{Tab:FishMomCoeffs}}
\begin{ruledtabular}
\begin{tabular}{lr|ccccc}
Term& Unit&PDE 1&PDE 2&PDE 3&PDE 4$\blacktriangleleft$&PDE 5\\\hline$f_{1}$$\mathbf{v}$&s$^{-1}$&--&--&\hphantom{-}2.97$\times$10$^{-2}$&\hphantom{-}3.80$\times$10$^{-2}$&\hphantom{-}3.08$\times$10$^{-2}$\\$f_{2}$$\rho \mathbf{v}$&m$^{2}$ s$^{-1}$&--&--&--&-5.94$\times$10$^{-5}$&-4.41$\times$10$^{-5}$\\$f_{3}$$\mathbf{v}_{\perp}$&s$^{-1}$&\hphantom{-}5.00$\times$10$^{-2}$&\hphantom{-}7.49$\times$10$^{-2}$&\hphantom{-}7.55$\times$10$^{-2}$&\hphantom{-}7.56$\times$10$^{-2}$&\hphantom{-}7.57$\times$10$^{-2}$\\$f_{4}$$|\mathbf{v}|^2 \mathbf{v}$&m$^{-2}$ s&--&--&-2.28$\times$10$^{0}$&-2.40$\times$10$^{0}$&-2.17$\times$10$^{0}$\\$f_{5}$$\nabla \rho$&m$^4$ s$^{-2}$&-3.26$\times$10$^{-5}$&-3.98$\times$10$^{-5}$&-3.96$\times$10$^{-5}$&-3.95$\times$10$^{-5}$&-3.94$\times$10$^{-5}$\\$f_{6}$$(\mathbf{v}\cdot \nabla) \mathbf{v}$&--&-2.54$\times$10$^{-1}$&-3.88$\times$10$^{-1}$&-3.98$\times$10$^{-1}$&-3.99$\times$10$^{-1}$&-4.02$\times$10$^{-1}$\\$f_{7}$$\nabla (\nabla \cdot \mathbf{v})$&m$^{2}$ s$^{-1}$&--&--&--&--&\hphantom{-}2.60$\times$10$^{-3}$\\$f_{8}$$\Delta \mathbf{v}$&m$^{2}$ s$^{-1}$&--&--&--&--&-1.30$\times$10$^{-3}$\\$f_{9}$$\nabla (\vert \mathbf{v}\vert^2)$&--&--&\hphantom{-}1.69$\times$10$^{-1}$&\hphantom{-}1.64$\times$10$^{-1}$&\hphantom{-}1.64$\times$10$^{-1}$&\hphantom{-}1.66$\times$10$^{-1}$\\$f_{10}$$(\nabla \cdot \mathbf{v}) \mathbf{v}$&--&--&--&--&--&--\\$f_{11}$$\Delta^2 \mathbf{v}$&m$^{4}$ s$^{-1}$&--&--&--&--&--\\
\end{tabular}
\end{ruledtabular}
}
\end{table*}

\begin{figure}
    \centering
    \includegraphics[width=0.9\textwidth]{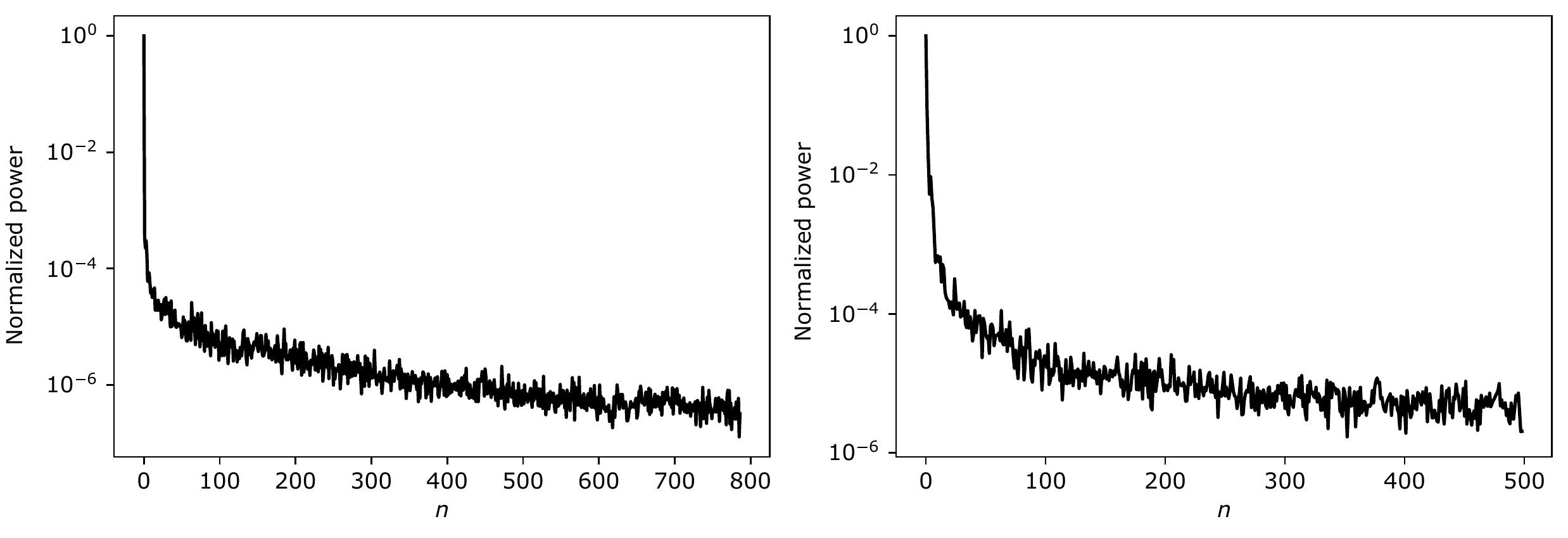}
    \markblue{\caption{Power spectra of coarse-grained data for the Quincke roller system (left; see Sec.~\ref{sec:AplExpData} in the main text) and for sunbleak fish (right; see SI~Sec.~\ref{App:Fish}), where $n$ denotes temporal Chebychev mode numbers. Similar to Fig.~\ref{Fig:MassCons}b (main text), we define the spatio-temporal power spectrum $S_{x;n,(\alpha,\beta)} = |\mathbf{e}_x \cdot \hat{\mathbf{v}}_{n, \alpha, \beta}|^2$, where $(\alpha, \beta)$ are the spatial Chebyshev mode numbers and the total spatial spectral power is given by $S_{x; n} = \sum_{\alpha, \beta}S_{x;n,(\alpha,\beta)}$. The normalized power shown in the two panels is given by $S_{x;n}/S_{x;n=0}$. For both the Quincke roller and the sunbleak fish data spectral powers decay exponentially with increasing temporal mode number $n$. The analog spatio-temporal power spectra $S_{y;n,(\alpha,\beta)}$ show similar behavior. Note, the use of a Chebyshev basis implies that the power spectrum $S_{x;n,(\alpha,\beta)}$ is related to real-space data by $\sum_{\alpha, \beta}\sum_nS_{x; n,(\alpha,\beta)}=\int dt d^2\mathbf{x}|v_x(t,\mathbf{x})|^2w_{\text{Ch}}(t)w_{\text{Ch}}(x) w_{\text{Ch}}(y)$, where $w_{\text{Ch}}$ is the Chebyshev weight function. \label{Fig:TempSpectraQuinckeFish}}}
\end{figure}

\markblue{
\section{\am{Temporal spectra of the coarse-grained data for the Quincke roller system and sunbleak fish}}
\label{App:TemporalSpectra}

The coarse-grained density $\rho(t,\mathbf{x})$ and velocity $\mathbf{v}(t,\mathbf{x})$ fields are projected onto Chebyshev basis functions in both space and time~(see Sec.~\ref{sec:QuinckeCoarseGraining}). To quantify how the energy decays with increasing frequency, we plot the summation of the squared mode amplitudes for each temporal mode in Fig.~\ref{Fig:TempSpectraQuinckeFish} to obtain a power spectrum similar to Fig.~\ref{Fig:MassCons}b. In all the three cases, we observe an exponential decay of the energy spectra with increasing frequency. This suggests that  significant fluctuations (\lq noise\rq) are absent in the coarse-grained field dynamics, typically visible in terms of slowly (algebraically) decaying temporal frequency spectra. Thus, our approach to model the dynamics using deterministic (instead of stochastic) partial differential equations is justified. 
}

\clearpage 
\twocolumngrid

\typeout{}
\bibliography{references}

\end{document}